\documentclass[fontsize = 12]{scrartcl}
\KOMAoption{listof}{totoc}
\DeclareNewTOC[%
  owner=\jobname,
  listname={\contentsname},
  setup=leveldown
]{atoc}

\makeatletter
\newcommand*{\useatocs}{%
  \renewcommand*{\ext@toc}{atoc}%
  \scr@ifundefinedorrelax{hypersetup}{}{% maybe you use hyperref and bookmarks
    \hypersetup{bookmarkstype=atoc}%
  }%
}
\makeatother 

\makeatletter
\renewcommand\maketitle{%
  \begingroup
  \begin{center}
    {\large\bfseries\@title\par\vspace{0.5em}}% Title
    {\normalsize\scshape\@author\par}% Author
    {\normalsize\@date\par}% Date
  \end{center}%
  \par\@thanks % <-- print the stored \thanks footnotes
  \endgroup
  \global\let\@thanks\@empty
}
\makeatother
\makeatletter
\renewenvironment{abstract}{%
  \small
  \begin{center}
    \bfseries \abstractname\vspace{-.5em}\vspace{0pt}
  \end{center}
  \list{}{%
    \setlength{\leftmargin}{1em}%  <— left indent
    \setlength{\rightmargin}{1em}% <— right indent
  }%
  \item\relax
}{%
  \endlist
}
\makeatother

\usepackage[style = authoryear, backend = biber, url = false, doi = false, isbn = false, eprint = false, maxcitenames=99, maxbibnames = 99, uniquename = false]{biblatex} 

\usepackage{microtype}
\usepackage{setspace}
\usepackage{amsmath, amssymb, mathtools}
\usepackage[dvipsnames]{xcolor}
\usepackage{enumitem}
\usepackage{booktabs, makecell, longtable, lscape}
\usepackage{tcolorbox}
\usepackage{tikz}
\usepackage{pgfplots}
\usepackage{pifont}
\usepackage{caption}
\usepackage[standard, amsmath]{ntheorem}
\usepackage{apptools}
\usepackage{float}
\usepackage{adjustbox}

\usetikzlibrary{arrows}
\usetikzlibrary{shapes}
\usetikzlibrary{positioning}
\pgfplotsset{compat=1.18}

\newcounter{algorithm}
\newenvironment{algorithm}[1][]{
	\refstepcounter{algorithm} \par\medskip \noindent 
	\begin{minipage}{\textwidth}
\rule{\linewidth}{1.0pt}\par
  \noindent \textbf{Algorithm~\thealgorithm: #1} 
  \par\vskip -1.5ex
  \noindent\rule{\linewidth}{0.2pt}
  \par\vskip -1.25ex
  \rmfamily}{\par\vskip -2.25ex\noindent\rule{\linewidth}{0.2pt}
  \medskip
\end{minipage}
\noindent}

\newtheorem{Assumption}{Assumption}
\AtAppendix{\setcounter{page}{1}}
\AtAppendix{\setcounter{figure}{0}}
\AtAppendix{\setcounter{table}{0}}
\AtAppendix{}
\AtAppendix{}
\AtAppendix{\renewtheorem{Lemma}{Lemma}[section]}
\AtAppendix{\singlespacing}

\DeclarePairedDelimiter\norm\lVert\rVert
\DeclarePairedDelimiter\abs\lvert\rvert

\newcommand{\E}{\mathbb{E}}

\DeclareMathOperator{\clo}{cl}
\DeclareMathOperator{\diag}{diag}

\DeclareMathOperator{\sign}{sign}

\title{Inference on effect size after multiple hypothesis testing}%\thanks{We are grateful for comments from Erik Hjalmarsson, Mikael Lindahl and participants at CFE-CMStatistics 2022, the 4th Aarhus Workshop in Econometrics, IPDC2025 in Montpellier, and Econometrics Forum 2025 in Awara. Dzemski acknowledges financial support from Jan Wallanders och Tom Hedelius stiftelse samt Tore Browaldhs stiftelse under grant P24-0135. Okui acknowledges financial support from the Japan Society for the Promotion of Science under KAKENHI Grant Nos.23K25501. Wang acknowledges financial support from the Singapore Ministry of Education Tier 1 grants RG104/21, RG51/24, and NTU CoHASS Research Support Grant. The authors used Grammarly to improve the readability and language of the manuscript. After using this service, we reviewed and edited the content as needed and take full responsibility for the content of the published article. All errors are our own.}}

\author{Andreas Dzemski\thanks{Department of Economics, University of Gothenburg, P.O. Box 640, SE-405 30, Gothenburg, Sweden. Email: andreas.dzemski@economics.gu.se}, Ryo Okui\thanks{Faculty of Economics, the University of Tokyo, Bunkyo-ku, Tokyo 113-0033, Japan. Email: okuiryo@e.u-tokyo.ac.jp}, and Wenjie Wang\thanks{Division of Economics, School of Social Sciences, Nanyang Technological University. HSS-04-65, 14 Nanyang Drive, Singapore 637332. Email: wang.wj@ntu.edu.sg.}}

\date{\today}

\begin{document}
\maketitle

\begin{abstract}
	% 166 words right now
	Significant treatment effects are often emphasised when interpreting and summarising empirical findings in studies that estimate multiple, possibly many, treatment effects. Under this kind of selective reporting, conventional treatment effect estimates may be biased, and their corresponding confidence intervals may undercover the true effect sizes. We propose new estimators and confidence intervals that provide valid inferences on the effect sizes of the significant effects after multiple hypothesis testing. Our methods are based on the principle of  selective conditional inference and complement a wide range of tests, including step-up tests and bootstrap-based step-down tests. Our approach is scalable, allowing us to study an application with over 370 estimated effects. 
	We justify our procedure for asymptotically normal treatment effect estimators.	
	We provide two empirical examples that demonstrate bias correction and confidence interval adjustments for significant effects. The magnitude and direction of the bias correction depend on the correlation structure of the estimated effects and whether the interpretation of the significant effects depends on the (in)significance of other effects.
	\par\vspace{0.5em}\noindent
\end{abstract}

\section{Introduction}

Many empirical studies across fields --- from public health and education to psychology and economics --- estimate multiple effects of interventions or treatments, whether by examining different outcome measures, types of interventions, or population subgroups \parencite{list2019multiple}. Estimating multiple treatment effects allows researchers to validate their methodological approach, understand how effects vary across contexts, and identify the underlying mechanisms at work.
These estimated effects are typically categorised as either statistically significant or insignificant using multiple hypothesis testing procedures. The significant effects are often interpreted as evidence for effective interventions, making them the primary focus of subsequent analysis. Researchers are particularly interested in their practical importance, as determined by the magnitude or size of these effects (for example, economists are concerned about an effect's ``economic significance'' and medical researchers about its ``clinical significance'').\footnote{See, for example, the special issue in the Journal of Socio-Economics, Volume 44, Issue 5 for the discussions regarding the difference between statistical and economic significance. In particular, see \textcite{ziliak2004size} and the comment articles on it.} However, when attention is restricted to the significant effects, their estimated magnitudes are subject to selection bias, which invalidates conventional statistical inference methods. 

This paper proposes new methods that provide valid statistical inference for significant effects by explicitly accounting for their selection based on statistical significance after multiple hypothesis testing. 
To determine significance, we consider a broad class of step-down and step-up multiple testing procedures, including classical methods such as those by \textcite{holm1979simple, benjamini1995controlling,benjamini2001control} and more powerful modern bootstrap-based methods such as \textcite{romano2005stepwise,list2019multiple}.
The proposed methods can be applied if the underlying treatment effect estimators are asymptotically jointly Gaussian with a consistently estimable covariance matrix. Given these mild assumptions, our methods apply broadly, including to standard causal inference designs with observational or experimental data.

We use the term ``selective reporting of significant results'' to refer to the common practice of emphasising statistically significant estimates. In particular, summaries of research findings in the abstract or introduction of a paper typically highlight the significant results. By contrast, insignificant results are relegated to less prominent parts of the paper, such as passing mentions in the empirical section or in the appendix.\footnote{%
For example, the introduction in \textcite{karlan2007does} --- on which one of our empirical illustrations builds --- summarises their subgroup analysis by reporting that revenue increased by 55\% in ``red'' states, while there was ``little effect'' in ``blue'' states. Their introduction thus highlights the subgroup with a significant effect.}
Further selection arises when the interpretation of a significant effect changes depending on the significance of other effects.%
\footnote{For instance, we may not interpret an effect as causal if we find the effect on a placebo outcome to be significant.}
One might argue that the solution is to avoid selective summaries and interpretations altogether. Yet this does not resolve the problem; it merely shifts the burden of synthesising the results onto the reader, who then faces a selective inference problem of their own.

Our new point estimators and confidence intervals are grounded in the principle of conditional selective inference \parencite{fithian_optimal_2017, lee2016exact}.
Specifically, we account for the fact that the same data are used for determining significance and estimating effect sizes by conditioning on a ``selection event'' that describes the  ``observed significance.'' Thereby, we control statistical error rates ``on average'' across all possible versions of a study that analyse the same effects based on observed significance.

Our conditioning approach allows researchers to apply standard multiple testing protocols based on the full sample to determine significance. In contrast, recently proposed selective inference methods introduce randomisation into the selection stage to facilitate post-selection inference \parencite{fithian_optimal_2017, tian_selective_2018, leiner2025data}. In the context of selection by significance, this reduces the power of the significance test, akin to using a smaller sample.\footnote{In data carving \parencite{fithian_optimal_2017}, selection is based on a split sample. Data fission \parencite{leiner2025data} and randomised response selection \parencite{tian_selective_2018} add noise at the selection stage.} Our approach is advantageous when establishing significance is the primary research goal and reporting effect sizes is of secondary concern. It also admits retrospective corrections of selection bias in existing studies that have already exhausted the full data for significance testing. Our methods can be applied ``out of the box'' to empirical analyses that already use multiple hypothesis testing. Randomisation-based methods require more careful planning, possibly requiring a pre-analysis plan that specifies the randomisation seed to safeguard against the randomness of selection results. 

A key feature of our approach is that it scales well with the number of treatment effects, greatly expanding the range of possible applications. In one of our data examples, we estimate fund-specific ``alphas'' for outperforming funds identified through multiple hypothesis testing among more than three hundred mutual funds. Other methods of post-selection inference are computationally infeasible at this scale; for example, \textcite{berk_valid_2013} state that their method can scale only to a dimension of ``20, or slightly larger'' (p. 16). Even alternative conditional inference methods, such as data carving \parencite{fithian_optimal_2017} and selection using randomised responses \parencite{tian_selective_2018}, face numerical difficulties when the number of hypotheses at the selection stage is large, since they carry out a complex (numerical) integration over randomised selection events.

We achieve scalability through two key ingredients: additional conditioning on a nuisance parameter and an algorithm that, for $m$ treatment effects, computes the conditional support in $\mathcal{O}(m^3 \log m)$ time. 
Developing this algorithm is a central technical contribution of the paper. Unlike in other selective inference settings \parencite[e.g.,][]{lee2016exact, tibshirani_exact_2016}, our problem admits no simple closed-form expression for the conditional support. The underlying difficulty is the complex geometry of the selection event. Step-up and step-down testing proceed sequentially, and the significance thresholds change depending on which effects have been found significant so far.

% In other settings of selective inference, the conditional support can often be computed from a closed-form expression \parencite[e.g.,][]{lee2016exact, sarfati2025post}. Such a closed-form computation is not tractable in the multiple testing setting due to the geometric complexity of the selection event. This complexity arises because, due to the stagewise nature of step-up and step-down testing, the threshold of significance for one effect can depend on whether other effects are significant. 
% The numerical challenges also appear with alternative conditional inference methods, such as data carving \parencite{fithian_optimal_2017} and selection using randomised responses \parencite{tian_selective_2018}. Specifically, these approaches require integrating the randomised component of the selection, which becomes complicated when dealing with many hypotheses at the selection stage. 

We prove the uniform asymptotic validity of our procedures under the assumption that the estimated treatment effects uniformly converge to a joint Gaussian distribution as the sample size approaches infinity. The uniformity of our result is important because it allows for local asymptotics where some treatment effects are on the margin of significance even in large samples. 
This ensures that our asymptotic result provides a relevant approximation of finite samples where we see relatively small $t$-statistics.

In Monte Carlo simulations, we demonstrate the validity of our procedures in finite samples. Even in non-Gaussian settings, our methods capture selection bias and yield confidence intervals with adequate (conditional) coverage. By contrast, conventional confidence intervals that ignore the selection problem exhibit significant undercoverage. Bonferroni-corrected simultaneous confidence intervals undercover when significance is ``marginal'' and are overly conservative when effects are highly significant.

To demonstrate the behavior of our methods when applied to real data, we provide two empirical applications. We find that the direction and magnitude of the bias correction is not easily anticipated, which highlights the relevance of a formal statistical approach to address the selection problem.
For marginally significant effects, our post-selection inference tends to correct the naive effect size estimates downward, potentially addressing concerns about inflated effect sizes in low-powered studies \parencite{ioannidis2017power,kvarven2020comparing}.

Our first application revisits the data from \textcite{karlan2007does} on the effects of a matching grant on charitable donations. We build on the earlier analysis by \textcite{list2019multiple}, who applied multiple testing procedures to identify significant treatments. Our methods provide inference on the effect sizes of these significant effects.

In our second application, we estimate mutual fund ``alphas,'' which measure performance beyond systematic risk factors. Using multiple testing, we identify funds with significantly positive alphas---commonly viewed as market outperformers---from a sample of more than 370 funds, highlighting that our method scales to settings with many estimated effects. We show that, even after accounting for selection based on significance, some outperformers still beat the market by an economically meaningful margin.

\paragraph*{Relation to the literature}

This paper is part of the selective inference literature.
This rapidly growing literature is reviewed in \textcite{zhang2022post}. A selection of relevant studies includes \textcite{berk_valid_2013, lockhart_significance_2014, fithian_optimal_2017, tibshirani_exact_2016, lee2016exact, bachoc_valid_2019, watanabe2021, jewell_testing_2022, terada2023, duy2023,sarfati2025post}.

In this paper, we adopt the conditional inference approach, which controls statistical error under a conditional distribution. Alternative approaches target the unconditional distribution \parencite{benjamini2005false,andrews_inference_2024,mccloskey2024hybrid, berk_valid_2013, zrnic2024locally}.
% Examples include the selective confidence intervals of \textcite{benjamini2005false}, the hybrid confidence intervals of \textcite{andrews2019identification,mccloskey2024hybrid}, and methods based on simultaneous inference \parencite{berk_valid_2013, zrnic2024locally}. 

The conditional inference approach is a natural fit for our problem since it allows us to control statistical error conditional on the observed significance. The observed significance provides the context in which the estimated effects are interpreted and determines the research questions that are emphasised when summarising the results. 
Inference based on the unconditional distribution averages statistical error over different contexts that focus on different research questions and hence attract different readers. \textcite{fithian_optimal_2017} observe that ``averaging our error rates across [\dots] two questions with two different interpretations, seems inappropriate'' (p. 31). 

% Second, the conditional inference approach --- particularly the polyhedral method by \textcite{lee2016exact}--- allows for scalable algorithms, overcoming the drawback of, for example, \textcite{berk_valid_2013} that we discussed above. A major technical contribution of this paper is the development of such an algorithm. 

Moreover, the conditional inference approach produces point estimates that capture selection bias. The unconditional approach proposed by \textcite{benjamini2005false} addresses only uncertainty, not bias. Therefore, their approach offers methods ``for confidence intervals but none for point estimation'' \parencite[][p. 714]{benjamini2010simultaneous}.

Targeting the unconditional distribution can increase power \parencite{goeman2024selection, zrnic2024locally}, but dismisses the important role of the selection event in the interpretation of the estimated effects. Additionally, we empirically show that conditional inference can yield tighter confidence intervals than simultaneous inference, demonstrating that the potential power gains from unconditional inference are not uniform. 

We use the ``polyhedral'' conditional inference method of \textcite{lee2016exact} as the basis for our procedures. Recently introduced randomisation-based methods \parencite{fithian_optimal_2017, tian_selective_2018, leiner2025data} provide alternative ways to carry out conditional selective inference and address some limitations of the polyhedral approach \parencite{kivaranovic2021, kivaranovic2024}. In addition, the data fusion approach of \textcite{leiner2025data} can even be applied in some non-Gaussian settings such as testing Poisson data. Randomisation methods rely on specific modifications at the selection stage, ruling out using out-of-the-box multiple testing procedures. 

% Recently, selective inference methods with randomization, such as Data carving \parencite{fithian_optimal_2017} and selection with randomised response \parencite{tian_selective_2018}, have gained popularity. They can address some limitations \parencite{kivaranovic2021, kivaranovic2024}. However, they encounter numerical issues as discussed earlier. More fundamentally, these methods require specific modifications at the selection stage and therefore depart from standard multiple-testing procedures. The data fission method \parencite{leiner2025data} can resolve this numerical issue, but still modifies the selection stage. Notably, they cannot be used to correct for selection bias in existing studies that rely on conventional significance testing. 

%There are alternative conditional inference procedures, such as data carving \parencite{fithian_optimal_2017} and selection with randomised response \parencite{tian_selective_2018}. These methods address some limitations of the polyhedral approach \parencite{kivaranovic2021, kivaranovic2024}, but they require specific modifications at the selection stage and therefore depart from standard multiple-testing procedures. They also suffer from numerical issues in the presense of many hypotheses in the selection stage. This numerical issue has been overcome by the data fission method in \textcite{leiner2025data}. In particular, they cannot be used to correct for selection bias in existing studies that rely on conventional significance testing. 

\paragraph*{Organisation of the paper}
Section~\ref{sec:setup} introduces our new procedures in a simplified setting. Section~\ref{sec:univariate example} provides some intuition based on a univariate example. Section~\ref{sec:what is significance} defines the multiple testing procedures covered by our methods. Section~\ref{sec:algorithm conditional support} describes the computational algorithm that makes our methods scalable. Section~\ref{sec:discussion} discusses some properties of our procedures and extends our procedures to more general settings. Section~\ref{sec:asymptotics} provides an asymptotic justification of our procedures. Simulation results are included in Section~\ref{sec:simulation}. Sections~\ref{sec:application list} and \ref{sec:application mutual funds} provide our empirical applications. Section~\ref{sec:conclusion} concludes. All proofs are in the online appendix.

\section{\label{sec:setup}Setup and procedures}

Our goal is to make inferences on the parameters that we find significant among multiple, possibly many, estimated parameters. This section defines our formal setting and introduces our proposed procedures.

\subsection{\label{sec:setting}Treatment effect parameters and significance testing}

We begin by outlining the context of our study. To introduce the method, we assume for now that the parameter estimator is normally distributed with a known variance-covariance matrix. We identify significant estimates based on $t$-statistics.

Let $\theta = (\theta_1, \dotsc, \theta_m)'$ denote a vector of parameters. 
For concreteness, we consider the parameters in $\theta$ to be treatment effects corresponding to $m$ specifications, which may involve different treatments, outcome variables, and/or subpopulations. 
% However, our approach is more general, and our procedures can be applied to any estimated parameters that meet our assumptions. 

We assume that there exists an estimator $\hat{\theta}$ for $\theta$ that is normally distributed with mean $\theta$ and known variance-covariance matrix $V$. We subsequently relax these assumptions and allow unknown $V$ (Section~\ref{sec:relax assumptions}) and settings where Gaussianity holds only asymptotically (Section~\ref{sec:asymptotics}).

Significance is determined based on the vector of $t$-statistics for the estimated effects, given by $X = \diag^{-1/2} (\mathbf{v}) \hat{\theta}$, where $\mathbf{v}$ is the diagonal of $V$. Note that $X\sim N(\diag^{-1/2} (\mathbf{v})\theta, \Omega)$, where
% \begin{align*}
	 $\Omega = \diag^{-1/2} (\mathbf{v}) V \diag^{-1/2} (\mathbf{v})$
% \end{align*}
is the correlation matrix of $\hat \theta$.

We assume that  the set of significant effects, denoted as $\hat{S} \subseteq \{1, \dotsc, m\}$, is determined by a step-down or step-up procedure that sequentially tests the significance of the estimated effects. A treatment effect $h$ is deemed significant and included in $\hat{S}$ if $\lvert X_h\rvert \geq \bar{x}_h$ (two-sided testing) or $X_h \geq \bar{x}_h$ (one-sided testing). The effect-specific critical values $\bar{x}_h$ depend on the data. We fully define the procedures covered by our methods in Section~\ref{sec:what is significance}. 

\subsection{\label{sec:summary} Conditional inference on significant effects}

Our approach follows the polyhedral method of \textcite{lee2016exact} and inverts a distribution function that conditions on significance and an additional nuisance parameter.

For a set $S$ of significant effects and a significant effect $s \in S$, we decompose $X$ as  
\begin{align*}
 X = \Omega_{\bullet, s} X_s + Z^{(s)},
\end{align*}
where $Z^{(s)} = X - \Omega_{\bullet, s} X_s$ and $\Omega_{\bullet, s}$ is the $s$-th column of $\Omega$.
By a property of the Gaussian distribution, $Z^{(s)}$ is independent of $X_s$, implying that the distribution of $X_s$ given $Z^{(s)}$ depends only on $\theta_s$ and not on the other elements of $\theta$. 

A selection event can be represented as a subset of the support of the $t$-statistics $X$. Specifically, we derive the set $\mathcal{X} (S) \subset \mathbb{R}^m$ such that $\hat{S} = S$ if and only if $X \in \mathcal{X} (S)$. As we show below, $\mathcal{X}(S)$ has a complex geometry and can be costly to compute except when $m$ is very small. We circumvent direct evaluation of $\mathcal{X}(S)$ by conditioning on $Z^{(s)} = z$. In particular, we observe the significant effects $S$ if $X_s$ is contained in the (marginal) conditional support $\mathcal{X}_s (z, S)$, where
\begin{align}
\label{eq:general definition marginal support}
 \mathcal{X}_s (z, S) = \left\{x \in \mathbb{R}: \Omega_{\bullet, s} x + z \in \mathcal{X} (S) \right\}.
\end{align}
Below, we develop an efficient algorithm that computes $\mathcal{X}_s(z, S)$ in polynomial time.

Let $F_s(x_s \mid z, \theta_s, S)$ denote the distribution function of $X_s$ conditional on $Z^{(s)} = z$, and $\hat{S} = S$, given by $F_s(x_s \mid z, \theta_s, S) = F_s(x_s \mid z, \theta_s, V, \bar{x}, S)$, where
\begin{align*}
    F_s(x_s \mid z, \theta_s, V, \bar{x}, S)
	% \\
	= \frac{\int_{\{\xi \in \mathbb{R}: \xi + V_{s, s}^{-1/2} \theta_s \in \mathcal{X}_s (z, S)\}} \mathbf{1} \left\{ \xi + V_{s, s}^{-1/2} \theta_s \leq x_s \right\} \, d\Phi(\xi)}{\int_{\{\xi \in \mathbb{R}: \xi + V_{s, s}^{-1/2} \theta_s \in \mathcal{X}_s (z, S)\}} \, d\Phi(\xi)}
\end{align*}
and $\Phi$ is the standard normal distribution function. This distribution is a truncated Gaussian distribution. Evaluating $F_s$ at the observed $x_s = X_s$, $z = Z^{(s)}$ and $S = \hat{S}$ gives the \textcite{rosenblatt1952remarks} transformation with the key property that 
\begin{align*}
	F_s (X_s \mid Z^{(s)}, \theta_s, S) \mid \left\{\hat{S} = S \right\} \sim \mathcal{U}[0, 1],
\end{align*}
where $\mathcal{U}[0, 1]$ denotes the uniform distribution on the unit interval.\footnote{We formally derive this property in Lemma~\ref{lem:rosenblatt_uniform_distribution} in Online Appendix \ref{sec:additional theory}.}

For $p \in (0, 1)$, let $\tilde{\theta}^{(p)}_s$ be defined as the unique solution to 
\begin{align}
	\label{eq:Rosenblatt equal to p}
	 F_s \left(X_s \mid Z^{(s)}, \tilde{\theta}^{(p)}_s, S\right) = p.
	\end{align}
Uniqueness of the solution is established in Lemma~\ref{lem:Rosenblatt_invertibility} in Online Appendix~\ref{sec:additional theory}. 
Our conditionally median-unbiased estimator is given by $\tilde{\theta}^{\text{ub}}_s = \tilde{\theta}^{(0.5)}_s$. Our conditional confidence interval with nominal coverage $1 - \alpha$ is given by $\left[ \tilde{\theta}^{(1 - \alpha/2)}_s, \tilde{\theta}^{(\alpha/2)}_s \right]$.

The conditionally median-unbiased estimator over- and underestimates the true effect $\theta_s$ with equal probability conditional on the selection event, as established in the following result.
\begin{theorem}[Median-unbiasedness]
	\label{thm:validity median-unbiased oracle}
    Suppose that $P(\hat{S} = S) > 0$ and $s \in S$. Then,
    \begin{align*}
        P \left( \tilde{\theta}^{\text{ub}}_s \geq \theta_s \mid \hat{S} = S \right) = 
        P \left( \tilde{\theta}^{\text{ub}}_s \leq \theta_s \mid \hat{S} = S \right) = 0.5.
    \end{align*}
\end{theorem}
The next result establishes that the conditional confidence interval covers the true effect $\theta_s$ with probability $1 - \alpha$ conditional on the selection event.
\begin{theorem}[Confidence set validity]
	\label{thm:validity CS oracle}
    Suppose that $P(\hat{S} = S) > 0$ and $s \in S$. Then,
    \begin{align*}
		P \left( \theta_s \in \text{CCI}_{\alpha}(\theta_s \mid S) \mid \hat{S} = S \right) = 1 - \alpha.
    \end{align*}
\end{theorem}
The conventional unconditional equal-tailed confidence interval for $\theta_s$, denoted $\text{CI}_\alpha(\theta_s)$, is given by $\hat{\theta}_s \pm \sqrt{\mathbf{v}_s} \Phi^{-1} \left(1 - \alpha/2 \right)$, where $\Phi^{-1}$ is the standard normal quantile function. Other selective confidence intervals modify the unconditional interval symmetrically by adjusting its width \parencite[e.g.][]{benjamini2005false,berk_valid_2013}. In contrast, our adjustment is asymmetric, allowing the confidence interval to reflect selection bias (see, e.g., Table~\ref{tab:LSXtab2} below).

We can use Bonferroni-correction to construct a joint conditional confidence set for all significant treatment effects. The previous theorem guarantees that 
\begin{align}
	\label{eq:joint confidence set}
	\left(\theta_s\right)_{s \in S} \in \bigtimes_{s \in S} \text{CCI}_{\alpha/\lvert S \rvert }(\theta_s \mid S)
\end{align}
with probability $1 - \alpha$ conditional on $\hat{S} = S$, where $\bigtimes$ denotes the Cartesian product.

\subsection{\label{sec:univariate example}Univariate example}
We now illustrate the selective inference problem and our solution in a simple setting with a single treatment effect ($m = 1$). 

Suppose that the significance of $\hat{\theta}_1$ is established by a $t$-test, that is, $\hat{S} = \{1\}$ if $X_1 \in \mathcal{X}(\{1\})$, where $\mathcal{X}(\{1\}) = \{x \in \mathbb{R}: \abs{x} \geq \Phi^{-1} (1 - \beta/2)\}$ for two-sided testing, and $\mathcal{X}(\{1\}) = \{x \in \mathbb{R}: x \geq \Phi^{-1} (1 - \beta)\}$ for one-sided testing. Here, $\beta$ denotes the nominal size of the test.
\begin{figure}
\centering
\input{figures/trunc_normal}
\caption{\label{fig:truncated normal distribution}Distribution of the $t$-statistic $X_1$. The dashed curve gives the unconditional distribution, and the solid curve shows the conditional distribution given two-sided (left panel) and one-sided (right panel) significance at the 10\% level. The shaded area represents $\mathcal{X}(\{1\})$, and the dashed vertical line gives $\theta_1$.}
\end{figure}
Figure~\ref{fig:truncated normal distribution} shows that the distribution of $X_1$ is a truncated Gaussian and typically asymmetric. This asymmetry implies that $\hat{\theta}_1$ is conditionally biased for the true effect size $\theta_1$. 

In the two-sided case, the median-unbiased estimator $\tilde{\theta}_1^{\text{ub}}$ that solves $F_1(X_1 \mid \tilde{\theta}_1^{\text{ub}}, \{1\}) = 0.5$ will always correct the conventional estimator $\hat{\theta}_1$ toward zero. This aligns with the intuition that significant effects are positively selected (``winner's curse''), and that their magnitudes are overestimated.

To verify this analytically, let $X_1 > \bar{x}_1 = \Phi^{-1}(1 - \beta/2)$ and note that 
\begin{align*}
F_1(X_1 \mid \hat{\theta}_1, \{1\}) = \frac{0.5 - \Phi(\bar{x}_1 - X_1) + \Phi(-\bar{x}_1 - X_1)}{1 - \Phi(\bar{x}_1 - X_1) + \Phi(-\bar{x}_1 - X_1)} < 0.5.
\end{align*}
Since $F_1(x \mid \theta_1, \{1\})$ is decreasing in $\theta_1$, $\tilde{\theta}_1^{\text{ub}}$ is smaller than $\hat{\theta}_1$.

\subsection{\label{sec:what is significance}Selection of significant effects by multiple testing}
This section defines the multiple-testing procedures used to determine the set $\hat{S}$ of jointly significant effects and characterises the set $\mathcal{X}(S)$ of $t$-statistics that lead to the selection of $\hat{S} = S$.
For exposition, we focus on step-down procedures in the main text. Step-up procedures are covered in Online Appendix~\ref{sec: appendix step-up rules}. 

The most commonly used step-down procedures control the family-wise error rate (FWER), that is, the probability of including at least one false positive in $\hat{S}$. For two-sided testing, FWER-control at level $\beta$ means that 
\begin{align*}
	\text{FWER} = P \left( \hat{S} \cap \{h: \theta_h = 0\} \neq \emptyset \right) \leq \beta.
\end{align*}
The corresponding definition for one-sided testing is obtained by replacing $\{h: \theta_h = 0\}$ with $\{h: \theta_h \leq 0\}$ in the previous display.

A step-down rule proceeds by successively comparing the ordered test statistics against a sequence of thresholds. 
Let $(>, j)$ denote the (data-dependent) index of the test statistic with the $j$th-largest magnitude, that is, $\lvert X_{(>, 1)} \rvert \geq \lvert X_{(>, 2)} \rvert \geq \dotsm \geq \lvert X_{(>, m)} \rvert$.
The thresholds are determined by a threshold function $\bar{x}$ that maps any $A \subseteq \{1, \dotsc, m\}$ to a positive number such that, for $A, B \subseteq \{1, \dotsc, m\}$ and $A \subseteq B$, $\bar{x} (A) \leq \bar{x} (B)$. 

For some rules, the threshold function can be written as $\bar{x} (A) = \Phi^{-1} \left(1 - \beta_{\lvert A \rvert} / 2 \right)$. For example, under the Bonferroni procedure we have $\beta_j = \beta/m$, and under the Holm procedure $\beta_j = \beta/(m + 1 - j)$, where $\beta$ is the nominal FWER.\footnote{Additional examples are provided in Table~\ref{tab:thresholds_simple_stepdown} of the Online Appendix.}  
For other rules, such as \textcite{romano2005stepwise}, $\bar{x}(A)$ depends on features of $A$ beyond its cardinality. This allows the procedure to exploit correlations among the test statistics, thereby increasing power.

A step-down procedure proceeds as follows:

{\singlespacing
\begin{algorithm}[Step-down testing]
\label{alg:generalized_step_down}
\begin{enumerate}[label = (\Alph*), ref = \Alph*]
	\item 
	Initialise the set of significant effects $\hat{S} \leftarrow \emptyset$ and the step counter $j \leftarrow 1$.
	\item 
	\label{alg:step_down:check}
	If $j = m$ or if $\lvert X_{>, j} \rvert < \bar{x} (\{(>, j), \dotsc, (>,m)\})$ then exit the algorithm and return the significant treatment effects $\hat{S}$. Otherwise, add $j$ to $\hat{S}$.
	\item Increment the step counter $j \leftarrow j + 1$ and go to Step~\ref{alg:step_down:check}.
\end{enumerate}
\end{algorithm}}
% The algorithm can be adapted to two-sided testing by replacing the test statistics $X_h$ by their absolute values and adjusting the threshold function $\bar{x}$. For the rules in Table~\ref{tab:thresholds_simple_stepdown}, this adjustment amounts to replacing $\alpha$ by half its value.
We denote the values of the test statistics $X$ for which Algorithm~\ref{alg:generalized_step_down} selects significant effects $S$ by $\mathcal{X}(S) \subseteq \mathbb{R}^m$.
\begin{figure}
\centering
\begin{tikzpicture}[scale=1.8]
    \def\vceilv{2.6}
    \def\vceilh{2.6}
    \def\vxbar{2}
    \def\vxxbar{1.6}
    % Draw axes
    \draw [<->,thick] (0,\vceilv + 0.1) node (yaxis) [above] {$x_2$}
        |- (\vceilh + 0.1,0) node (xaxis) [right] {$x_1$};
    % set select one 
    \draw [draw = none, fill=black!10] (0, \vceilv) rectangle (\vxxbar, \vxbar) node[midway] {$\mathcal{X}(\{2\})$};
    \draw [draw = none, fill=black!10] (\vceilh, 0) rectangle (\vxbar, \vxxbar) node[midway, rotate = 90] {$\mathcal{X}(\{1\})$};
    \draw [draw = none, fill = black!20] (\vxxbar,\vceilv) -- (\vxxbar,\vxbar) -- 
        (\vxbar, \vxbar) -- (\vxbar, \vxxbar) -- 
        (\vceilh, \vxxbar) -- (\vceilh, \vceilv) -- cycle;
    % add tick marks 
    \draw (-0.035, \vxxbar) -- (0.035, \vxxbar) node[left = 0.1]{$\bar{x}_2$};
    \draw (-0.035, \vxbar) -- (0.035, \vxbar) node[left = 0.1]{$\bar{x}_1$};
    \draw (\vxxbar, -0.035) -- (\vxxbar, 0.035) node[below = 0.1]{$\bar{x}_2$};
    \draw (\vxbar, -0.035) -- (\vxbar, 0.035) node[below = 0.1]{$\bar{x}_1$};
    % Draw boundary of set 
    \draw (\vxxbar,\vceilv) -- (\vxxbar,\vxbar) -- 
        (\vxbar, \vxbar) -- (\vxbar, \vxxbar) -- 
        (\vceilh, \vxxbar);
    % Draw diagonal
    \draw [dashed] (0,0) -- (\vceilh, \vceilv);
    \draw [draw = none] (\vxbar, \vxbar) -- (\vceilh, \vceilv) node [pos = 0.3, rotate = -45] {$\mathcal{X}(\{1, 2\})$};
\end{tikzpicture}
\caption{\label{fig:step-down geometry} Geometry of step-down testing with $m=2$ effects in $t$-statistics space. For ease of presentation, we depict the pattern in the positive quadrant. The full pattern is obtained by axial symmetry.}
\end{figure}
For the case of two effects, Figure~\ref{fig:step-down geometry} illustrates the region $\mathcal{X} (S)$ for $S = \{1\}$ (only first effect is significant), $S = \{2\}$ (only second effect is significant) and $S = \{1, 2\}$ (both effects are significant).

We introduce additional notation to prepare for a general characterisation of $\mathcal{X}(S)$. For $S \subseteq \{1, \dotsc, m\}$, let $\mathcal{E}(S)$ denote the class of bijections from $S$ onto $\{1, \dotsc, \lvert S \rvert\}$.
For $\sigma \in \mathcal{E}(S)$, $\sigma(h)$ maps effect $h$ to its ``rank'' under the ordering $\sigma$. For example, the pre-image $\sigma^{-1} (\{1, 2\})$ gives the treatment effects with ranks $1$ and $2$. $S^{\mathsf{c}}$ is the complement of $S$ in $\{1, \dotsc, m\} $.
The following lemma characterises $\mathcal{X}(S)$.
\begin{Lemma}
\label{lem:geometry generalized step-down}
Algorithm~\ref{alg:generalized_step_down} selects $\hat{S} = S$ if and only if $X \in \mathcal{X}(S)$ with 
\begin{align}
	\label{eq:uncond support X}
	\mathcal{X}(S) =  \bigcup_{\sigma \in \mathcal{E}(S)} 
		\bigg(
		\begin{aligned}[t]
			& \bigcap_{s \in S}
			\left\{x_s \in \mathbb{R}:	\lvert x_s \rvert \geq \bar{x} \left(\sigma^{-1}(\{\sigma (s), \dotsc, \lvert S \rvert \}) \cup S^{\mathsf{c}}\right) \right\}
			\\
			& \cap \bigcap_{r \in S^{\mathsf{c}}} \left\{ x_r \in \mathbb{R} : \lvert x_r \rvert < \bar{x} \left( S^{\mathsf{c}}\right) \right\}
		\bigg).
		\end{aligned}
\end{align}
\end{Lemma}
% A corresponding result for two-sided testing is stated in Online Appendix~\ref{sec:additional theory}.
We use characterisation \eqref{eq:uncond support X} of $\mathcal{X}(S)$ in our theoretical analysis. Due to computational complexity, we do not directly evaluate $\mathcal{X}(S)$ when implementing our procedures. Our solution is discussed in the next section.

\subsection{\label{sec:algorithm conditional support}Algorithm for computing the conditional support}

This section describes an efficient algorithm for computing the conditional support $\mathcal{X}_s(z, S)$ without evaluating $\mathcal{X}(S)$. 
Given that $\mathcal{X} (S)$ is a complex set, defined by iterating over $2^m$ permutations, and potentially non-convex set (see Figure~\ref{fig:step-down geometry}), the existence of such an algorithm is not obvious.

Our algorithm has only polynomial complexity and evaluates $\mathcal{X}_s(z, S)$ in $\mathcal{O}(m^3 \log m)$ time, enabling applications with many effects, such as $m=371$ in our mutual fund study. In Online Appendix~\ref{sec:appendix initial bounds}, we provide an improved algorithm based on the \textcite{bentley1979algorithms} algorithm that reduces the computational cost to $\mathcal{O}(\tilde{n}_I \lvert S \rvert \log \lvert S \rvert)$, where $\lvert S \rvert$ is the number of significant effects and $\tilde{n}_I < S$ is a number that depends on the data.

For exposition, we present a simple version of the algorithm for two-sided step-down rules and the selection event $\hat{S} = S$. Improvements and generalisations of the algorithm are described in Online Appendix~\ref{sec:more algorithms}. The algorithm approximates $\mathcal{X}_s(z, S)$ by its closure.
The approximation error has Lebesgue measure zero and does not affect any of the statistics that we consider. 

We first introduce some notation. Let $\mathbf{x}_z (x_s) = \Omega_{\bullet, s} x_s + z$ denote the vector-valued function of $x_s$ that gives the $t$-statistics if $X_s = x_s$ and $Z^{(s)} = z$. For $\sigma \in \mathcal{E}(S)$, let $\bar{x}_{\sigma, h} = \bar{x} \left(\sigma^{-1}(\{\sigma (h), \dotsc, \lvert S \rvert \}) \cup S^{\mathsf{c}} \right)$ for $h \in S$ and $\bar{x}_{\sigma, h} = \bar{x} \left( S^{\mathsf{c}}\right)$ for $h \notin S$. 

Our algorithm is based on intervals $I$ over which the linear functions $\mathbf{x}_{z,h}(x_s)$, $h=1, \dotsc, m$, can be unambiguously ordered --- that is, they neither intersect nor change sign. These intervals are obtained by partitioning the real line at points where a function $\mathbf{x}_{z,h}(x_s)$ crosses the horizontal axis or where two functions $\mathbf{x}_{z,h}(x_s)$ and $\mathbf{x}_{z,h'}(x_s)$ intersect. In the online appendix, we describe refinements that avoid computing all such intersection points. The key observation underlying our algorithm is that the ordering of the $t$-statistics $\mathbf{x}_z(x_s)$ remains constant within each interval $I$. By Lemma~\ref{lem:geometry generalized step-down}, whether $x_s \in \mathcal{X}_s(z, S)$ depends only on the permutation that reorders the components of $\mathbf{x}_z(x_s)$ by their absolute values. Consequently, the conditional support can be computed without enumerating all permutations in $\mathcal{E}(S)$. The number of permutations that must be considered is bounded by the number of intervals $I$, which is quadratic.

Our algorithm proceeds as follows:

{\singlespacing
\begin{algorithm}[Compute $\mathcal{X}_s(z, S)$]
\label{alg: simple conditional support}
\begin{enumerate}[label = (\Alph*), ref = \Alph*]
	\item 
	Find the intervals $I$ by computing all intersection points of the linear functions $\mathbf{x}_{z, h} (x_s)$, $h = 1, \dotsc, m$ with each other or with the horizontal axis.
	\item For each interval $I$:
	\begin{enumerate}[label = \roman*, ref = \roman*]
		\item Let $\tilde{x}$ denote a value in the interior of $I$ and find the unique permutation $\sigma^*_I$ that orders the components of $\{\lvert \mathbf{x}_{z, h}(\tilde{x})\rvert \}_{h=1, \dotsc, m}$ in descending order. 
			\item Let $\chi_A(h) = (-1)^{\mathbf{1}\{h \notin A\}}$ denote the signed indicator for $h \in A$ and $N = \{h: \mathbf{x}_{z, h}(\tilde{x}) < 0\}$ the hypotheses with negative $\mathbf{x}_{z, h}$ on $I$. Let
			\begin{align*}
				\ell(I) =& \max_{h: \chi_S(h) \chi_N(h) \Omega_{h, s} > 0} \frac{\chi_N(h) \bar{x}_{\sigma^*_I, h} - z_h}{\Omega_{h, s}}, 
				\\
				u(I) =& \min_{h: \chi_S(h) \chi_N(h) \Omega_{h, s} < 0} \frac{\chi_N(h) \bar{x}_{\sigma^*_I, h} - z_h}{\Omega_{h, s}} .
			\end{align*}
			% $H_{\ell, +} = \{h \in S: \Omega_{h, s} > 0, \bar{x}_{z, h}(\tilde{x}) > 0\} \cup \{h \notin S: \Omega_{h, s} < 0, \bar{x}_{z, h}(\tilde{x}) > 0\}$, 
			% $H_{\ell, -} = \{h \in S: \Omega_{h, s} < 0, \bar{x}_{z, h}(\tilde{x}) < 0\} \cup \{h \notin S: \Omega_{h, s} > 0, \bar{x}_{z, h}(\tilde{x}) < 0\}$, 
			% $H_{u, +} = \{h \in S: \Omega_{h, s} < 0, \bar{x}_{z, h}(\tilde{x}) > 0\} \cup \{h \notin S: \Omega_{h, s} > 0, \bar{x}_{z, h}(\tilde{x}) > 0\}$, 
			% $H_{u, -} = \{h \in S: \Omega_{h, s} > 0, \bar{x}_{z, h}(\tilde{x}) < 0\} \cup \{h \notin S: \Omega_{h, s} < 0, \bar{x}_{z, h}(\tilde{x}) < 0\}$ and 
			% \begin{align*}
			% 	\ell (I) =& \max\left\{ \max_{h \in H_{\ell, +}} \frac{\bar{x}_{\sigma^*_I, h} - z_h}{\Omega_{h, s}}, \max_{h \in H_{\ell, -}} \frac{-\bar{x}_{\sigma^*_I, h} - z_h}{\Omega_{h, s}} \right\},
			% 	\\
			% 	u(I) =& \min\left\{\min_{h \in H_{u, +}} \frac{\bar{x}_{\sigma^*_I, h} - z_h}{\Omega_{h,s}}, \min_{h \in H_{u, -}} \frac{-\bar{x}_{\sigma^*_I, h} - z_h}{\Omega_{h,s}} \right\}.	
			% \end{align*}
	\end{enumerate}
	\item Return $\bigcup_{I} I \cap \left[\ell(I), u(I)\right]$.
\end{enumerate}
\end{algorithm}}

The following result establishes the validity and efficiency of our algorithm.
\begin{theorem}
	\label{thm: validity algorithm}
	Algorithm~\ref{alg: simple conditional support} computes the closure of $\mathcal{X}_s(z, S)$ in $\mathcal{O}(m^3 \log m)$ time. 
\end{theorem}

An immediate corollary to Theorem~\ref{thm: validity algorithm} is that the conditional support is a union of intervals.
By merging overlapping intervals, this union can be written as a union of disjoint intervals $[u_k, \ell_k]$, $k = 1, \dotsc, n_I$ with $n_I \leq m(m+1)/2 + 1$. In practice, the number of disjoint intervals $n_I$ is typically very small. In Online Appendix~\ref{sec:appendix bound on n_I}, we characterise scenarios in which $n_I = 1$ (i.e., a connected interval).

\section{\label{sec:discussion}Discussion and extensions}

In this section, we examine when conditional inference coincides with conventional methods and how to adapt it to estimated variance–covariance matrices or threshold functions. We also discuss which selection events provide an appropriate summary of ``observed significance.''

\subsection{Relationship between conditional and unconditional inference\label{sec:relationship unconditional CI}}

In certain situations, conditional and unconditional inference yield similar results. Very roughly speaking, this happens when the effects are ``highly significant.'' 
In a univariate setting, this means that the $t$-statistic of the effect of interest is large. In a multivariate setting, significance of an effect depends also on the $t$-statistics of other effects and a more careful examination of what ``highly significant'' means is needed.

% In a univariate setting, the difference between conditional and unconditional inference depends only on the magnitude of the $t$-statistic of the effect of interest. By contrast, in a multivariate setting, it potentially depends also on the $t$-statistics of other effects.

% The difference between the unconditional confidence interval $\text{CI}_\alpha (\theta_s)$ and the conditional interval $\text{CCI}_\alpha (\theta_s \mid S)$ is small when the distribution function of $X_s$ evaluated close to the observed $X_s$ does not change much whether we condition on $X_s \in \mathcal{X}_s (Z^{(s)}, S)$ or not. In a univariate setting, this depends only on the magnitude of $s$'s own $t$-statistic $X_s$. With multiple treatment effects, it can also depend on the magnitude of the $t$-statistics of other effects that are correlated with $s$.

The following theorem gives sufficient conditions for the convergence of the conditional confidence interval to the unconditional confidence interval. It holds for selection events $\hat{S} = S$ with a positive, but not necessarily high, probability in the limit.\footnote{In a different context concerning conditional inference, \textcite{andrews_inference_2024} demonstrate that their conditional confidence intervals converge to unconditional confidence intervals when the selection event's probability approaches one (see Proposition~3 in their paper). Our framework allows for a similar conclusion to be drawn as a corollary to Theorem~\ref{thm:convergence to unconditional CI}.}

\begin{theorem}[Convergence to unconditional confidence interval]
    \label{thm:convergence to unconditional CI}
 Consider a sequence of effect estimators $\hat{\theta}_n \sim N(\theta_n, V_n)$ and $s \in S \subseteq \{1, \dotsc, m\}$ such that $P_{\theta_n, V_n} (\hat{S} = S)$ is bounded away from zero. Let $V_{n, h, s}$ denote the $(h, s)$-th entry of $V_n$ and let $\mathbf{v}_{n, h} = V_{n, h, h}$. Let $d_H (A, B)$ denote the Hausdorff distance between sets $A$ and $B$. Suppose that one of the following conditions holds:
    \begin{enumerate}
        \item 
 The selection rule is a 1-sided step-down rule. For all $h$ with $\limsup_{n \to \infty} \lvert V_{n, h, s} \rvert > 0$ we have either $h \in S$ and $\mu_{n, h} = \theta_{n, h} / \sqrt{\mathbf{v}_{n, h}} \to \infty$ or $h \notin S$ and $\mu_{n, h} \to -\infty$.
        \item The selection rule is a 2-sided step-down rule. For all $h$ with $\limsup_{n \to \infty} \lvert V_{n, h, s} \rvert > 0$ we have $h \in S$ and $\lvert \mu_{n, h} \rvert \to \infty$. 
    \end{enumerate}
 Then, for all $\epsilon > 0$,
 $P_{\theta_n, V_n} \big( d_H \left( \text{CI}_\alpha (\theta_s), \text{CCI}_\alpha (\theta_s \mid S) \right) > \epsilon \mid \hat{S} = S\big) \to 0$ as $n \to \infty$.
\end{theorem}
For a setting with (asymptotically) uncorrelated estimators, such as subgroup analysis, Theorem~\ref{thm:convergence to unconditional CI} implies that conditional and unconditional confidence interval for a significant effect $s$ converge if we expect the $t$-statistic $X_s$ to be large. 

In settings with correlated estimators, Theorem~\ref{thm:convergence to unconditional CI} imposes more stringent conditions. 
Suppose that we conduct inference on $s$ conditional on $\hat{S} = S$ and that the effect $h \neq s$ is correlated with $s$. 
The theorem requires that the significance status of $h$ becomes asymptotically certain: if $h \in S$ then $h \in \hat{S}$ with probability approaching one and if $h \notin S$ then $h \notin \hat{S}$ with probability approaching one. For two-sided testing this rules out $h \notin S$ since the rejection probability of a two-sided test is bounded away from zero regardless of the true effect size.
% Suppose that $\hat{\theta}_s$ and $\hat{\theta}_h$ are correlated. Theorem~\ref{thm:convergence to unconditional CI} requires that the significance status of $h$ becomes asymptotically certain. That is, if $h \in S$ then $h \in \hat{S}$ with probability approaching one and if $h \notin S$ then $h \notin \hat{S}$ with probability approaching one. For two-sided testing we rule out $h \notin S$. This additional restriction is required because, for a two-sided test, the null hypothesis contains only a single parameter value at which we reject with positive probability. Therefore, insignificance of $h$ cannot be almost certain and will affect conditional inference on $s$.

A straightforward corollary to Theorem~\ref{thm:convergence to unconditional CI} is that the conditional median-unbiased estimator converges to the unconditional point estimator under the same conditions.

\subsection{\label{sec:relax assumptions}Estimated covariance matrix and thresholds}

In this section, we accommodate unknown variance-covariance matrices and data-dependent threshold functions. This requires two modifications. First, we replace the test statistics $X$ by studentised $t$-statistics based on a consistent estimator $\hat{V}$ of the covariance matrix. Second, we replace the fixed threshold function $\bar{x}$ by a data-dependent threshold function $\hat{\bar{x}}$. The latter modification is important because data-dependent threshold functions, such as those in \textcite{romano2005stepwise,list2019multiple}, yield more adaptive and potentially more powerful tests.

The studentised $t$-statistics are defined as $\hat{X} = \diag^{-1/2} (\hat{\mathbf{v}}) \hat{\theta}$, where $\hat{\mathbf{v}}$ is the diagonal of $\hat{V}$.
Our inference is based on an estimate of the true distribution function $F_s(x_s \mid z, \theta_s, S)$ given by
\begin{align*}
\widehat{F}_s (x_s \mid z, \theta_s, S) = F_s (x_s \mid z, \theta_s, \hat{V}, \hat{\bar{x}}, S),
\end{align*}
where $\hat{Z}^{(s)} = \hat{X} - \hat{\Omega}_{\bullet, s} \hat{X}_s$ and  
	$
	\hat{\Omega} = \diag^{-1/2}(\hat{\mathbf{v}}) \hat{V} \diag^{-1/2}(\hat{\mathbf{v}})
	$.
We now define our conditional inference procedures as in Section~\ref{sec:summary}, with $\widehat{F}_s$ replacing $F_s$. 
Let $\hat{\theta}_s^{(p)}$ denote the unique solution to
\begin{align*}
	\widehat{F}_s \left(\hat{X}_s \mid \hat{Z}^{(s)}, \hat{\theta}_s^{(p)}, S\right) = p.
\end{align*}
The estimator $\hat{\theta}_s^{\text{ub}}$ and the confidence interval $\widehat{\text{CCI}}_{\alpha}(\theta_s \mid S)$ are now defined analogously to $\tilde{\theta}_s^{\text{ub}}$ and $\text{CCI}_{\alpha}(\theta_s \mid S)$ above, with $\tilde{\theta}_s^{(p)}$ replaced by $\hat{\theta}_s^{(p)}$.
The theoretical analysis of the modified procedure is given in Section \ref{sec:asymptotics}.

\subsection{\label{sec:conditioning event}Choice of selection event}

What event describes ``observed significance'' in a multivariate setting is not obvious. We now motivate our focus on the event $\hat{S} = S$ and discuss alternative selection events that may be relevant in some applications.

By conditioning on $\hat{S} = S$, we fully take into account the outcome of the multiple testing procedure, including both which effects are significant and which are not.

This selection event is appropriate if the interpretation and framing of a significant effect $s \in S$ depends on whether other effects $h \neq s$ are found significant or insignificant, as is often the case in empirical studies. 

For example, in Section~\ref{sec:application list} below, we partially replicate the study of \textcite{karlan2007does} on the effects of a matching grant on charitable giving. 
An effect on the response rate captures whether the matching grant motivates more people to donate, whereas effects on the donation amount capture also the size of donations.
If the response rate is insignificant, we may interpret the additional donations as coming from the existing donor pool. Other examples where insignificant effects shape the interpretation of empirical results include placebo tests (see our application in Online Appendix~\ref{sec:application Couttenier}). 

In some applications, conditioning on $\hat{S} \supseteq S $ is more suitable. This choice removes information about insignificance from the selection event. This is desirable, for example, in our mutual fund application. Funds with significantly positive alphas are funds that systematically outperform the market. The reporting on outperforming funds is unlikely to change when finding insignificant alphas for other funds. 
The event $\hat{S} \supseteq S$ accounts for the funds in $S$ to be classified as outperformers, but it is agnostic about the performance of the other funds. Even though $\hat{S} \supseteq S$ may appear to be an even more complex event than $\hat{S} = S$, our polynomial-time algorithm readily extends to this case (see Online Appendix~\ref{sec:appendix support alternative}).

One may also consider the selection event $s \in \hat{S}$ for a given significant effect $s \in S$. This is the most direct analogue of conditioning on $\hat{S} = \{1\}$ in the univariate case. 
However, this defines different selection events for different significant effects, making it difficult to conduct joint conditional inference on all significant effects. We mention this event for completeness but do not pursue it further in this paper.

\section{\label{sec:asymptotics}Asymptotic properties}

We now provide asymptotic counterparts to Theorem~\ref{thm:validity median-unbiased oracle} and Theorem~\ref{thm:validity CS oracle} with estimated covariance matrices, data-dependent thresholds and asymptotically Gaussian estimators of the treatment effects. We allow all population quantities to depend on the probability measure and prove asymptotic results that are uniform over sequences of probability measures. 

Let $\lVert \cdot \rVert$ denote the $L_2$-norm. For a sequence of classes of probability measures $\mathcal{P}_n$, assume: 
\begin{Assumption} 
\label{assume:asymp}
\begin{enumerate}
\item (Asymptotic normality)
\label{assume:asymp_local:normality}
Let $BL_1$ denote the class of Lipschitz functions that are bounded by one in absolute value and have Lipschitz constant bounded by one. Let $\mathcal{V}_m$ denote the space of semi-definite and symmetric $m \times m$ matrices. For $P \in \mathcal{P}_n$, there is $\theta (P) \in \mathbb{R}^m$ and $V(P) \in \mathcal{V}_m$ such that for $\mathbf{v}(P) = (V_{1,1}(P), \dotsc, V_{m,m}(P))$, $\Omega = \diag^{-1/2} (\mathbf{v}(P)) V(P) \diag^{-1/2} (\mathbf{v}(P))$ and $\xi_{\Omega(P)} \sim N(0, \Omega(P))$, we have
\begin{align*}
    \lim_{n \to \infty} \sup_{P \in \mathcal{P}_n} \sup_{f \in BL_1} \left\lvert \mathbb{E}_P f \left( \diag^{-1/2} (\mathbf{v}(P)) \big( \hat{\theta}  - \theta(P)\big)\right) - \E f(\xi_{\Omega(P)}) \right\rvert = 0.
\end{align*}
\item (Consistent variance estimation)
\label{assume:asymp_local:vhat}
For all $\epsilon > 0$, we have
\begin{align*}
	\lim_{n \to \infty} \sup_{P \in \mathcal{P}_n} \max_{j = 1, \dotsc, m}
	P \left(
		\left\lvert {\hat{\mathbf{v}}_j}/{\mathbf{v}_j (P)} - 1 \right\rvert > \epsilon
	\right)
	=& 0 \quad \text{and}
	\\
    \lim_{n \to \infty} \sup_{P \in \mathcal{P}_n} 
    P \left(
        \left\lVert \hat{V} - V(P) \right\rVert > \epsilon
    \right) =& 0.
\end{align*}
\item (Consistent estimation of threshold function)
\label{assume:asymp_local:xbar}
There exist an upper bound $C_{\bar{x}}$ and a mapping of $P \in \mathcal{P}_n$ to bounded functions $\lvert \bar{x} (P) \rvert \leq C_{\bar{x}}$ such that, for all $\epsilon > 0$,
\begin{align*}
	\lim_{n \to \infty} \sup_{P \in \mathcal{P}_n} \sup_{A \subset \{1, \dotsc, m\}} P \left(\norm{\hat{\bar{x}} (A) - \bar{x} (P) (A)} > \epsilon \right) = 0. 
\end{align*}
\end{enumerate}
\end{Assumption}
Assumption~\ref{assume:asymp}.\ref{assume:asymp_local:normality} weakens the assumption that the treatment effect estimators are exactly Gaussian to uniform asymptotic Gaussianity \parencite[see, e.g.,][]{kasy2019uniformity}. Assumptions~\ref{assume:asymp}.\ref{assume:asymp_local:vhat} and \ref{assume:asymp}.\ref{assume:asymp_local:xbar} require the estimated variance-covariance matrix and the data-dependent threshold function to be uniformly consistent for their respective population counterparts.

The following theorem is an asymptotic counterpart to Theorem~\ref{thm:validity median-unbiased oracle}.
\begin{theorem}[Asymptotic median-unbiasedness]
\label{thm:asymp median-unbiased}
Under Assumption~\ref{assume:asymp},  
\begin{align*}
	\lim_{n \to \infty}
	\sup_{P \in \mathcal{P}_n} \left\lvert P \left( \hat{\theta}^{\text{ub}}_s \leq \theta_s (P) \mid \hat{S} = S \right) - 0.5 \right\rvert P (\hat{S} = S) = 0. 
\end{align*}
\end{theorem}
This result establishes that $\hat{\theta}^{\text{ub}}_s$ is asymptotically median-unbiased for $\theta_s$ uniformly over $\mathcal{P}_n$ with the possible exception of sequences $P_n \in \mathcal{P}_n$ along which $P_n (\hat{S} = S)$ vanishes. 

The uniformity of our result is important because it covers local-to-zero signal-to-noise ratios, where effective treatments are not always detected in the limit. In such cases, selectivity cannot be ignored, and conventional confidence intervals are invalid. Under conventional ``pointwise'' asymptotics, by contrast, effective treatments have diverging $t$-statistics and are therefore highly ``significant'' in the limit.  This limit is a poor approximation of typical finite-sample settings, where $t$-statistics are often quite small even for effective treatments.
% Therefore, under the conditions of Theorem~\ref{thm:convergence to unconditional CI},  the conventional unconditional confidence intervals are also conditionally valid and there is not need to implement our methods.
% Pointwise asymptotics are therefore not reflective of typical finite sample settings where $t$-statistics are often quite small, even for effective treatments. Such settings are better approximated by uniform asymptotics.

The statement of Theorem \ref{thm:asymp median-unbiased} holds trivially when $P(\hat S=S)=0$. By focusing on cases where $ P(\hat S = S)$ is positive, we ensure that we do not condition on an event with zero probability in the asymptotic limit.  Similar assumptions are made in other applications of conditional inference \parencite[e.g.][]{andrews_inference_2024}. 

The next result establishes that the conditional confidence interval $\widehat{\text{CCI}}_\alpha(\theta_s \mid S)$ has an asymptotic conditional coverage probability of $1-\alpha$. 
\begin{theorem}[Asymptotic confidence set validity]
\label{thm:asymp CS validity}
Under Assumption~\ref{assume:asymp},  
\begin{align*}
	\lim_{n \to \infty} \sup_{P \in \mathcal{P}_n} \left\lvert P \left(
		\theta_s (P) \in \widehat{\text{CCI}}_{\alpha} (\theta_s \mid S) \mid \hat{S} = S 
	\right)
	- (1 - \alpha) \right\rvert P (\hat{S} = S) = 0.
\end{align*}
\end{theorem}
This result immediately implies the \emph{unconditional} validity of our conditional confidence interval. For example, for the joint confidence set \eqref{eq:joint confidence set}, it holds that 
\begin{align*}
	\lim_{n \to \infty} \sup_{P \in \mathcal{P}_n} P \left(
		\left(\theta (P)\right)_{s \in \hat{S}} \in \bigtimes_{s \in \hat{S}} \widehat{\text{CCI}}_{\alpha/\lvert \hat{S} \rvert} (\theta_s \mid \hat{S})
	\right) \geq 1 - \alpha.
\end{align*}

\section{\label{sec:simulation}Monte Carlo simulations}

We use Monte Carlo simulations to assess the finite-sample performance of our conditional confidence interval and conditional median-unbiased estimator.

We consider two designs: one with normally distributed data (``Normal'') and one with chi-squared distributed data (``Chi-squared''). In both, we test $m = 5$ parameters with population values $\theta = (0.05, 0.03, 0.01, 0, 0)'$ and focus on inference for the first parameter when it is significant. Significance is determined using the Holm step-down procedure at a FWER of $\beta = 10\%$, based on $t$-tests computed from $n$ i.i.d. observations of an $m$-vector $Y_i$. The test statistic is $X_k = \sqrt{n} \bar{Y}_k / S_k$, where $\bar{Y}_k$ and $S_k^2$ denote the sample mean and variance. Components of $Y_i$ have pairwise correlation $\rho = 0.5$. The two designs differ only in the marginal distribution of $Y_i$.

In the ``Normal'' design, $Y_i$ follows a multivariate normal distribution with mean $\theta$ and a variance–covariance matrix with ones on the diagonal and $\rho$ off-diagonal. In the ``Chi-squared'' design, $Y_{ik} = (U_{ik}^2 - 1)/\sqrt{2} + \theta_k$, where $U_i$ is multivariate normal with mean zero and a variance–covariance matrix with ones on the diagonal and $\sqrt{\rho}$ off-diagonal, yielding correlation $\rho$ between $Y_{ik}$ and $Y_{ik'}$ for $k \neq k'$.

We consider $n \in \{100, 300, 500, 700, 900\}$. Each design uses 5{,}000 replications, increased to 20{,}000 for $n = 100$ to ensure enough cases in which the first parameter is selected. We compute the conditional, unconditional (``naive''), and Bonferroni-corrected confidence intervals, each with nominal coverage $(1 - \alpha) = 90\%$.

We simulate confidence interval coverage and estimator bias conditional on the first parameter being significant, that is, on the event $\{1\} \subseteq \hat{S}$. Table~\ref{tab:simu_1} reports results for two-sided tests; additional results for one-sided tests appear in Online Appendix~\ref{sec:further_simu}. Selection probabilities---the chances that the first parameter is found significant---range from 2\% to 21\%.

\begin{table}
	\centering 
	\footnotesize
    \begin{tabular}{lrrrrrrrrrr}
              \toprule 
          & \multicolumn{5}{c}{\textbf{Normal} }                &       \multicolumn{5}{c}{\textbf{Chi-squared}} \\ 
          \hline
   \hfill $n$   & 100   & 300  & 500  & 700  & 900    & 100   & 300  & 500   & 700  & 900  \\ 
Sel Prob &  0.040   & 0.082   &  0.116    & 0.155    &  0.208    &  0.023  &  0.050  & 0.092  & 0.137  & 0.191  \\
\\
          & \multicolumn{10}{c}{\textbf{Conditional Coverage}} \\ 
Cond CI  &  0.950  & 0.910  & 0.919  & 0.901 & 0.911  & 0.588  & 0.920  & 0.932   & 0.915  & 0.899  \\
Naive CI     & 0.005  & 0.351   & 0.569   & 0.712  & 0.767  & 0.002  & 0.188   & 0.578   & 0.729   & 0.780   \\
Bonf CI   & 0.660 & 0.866  & 0.912  & 0.937 & 0.955  & 0.405  & 0.872  & 0.921  & 0.962  & 0.964  \\
\\
           & \multicolumn{10}{c}{\textbf{Conditional Median CI Length}} \\ 
Cond CI  & 0.416 & 0.245 & 0.196  & 0.163  & 0.145 & 0.346 & 0.280 & 0.204  & 0.173  & 0.153  \\
Naive CI           &  0.327 & 0.189  & 0.147  & 0.124  & 0.109 & 0.319  & 0.187  & 0.146  & 0.123  & 0.109  \\
Bonf CI    & 0.463  & 0.268  & 0.207  & 0.175  & 0.155 & 0.451  & 0.265  & 0.206  & 0.175 & 0.154 \\
\\
          & \multicolumn{10}{c}{\textbf{Conditional Median Bias}} \\ 
Cond Est  & -0.015  & -0.014  & -0.013   & -0.015  & -0.013  &  -0.045  & -0.015  & -0.030 & -0.021  & -0.019   \\
Naive Est  &  0.084  & 0.083  & 0.061   & 0.045   & 0.035 & -0.040  & 0.032  & 0.051  & 0.046  & 0.037   \\
\bottomrule
    \end{tabular}
	\caption{\label{tab:simu_1}%  
		            Coverage properties of various confidence intervals. The selection is made by the two-sided Holm procedure. The nominal level for all confidence intervals is 90\%. ``Sel Prob'' is the probability that the first parameter is significant. ``Cond CI'' and ``Cond Est'' denote the conditional CI and median-unbiased estimator. ``Naive CI'' and ``Naive Est'' denote the conventional (unconditional) CI and estimator,  respectively,  while ``Bonf CI'' denotes the conventional (unconditional) CI with Bonferroni correction.}
\end{table}

We first examine coverage. The conditional intervals consistently achieve coverage close to the nominal 90\% level. In the ``Chi-squared'' design, we observe light undercoverage at the smallest sample size ($n=100$), suggesting that, in this setting, somewhat larger samples are required for the asymptotic approximation to be fully accurate. For sample sizes of 300 or more, empirical coverage is satisfactory even with non-normal data. By contrast, naive confidence intervals undercover across all designs and sample sizes, particularly when selection is infrequent. Bonferroni intervals are robust to selection but reach nominal coverage only at large $n$ and tend to be overly conservative. In terms of precision, the conditional intervals are wider than the naive ones yet substantially shorter than the Bonferroni intervals, highlighting their efficiency gains.

Beyond providing valid confidence intervals, our approach also mitigates selection bias. The conditional bias of the conventional estimator ($0.084$ in the ``Normal'' design at $n=100$) is reduced to $-0.015$ by our conditional median-unbiased estimator. A possible explanation for the Bonferroni interval's conditional undercoverage is that it is centred on the conventional estimator, which exhibits substantially larger conditional bias (e.g., $0.084$ in the ``Normal'' design at $n=100$).

\section{\label{sec:application list}Empirical application: Charitable giving}

We illustrate our selective conditional inference methods by revisiting the field experiment conducted by \textcite{karlan2007does}, which studies the effectiveness of matching grants on charitable giving. The experiment is also reanalysed by \textcite{list2019multiple} using step-down multiple testing. We add to their analysis by inferring the effect sizes of the significant results.
We focus on the binary treatment whether a potential donor was offered a matching grant and ignore heterogeneous characteristics of the matching grants.\footnote{See Section 4 of \textcite{list2019multiple} for details on the different grants.}  

The details of the estimation procedures are given in Online Appendix~\ref{sec: diff in means by regression}. 
To detect significant effects, we use three FWER-controlling procedures: the Bonferroni correction, the Holm procedure, and a bootstrap procedure based on \textcite[][, henceforth RW2005]{romano2005stepwise}. Our implementation of the RW2005 procedure is described in Online Appendix~\ref{sec:RW bootstrap}.
We set the nominal FWER to 10\%. All reported confidence intervals use nominal coverage levels of 95\%.

We study treatment effects corresponding to four different outcomes: the response rate, dollars given not including match, dollars given including match, and the change in the amount donated from the previous elicitation.\footnote{In Appendix~\ref{sec:appendix additional results list}, we consider additional specifications that also comprise multiple subgroups and treatments.}
For comparability, we measure effect size for all outcomes in terms of standard deviations.\footnote{For the response rate, this standard deviation is defined above and equal to 13\%. For the other outcomes, we use $\$8.18$, the standard deviation of the amount donated in the control group.}
Given that the signs of the effects are not obvious, we use two-sided testing.\footnote{For example, a matching grant may increase the amount the charity receives but decrease individual donations due to crowding out.}

The outcomes ``Response rate'' and ``Dollars given incl match'' are found to be significant by all three procedures. The RW2005 procedure detects an additional significant effect for ``Dollars given not incl match.'' 

Our conditional inference results are summarised in Table~\ref{tab:LSXtab1}.
\begin{table}
	\centering	
	\footnotesize
	
\begin{tabular}{lllll}
\toprule
\multicolumn{1}{c}{} & \multicolumn{1}{c}{} & \multicolumn{3}{c}{Conditional} \\
\cmidrule(l{3pt}r{3pt}){3-5}
Outcome & Unconditional & RW2005 & Holm & Bonferroni\\
\midrule
\addlinespace[0.3em]
\multicolumn{5}{l}{\textbf{Conditional on $\hat{S} = S$}}\\
\hspace{1em}Response rate & \makecell{0.032 \\ (0.010, 0.054)} & \makecell{0.016 \\ (-0.001, 0.050)} & \makecell{0.150 \\ (0.017, 0.667)} & \makecell{0.042 \\ (-0.001, 0.115)}\\
\hspace{1em}Dollars given not incl match & \makecell{0.019 \\ (-0.003, 0.041)} & \makecell{-0.004 \\ (-0.019, 0.036)} & - & \vphantom{1} -\\
\hspace{1em}Dollars given incl match & \makecell{0.255 \\ (0.211, 0.299)} & \makecell{0.216 \\ (0.170, 0.292)} & \makecell{0.534 \\ (0.242, 1.743)} & \makecell{0.286 \\ (0.206, 0.449)}\\
\hspace{1em}Amount change & \makecell{0.774 \\ (-2.908, 4.456)} & - & - & \vphantom{1} -\\
\addlinespace[0.3em]
\multicolumn{5}{l}{\textbf{Conditional on $\hat{S} \supseteq S$}}\\
\hspace{1em}Response rate & \makecell{0.032 \\ (0.010, 0.054)} & \makecell{0.016 \\ (-0.001, 0.050)} & \makecell{0.029 \\ (-0.000, 0.051)} & \makecell{0.028 \\ (-0.001, 0.051)}\\
\hspace{1em}Dollars given not incl match & \makecell{0.019 \\ (-0.003, 0.041)} & \makecell{-0.004 \\ (-0.019, 0.036)} & - & -\\
\hspace{1em}Dollars given incl match & \makecell{0.255 \\ (0.211, 0.299)} & \makecell{0.216 \\ (0.170, 0.292)} & \makecell{0.254 \\ (0.204, 0.294)} & \makecell{0.253 \\ (0.200, 0.294)}\\
\hspace{1em}Amount change & \makecell{0.774 \\ (-2.908, 4.456)} & - & - & -\\
\bottomrule
\end{tabular}
	\caption{\label{tab:LSXtab1}Multiple treatments \parencite[Specification from Table~1 in][]{list2019multiple}. Confidence intervals (CI) are given in parentheses.}
\end{table}
Under the selection event $\hat{S} = S$, our conditional inference post RW2005-selection carries out minor downward correction, consistent with positively selected significant effects.
Inference after Holm and Bonferroni detection is more subtle, indicating a negative bias for the effect on ``Dollars given incl match.'' This outcome is positively correlated with the effect on ``Dollars given not incl match,'' which is insignificant. 
Test statistics and therefore significance depend on the true effect sizes and intrinsic randomness. Since the two outcomes are correlated, they share some of their randomness. To reconcile observing one effect as significant and the other as insignificant, the true effect for the significant effect should be larger than its unconditional estimate.
The upward correction for the effect on ``Dollars given incl match'' is more pronounced for the Holm procedure than the Bonferroni procedure. This is because ``Dollars given not incl match'' is closer to the threshold of significance under the Holm procedure than under the Bonferroni procedure.\footnote{The Bonferroni $p$-value for ``Dollars given not incl match'' is 20\%, the Holm $p$-value is just above 10\%.} 

The correlation structure among the outcomes is crucial for explaining the upward correction. For uncorrelated treatment effect estimators, the bias correction will always be downward, consistent with the positive ``winner's curse'' bias in the univariate case (see our results for multiple subgroups in Online Appendix~\ref{sec:list subgroups}).

In contrast to the selection event $\hat{S} = S$, the selection event $\hat{S} \supseteq S$ does not include information about insignificant effects. This information is crucial for the explanation of a possible upward correction.
The results in Table~\ref{tab:LSXtab1} demonstrate that all bias corrections are downward when we condition on $\hat{S} \supseteq S$. The downward corrections behave differently from the corresponding correction in a univariate setting where a more powerful test is associated with a smaller positive bias. In our multivariate setting, we observe a steeper downward correction for the Holm procedure than for the Bonferroni procedure, even though the former is more powerful. This result arises because the Holm procedure provides a threshold closer to the $t$-statistic for ``Dollars given not including match'' than the Bonferroni procedure.

\section{\label{sec:application mutual funds}Application: Outperforming mutual funds}

Our second application uses our method to infer the performance of mutual funds identified as market outperformers through multiple hypothesis testing among hundreds of funds, illustrating the computational scalability of our approach.

The multiple testing challenge of detecting outperforming funds among numerous candidates has been examined by \textcite{giglio2021thousands}, and similar issues arise in other areas of finance \parencite[e.g.,][]{harvey2016and, harvey2020evaluation, heath2023reusing}.

We examine fixed-income mutual funds using the CRSP Survivor-Bias-Free U.S. Mutual Fund Database available through Wharton Research Data Services (WRDS). Our sample covers the period from January 2000 to April 2024. After excluding funds with missing data, we obtain monthly returns for 371 funds.

A mutual fund is said to outperform the market if it exhibits a significantly positive ``alpha,'' representing excess returns unexplained by exposure to systematic risk factors.
Each fund's alpha is estimated using the Fama-French five-factor model \parencite{fama2015five},
\begin{align}
R_{i,t} = \alpha_i + b_i R_{m,t} + s_i R_{smb,t} + h_i R_{hml,t} + r_i R_{rmw,t} + c_i R_{cma,t} + \epsilon_{i,t},
\end{align}
where $R_{i,t}$ denotes the fund's excess return over the risk-free rate, proxied by the beginning-of-month 30-day T-bill yield. The factors include the market excess return ($R_{m,t}$) and four zero-investment portfolios capturing size ($R_{smb,t}$), value ($R_{hml,t}$), profitability ($R_{rmw,t}$), and investment ($R_{cma,t}$) effects.

This model estimates 371 alphas and $371 \times 5$ factor loadings. To address the high dimensionality of the resulting covariance matrix, we employ the principal orthogonal complement thresholding (POET) method of \textcite{fan2013large}. The POET estimator is obtained by thresholding the remaining components of the sample covariance matrix (the principal orthogonal complement) after removing the first $K$ principal components. For our dataset, $K=1$ is selected using the criterion of \textcite{bai2002determining}, and we use the default soft-thresholding implementation in the R package \texttt{POET} \parencite{Rpoet2016}. This covariance matrix is used to compute $t$-statistics and subsequently for implementing our conditional inference procedures.

Following \textcite{giglio2021thousands}, we apply multiple hypothesis testing to identify funds with significantly positive alphas, interpreted as market outperformers. We employ the Holm (FWER control) and \textcite[][henceforth BY]{benjamini2001control} (FDR control) procedures, setting the nominal FWER and FDR levels to 10\%. Both procedures detect five funds with significantly positive alphas (see Table~\ref{tab: funds}).

\begin{table}
	\centering
	\footnotesize
	\begin{tabular}{lcccc}
		\toprule
		\multicolumn{1}{c}{} & \multicolumn{1}{c}{} & \multicolumn{3}{c}{Conditional} \\
		\cmidrule(l{3pt}r{3pt}){3-5}
		Fund Ticker & Unconditional & Holm & BY  &   \\
		\midrule
		ACEIX & \makecell{0.399 \\ (0.088, 0.711)} 
		      & \makecell{0.383 \\ (0.047, 0.598) [0.884]} 
		      & \makecell{0.392 \\ (0.108, 0.598) [0.787]} &  \\
		AMECX & \makecell{0.509 \\ (0.131, 0.887)} 
		      & \makecell{0.497 \\ (0.135, 0.750) [0.813]} 
		      & \makecell{0.504 \\ (0.188, 0.751) [0.743]} &  \\
		FKINX & \makecell{0.528 \\ (0.244, 0.813)} 
		      & \makecell{0.528 \\ (0.341, 0.710) [0.649]} 
		      & \makecell{0.528 \\ (0.344, 0.710) [0.644]} &  \\
		NOIEX & \makecell{0.481 \\ (0.189, 0.774)} 
		      & \makecell{0.481 \\ (0.273, 0.668) [0.674]} 
		      & \makecell{0.481 \\ (0.282, 0.668) [0.660]} &  \\
		VWINX & \makecell{0.435 \\ (0.129, 0.741)} 
		      & \makecell{0.430 \\ (0.167, 0.630) [0.756]} 
		      & \makecell{0.433 \\ (0.195, 0.630) [0.711]} &  \\ 
		\bottomrule
	\end{tabular}
\caption{\label{tab: funds} 
Unconditional and conditional estimates of fund alphas with 90\% confidence intervals (in parentheses). 
Unconditional intervals are Bonferroni-adjusted for all 371 funds, and conditional intervals for the five selected funds. 
Ratios of conditional to unconditional interval lengths are shown in brackets.}
\end{table}

Having identified five significant funds, we now use our methods to quantify the extent to which they outperform the market. Because these few outperformers are found significant among a large pool of 371 funds, their estimated alphas are likely to be positively selected. Our procedures explicitly account for this issue and provide unbiased inference on the outperformers' alphas. Importantly, our algorithm can handle the complex selection event that arises from testing all 371 alphas jointly.

We condition on the event $\hat{S} \supseteq S$, that is, on finding the five outperformers but not on finding the remaining funds insignificant. This choice is natural because, as discussed in Section~\ref{sec:conditioning event}, for which funds we find insignificant alphas does not affect our interpretation of the outperformers.

Table~\ref{tab: funds} reports conditionally median-unbiased alphas and conditional confidence intervals for the outperforming funds, alongside conventional (unconditional) estimates and intervals for comparison. The nominal coverage level for all confidence intervals is set to $1 - \alpha = 0.9$. Conventional intervals are Bonferroni-corrected for the total number of funds ($m=371$), whereas conditional intervals are Bonferroni-corrected for the number of selected funds ($\lvert \hat{S} \rvert = 5$).

The median-unbiased estimates are only slightly smaller than their unconditional counterparts, suggesting that positive selection bias is of limited practical importance in this application. Conditional confidence intervals are generally narrower than their unconditional analogues, with length reductions of roughly 12-35\% under the Holm-based procedure and 21-36\% under the BY-based procedure, reflecting the reduced Bonferroni correction under conditional inference.

Testing alphas against zero identifies statistical rather than economic significance. We assess economic significance based on the lower bounds of the joint conditional confidence intervals, which exceed 0.135 for four of the five outperforming funds, indicating economically meaningful outperformance.

\section{\label{sec:conclusion}Conclusion}

% In this paper, we have developed methods for conducting statistical inference following multiple hypothesis testing. Our proposed algorithm effectively handles cases where numerous hypotheses are tested simultaneously. This scalability is essential because empirical studies can involve many statistical analyses, with only a few selected results receiving emphasis. Such selective reporting can occur even in studies with careful research designs and pre-registration or pre-analysis plans. A further complexity lies in determining the appropriate conditioning for selection: should we condition on the realization of all significant results or only on the event that specific results are selected? Our procedure accommodates various definitions of selection events.
In this paper, we have proposed new methods for conducting inference on the magnitude of significant effects detected through multiple hypothesis testing. Our methods support a wide range of multiple testing procedures, scale to scenarios involving many estimated effects, and accommodate different selection events that operationalise the notion of ``observed significance.''

While our primary focus is on inference after multiple hypothesis testing, our procedure can also be applied to conduct inference conditional on pre-test and placebo results. For instance, a pre-test is considered passed if the effects of one or more placebo outcomes are found to be insignificant. This type of test is commonly used, for example, when assessing parallel trends in a difference-in-differences design \parencite{roth2022pretest}. In the online appendix, we provide another example where property crime serves as a placebo outcome to test an empirical specification for estimating an effect on violent crime.

Our approach follows the multiple testing literature in assuming that the full set of tested hypotheses is known. Addressing situations where insignificant results are hidden requires a different framework \parencite{andrews2019identification, berk_valid_2013}. As shown by \textcite[Proposition 4]{sarfati2025post}, limiting attention to a pre-specified set of possible hypotheses is essential for valid conditional inference. This requirement implies that the framework cannot accommodate unrestricted data mining.

%An interesting avenue for future research is to explore alternative conditional inference procedures, such as data carving \parencite{fithian_optimal_2017} and selection with randomised response \parencite{tian_selective_2018}. These methods address some limitations of the polyhedral approach \parencite{kivaranovic2021, kivaranovic2024}, but they require specific modifications at the selection stage and therefore depart from standard multiple-testing procedures. In particular, they cannot be used to correct for selection bias in existing studies that rely on conventional significance testing. 

Finally, our numerical examples demonstrate the strong power properties of our procedure, and it would be worthwhile to explore these aspects further from a theoretical perspective. Our empirical illustrations in Sections \ref{sec:application list} and \ref{sec:application mutual funds} suggest that our procedure can produce informative confidence intervals. Additionally, the simulation results presented in Section \ref{sec:further_simu} confirm its strong power characteristics. It would be interesting to investigate whether the optimal selective inference discussed by \textcite{fithian_optimal_2017} can be applied in this context.

\section*{Acknowledgments and Funding}

We are grateful for comments from Erik Hjalmarsson, Mikael Lindahl and participants at CFE-CMStatistics 2022, the 4th Aarhus Workshop in Econometrics, IPDC2025 in Montpellier, and Econometrics Forum 2025 in Awara. Dzemski acknowledges financial support from Jan Wallanders och Tom Hedelius stiftelse samt Tore Browaldhs stiftelse under grant P24-0135. Okui acknowledges financial support from the Japan Society for the Promotion of Science under KAKENHI Grant Nos.23K25501. Wang acknowledges financial support from the Singapore Ministry of Education Tier 1 grants RG104/21, RG51/24, and NTU CoHASS Research Support Grant.
The authors made use of the AI tools Grammarly and ChatGPT for language editing. All AI-generated suggestions were reviewed and verified by the authors, who take full responsibility for the content of the article. All errors remain our own.

\printbibliography

\singlespacing
\pagebreak 
\appendix 
\begin{refsection}

\begin{center}
    {\large\bfseries Online appendix for ``Inference on effect size after multiple hypothesis testing''\par\vspace{0.5em}}% Title
    {\normalsize\scshape Andreas Dzemski \quad Ryo Okui \quad Wenjie Wang\par\vspace{0.5em}}% Author
    {\normalsize\today}% Date
  \end{center}

\useatocs
\listofatocs
\pagebreak
\section*{Overview of Online Appendix}

The appendix is organised into six sections. Appendix \ref{sec:appendix application} presents an additional empirical example that examines the impact of exposure to conflict on violent crime. This section also includes further analyses related to the example in Section \ref{sec:application list}. 
Appendix \ref{sec: app multiple testing} supplements the discussion of multiple testing protocols. In particular, Appendix~\ref{sec: app more step-down rules} gives an overview of different step-down methods, Appendix \ref{sec: appendix step-up rules} discusses step-up testing, and Appendix \ref{sec:RW bootstrap} describes our implementation of the RW2005 step-down procedure. Appendix \ref{sec:more algorithms} details the algorithm used to compute selection events. Appendix \ref{sec:additional theory} offers supplementary theoretical results, focusing on the properties of the Rosenblatt transformation and alternative definitions of the selection event. Appendix \ref{S: A: proof} contains the proofs of the theorems, while technical lemmas are provided in Appendix \ref{sec:lemmas}.

\section{\label{sec:appendix application}Empirical applications}

Section~\ref{sec: diff in means by regression} describes how we implement the difference-in-means estimators by regression to obtain their estimated covariance matrix. In Section~\ref{sec:application Couttenier}, we revisit the study by \textcite{couttenier2019violent}. Section~\ref{sec:list many} examines the situations with many hypotheses in the context of the empirical research by \textcite{karlan2007does}.

\subsection{\label{sec: diff in means by regression}Implementing the difference-in-means estimator by regression}

Here we discuss the estimation of treatment effects in our replication of the studies by \textcite{karlan2007does} and \textcite{list2019multiple} in Section~\ref{sec:application list} of the main text. We use their difference-in-means estimators but implement them by linear regression. This approach allows us to estimate all treatment effects simultaneously and to obtain the estimated variance-covariance matrix of the treatment effect estimators.

\paragraph{Multiple subgroups}
We use the same difference-in-means estimator as \textcite{list2019multiple} with the control group given by the participants not being offered a matching grant. It is implemented by linear regression. 
We label these subgroups as $z = 1, \dotsc, 4$ and denote participant $i$'s subgroup as $Z_i$.
The binary treatment $D_i$ represents the random assignment to one of 36 possible matching grants, while the outcome $Y_{i}$ indicates whether participant $i$ donated (response rate).
The difference-in-means estimator for subgroup $z$ is given by an estimate of the coefficient $\theta_{z}$ in the linear regression
\begin{align*}
 Y_{i} = \sum_{z = 1}^4 \alpha_z \mathbf{1} \{Z_i = z\} + \sum_{z = 1}^4 \theta_{z} \mathbf{1} \{Z_i = z, D_i = 1\} + \epsilon_i,
\end{align*}
where $\epsilon_i$ is an error term. The lower-right $4 \times 4$ block of the estimated variance-covariance matrix for the regression provides the estimate $\hat{V}$ required by our conditional inference procedure. 

\paragraph{Multiple outcomes}
We label the outcomes as $k = 1, \dotsc, 4$ and denote the outcome for participant $i$ as $Y_{k, i}$. The binary treatment $D_i$ is defined in the same way as described previously. Implementing the difference-in-means estimator for multiple outcomes requires first reshaping the data such that each row corresponds to a unique combination of unit and outcome.
Enumerate the rows in this format as $j = 1, \dotsc, 4n$. For a generic row $j$ corresponding to unit $i$ and outcome $k$, define the variables $I_j = i$, $K_j = k$, $Y_j = Y_{k, i}$ and $D_j = D_i$. 
To obtain the estimated treatment effects, fit the regression 
\begin{align*}
 Y_j = \sum_{k = 1}^4 \alpha_k \mathbf{1} \{K_j = k\} + \sum_{k = 1}^4 \theta_{k} \mathbf{1} \{K_j = k, D_j = 1\} + \epsilon_j,
\end{align*}
where $\epsilon_j$ is an error term. The OLS estimate of $\theta_k$ is the difference-in-means estimator for outcome $k$. We cluster at the unit level to ensure that the lower-right $4 \times 4$ block of the variance-covariance matrix for the regression is a valid estimate of $V$.

\subsection{\label{sec:application Couttenier}Additional empirical application: Exposure to conflict and violent crime}

We consider an additional empirical application in which considering both insignificant and significant results is crucial in interpreting the results. 
This application is based on \textcite{couttenier2019violent}, who study the effect of childhood exposure to a violent conflict on violent crime. They use data on male asylum seekers in Switzerland and identify the effect of conflict exposure from variation in crime propensity between cohorts born before and after a conflict in their country of origin. Male asylum seekers are observed at an aggregated level with the unit of observation defined by the interaction of country of origin (denoted by subscript $n$), year of birth (denoted by subscript $a$), and year of observation (denoted by subscript $t$). 

The main empirical specification is 
\begin{align}
    \label{eq:Couttenier specification}
 \text{CP}_{k, n, a , t} = \theta_k \texttt{kid}_{n, a, t} + \sum_{j \geq 13} \beta_{k, j} \texttt{expo}_{n, a, t, j} + \mathbf{FE}_{k, n, t} + \mathbf{FE}_{k, a} + \epsilon_{k, n, a, t},
\end{align}
where $\text{CP}_{k, n, a, t}$ is the propensity to commit a crime in category $k$ for individuals from country $n$ born in year $a$ and observed in year $t$. The variable $\texttt{kid}_{n, a, t}$ is a binary indicator for individuals exposed to violent conflict in their country of origin before age 13. The coefficient on this variable is the effect of childhood exposure to conflict on crime propensity. 
The variables, $\texttt{expo}_{n, a, t, j}$, control for conflict exposure at a later age $j$. Additional controls are given by fixed effects for age ($\mathbf{FE}_{k, a}$) and fixed effects of country of origin and year of observation ($\mathbf{FE}_{k, n, t}$). Finally, $\epsilon_{k, n, a, t}$ is an error term.

The key finding presented in \textcite{couttenier2019violent} is derived from both insignificant and significant results. They argue that their estimation results provide evidence for conflict exposure as a causal mechanism that perpetuates violence. This interpretation rests on finding an effect on violent crimes but not other crime categories \parencite[p4398]{couttenier2019violent}.
Passing this placebo test is a key part of their analysis that leads them to conclude in their abstract that
\begin{quote}
 [\dots] cohorts exposed to civil conflict/mass killings during childhood are 35 percent more prone to violent crime than the average cohort.
\end{quote}

The estimated effect of a 35\% increase is potentially subject to bias from multiple hypothesis testing due to carrying out a placebo test.
We provide unbiased inference by conditioning on observing $\hat{S} = S$, i.e., we condition on both significance and insignificance. This conditioning corresponds to passing the placebo test (observing an insignificant effect on property crimes) and then finding a significant effect on the primary outcome (violent crime).  

We consider three procedures to test the significance of the two outcomes. The first procedure is to conduct two consecutive $t$-tests. This approach fits into our framework\footnote{The two $t$-tests can be seen as a step-down rule with invariant thresholds.} but is not a valid multiple-hypothesis testing procedure to control the overall size. However, it has good power for detecting a potential placebo effect, which may be the primary concern in practice. In addition to the consecutive $t$-tests, we consider the Holm and RW2005 procedures.

We set the nominal size of the consecutive $t$-tests to 10\% and the FWER level of the Holm and RW2005 procedures to 10\%. We carry out both one-sided and two-sided tests. For simplicity, we choose the same type of hypothesis for both outcomes.\footnote{Alternatively, one would use a one-sided test for the main outcome and a two-sided test for the placebo outcome. It is straightforward to extend our framework to accommodate this.}

\begin{table}
\centering 
\footnotesize

\begin{tabular}{lccccc}
\toprule
\multicolumn{1}{c}{ } & \multicolumn{3}{c}{Violent Crime} & \multicolumn{2}{c}{Property Crime} \\
\cmidrule(l{3pt}r{3pt}){2-4} \cmidrule(l{3pt}r{3pt}){5-6}
 & uncond & signif & cond & uncond & signif\\
\midrule
\addlinespace[0.3em]
\multicolumn{6}{l}{\textbf{1-sided}}\\
\hspace{1em}T-test & \makecell{0.35\\ (0.06, 0.64)} & \checkmark & \makecell{0.47\\ (-0.05, 1.19)} & \makecell{0.15\\ (-0.17, 0.47)} & \ding{55}\\
\hspace{1em}Holm & \makecell{0.35\\ (0.06, 0.64)} & \checkmark & \makecell{0.42\\ (-0.24, 1.19)} & \makecell{0.15\\ (-0.17, 0.47)} & \ding{55}\\
\hspace{1em}RW2005 & \makecell{0.35\\ (0.06, 0.64)} & \checkmark & \makecell{0.41\\ (-0.10, 1.02)} & \makecell{0.15\\ (-0.17, 0.47)} & \ding{55}\\
\addlinespace[0.3em]
\multicolumn{6}{l}{\textbf{2-sided}}\\
\hspace{1em}T-test & \makecell{0.35\\ (0.06, 0.64)} & \checkmark & \makecell{0.31\\ (-0.26, 0.79)} & \makecell{0.15\\ (-0.17, 0.47)} & \ding{55}\\
\hspace{1em}Holm & \makecell{0.35\\ (0.06, 0.64)} & \checkmark & \makecell{0.18\\ (-0.68, 0.76)} & \makecell{0.15\\ (-0.17, 0.47)} & \ding{55}\\
\hspace{1em}RW2005 & \makecell{0.35\\ (0.06, 0.64)} & \checkmark & \makecell{0.27\\ (-0.34, 0.74)} & \makecell{0.15\\ (-0.17, 0.47)} & \ding{55}\\
\bottomrule
\end{tabular}
\caption{\label{tab:crime types}Estimated effects of childhood exposure to conflict on violent and property crimes. The first column gives the tested null hypothesis and choice of step-down procedure. ``T-test'' refers to the procedure that performs two consecutive univariate $t$-tests. Columns labeled ``uncond'' give unconditional point estimates and confidence intervals in parentheses. Columns labeled ``cond'' give conditionally median-unbiased point estimates and conditional confidence intervals in parentheses. The nominal level for all confidence intervals is 10\%. ``signif'' indicates the significance of the outcome under the given multiple testing procedure.}
\end{table}
To implement our approach, we must obtain an estimate $\hat{V}$ of the variance-covariance matrix of the estimated treatment effects. We accomplish this by estimating the fixed-effect regression~\eqref{eq:Couttenier specification} simultaneously for both outcomes by interacting the right-hand side with the indicator variable for the different outcomes, similar to the approach used for the specification with multiple outcomes described in Section~\ref{sec: diff in means by regression} above. Following \textcite{couttenier2019violent}, we cluster the standard errors by country of origin.

Our inferential results are summarised in Table~\ref{tab:crime types} where we express the effect sizes as a percentage of the corresponding population average to facilitate comparison with the reported 35\% increase.
Regardless of the choice of step-down test, we find the effect on violent crimes to be significant and the effect on property crimes to be insignificant, confirming the finding of \textcite[Table~5a and Table~6]{couttenier2019violent}. 

For one-sided testing, the conditionally median-unbiased estimator proposes an upward correction of the estimated effect on violent crimes. For two-sided testing, the bias correction is downward. The magnitude of the bias correction depends on the specific testing procedure but is generally moderate compared to the error margins.

For all testing procedures, conditional confidence intervals are wider than their unconditional counterparts and do not allow us to bound the effect size away from zero. This result is not surprising since unconditional inference ignores statistical uncertainty arising from placebo testing and establishing the significance of the main effect. We do not interpret this observation as casting doubt on the conclusions in \textcite{couttenier2019violent}, but rather as reflecting a broader issue of power considerations in empirical research, not reflecting sources of uncertainty ignored by the state-of-the-art approach to statistical inference.

\subsection{\label{sec:appendix additional results list}Additional results for application to charitable giving}

\subsubsection{\label{sec:list subgroups}Multiple subgroups}

\textcite{karlan2007does} conduct a subgroup analysis by classifying participants according to the political leaning of their home county and state. A county or state is designated as ``red'' if George Bush won it in the 2004 presidential election and as ``blue'' otherwise. 
We consider four subgroups: blue counties in blue states, blue counties in red states, red counties in blue states, and red counties in red states. 
Effect size is measured in terms of the standard deviation of the response rate in the control group (=13\%). 

To detect significantly positive effects, we use the one-sided versions of the three multiple testing procedures.  All three step-down procedures detect two subgroups with significant effects: ``blue county in a red state'' and ``red county in a red state'' with estimated effect sizes of 0.053 and 0.074 standard deviations, respectively. 

Our inference results conditional on the selection event $\hat{S} = S$ are summarised in Table~\ref{tab:LSXtab2}.
\begin{table}
	\centering	
	\footnotesize
	
\begin{tabular}{lcccc}
\toprule
\multicolumn{1}{c}{} & \multicolumn{1}{c}{} & \multicolumn{3}{c}{Conditional} \\
\cmidrule(l{3pt}r{3pt}){3-5}
Subgroup & Unconditional & RW2005 & Holm & Bonferroni\\
\midrule
blue county in blue state & \makecell{0.011 \\ (-0.025, 0.048)} & - & - & -\\
blue county in red state & \makecell{0.053 \\ (-0.004, 0.111)} & \makecell{-0.007 \\ (-0.300, 0.094)} & \makecell{-0.018 \\ (-0.354, 0.093)} & \makecell{-0.109 \\ (-0.826, 0.082)}\\
red county in blue state & \makecell{-0.000 \\ (-0.049, 0.049)} & - & - & -\\
red county in red state & \makecell{0.072 \\ (0.031, 0.113)} & \makecell{0.072 \\ (0.031, 0.107)} & \makecell{0.072 \\ (0.031, 0.107)} & \makecell{0.071 \\ (0.029, 0.107)}\\
\bottomrule
\end{tabular}
	\caption{\label{tab:LSXtab2}Multiple subgroups \parencite[Specification from Table~2 in][]{list2019multiple}. Effect size is measured in standard deviations. Confidence intervals (CI) are given in parentheses. Unconditional CI are Bonferroni-adjusted for the total number of subgroups (=4), conditional CI are adjusted for the number of significant effects (=2).}
\end{table}
The conditional confidence intervals are Bonferroni-adjusted for the number of significant effects (=2) as in equation \eqref{eq:joint confidence set}. The unconditional intervals are adjusted for the total number of subgroups (=4).

Both unconditional and conditional confidence intervals reveal substantial uncertainty for the group ``blue county in red state.'' 
For this group, our inference does not pin down a range of possible effect sizes that is clearly above zero and, therefore, does not provide conclusive evidence of economic gains. The lower bound of the conditional confidence interval is equal to -0.3 standard deviations or less, and the median-unbiased effect size estimate is negative for all three multiple testing procedures.

For the group ``red county in red state,'' the effect size is statistically well separated from zero, and the bias correction is minor. The conditional confidence interval is narrower than the unconditional simultaneous interval, demonstrating possible power gains through conditional inference. These gains affect mainly the upper bound, leaving the lower bound virtually unchanged. This asymmetric adjustment reflects the positive bias in positively selected significant effects. 

\subsubsection{\label{sec:list many}Multiple treatments, outcomes and subgroups}

\textcite{list2019multiple} study a specification with multiple treatments, outcomes, and subgroups. In addition to the outcomes considered in \ref{sec:application list} and the four subgroups from Section~\ref{sec:list subgroups}, they define three treatments corresponding to the randomised matching ratios 1:1, 1:2, and 1:3. This gives rise to 48 treatment effects corresponding to the unique combinations of matching ratio, outcome, and subgroup. 

The step-up procedure of \textcite[][henceforth BY2001]{benjamini2001control} controls the False Discovery Rate (FDR). It offers a weaker coverage guarantee than FWER control but may be potentially more powerful. We apply the BY2001 and RW2005 procedures to conduct two-sided significance testing. We set the nominal FDR level to 10\% for the BY2001 procedure and the FWER level to 10\% for the RW2005 procedure.

The BY2001 procedure detects 12 significant effects, while the RW2005 procedure detects 15. To ensure that our comparison of the conditional confidence sets is not affected by the different number of selected effects, we report univariate confidence intervals without Bonferroni adjustment. Table~\ref{tab:list many} below gives full selection and conditional inference results. Here, we focus on the results for inference conditional on the event $\hat{S} = S$ in more detail.

Figure~\ref{fig:tab5 Holm vs BY} plots the bias correction and the width of the conditional confidence intervals for the significant effects against the absolute value of their t-statistics. The bias correction is calculated as the difference between the conditional and unconditional point estimates.
\begin{figure}
	\centering
	\input{figures/fig_tab5_HolmvsBY_bias}
	\caption{\label{fig:tab5 Holm vs BY}Comparison of RW2005 procedure and BY2001 for selection among 48 effects, conditional on the $\hat{S} = S$ event. The upper panel gives the bias correction calculated as the difference between the conditional and unconditional point estimate. The lower panel shows the width of the conditional confidence interval.}
\end{figure}
The figure shows a positive correlation between the amount of bias correction and the width of the conditional confidence interval. The following observation explains this correlation. Both bias and confidence interval reflect a deviation from Gaussianity in the conditional distribution. This deviation becomes more pronounced the closer the observed effects are to the boundary of the selection event. In a relevant sense, what effects are ``close to the boundary'' is not immediately obvious and depends on the correlation structure of the estimated effects. For example, our procedure proposes a substantial downward correction for the outcome ``Dollars given including match'' for individuals in the subgroup ``blue county in red state'' facing the 3:1 matching ratio, even though this effect is estimated with a relatively large $t$-statistic of 3.89. The smaller $ t$-statistics explain the bias correction for the highly correlated outcomes ``Dollars given not including match'' ($t$-statistic of 1.69) and ``response rate'' ($t$-statistic of 2.22).

Figure~\ref{fig:tab5 Holm vs BY} also reveals that conditional inference on the significant effects is sensitive to the choice of multiple-testing procedure. For some effects, the bias correction after the BY2005 selection has the opposite sign of the bias correction after the RW2005 selection. 
Here, the two multiple-testing procedures detect different sets of significant effects. 
Even if two procedures detect the same set of significant effects, conditional inference will depend on the specific procedure used for selection. 
\begingroup
\footnotesize
\begin{landscape}
\begin{longtable}{llcccccc}
\caption{\label{tab:list many}Inference after multiple hypothesis testing for Table~5 in \textcite{list2019multiple} based on two-sided testing and the BY2001 or the RW2005 procedures. For conditional inference, the point estimate is the conditionally median-unbiased estimator, and the 90\% conditional confidence interval is given in parentheses. The column labeled ``Uncond'' gives the conventional point estimator and confidence interval.} \\
\multicolumn{1}{c}{ } & \multicolumn{1}{c}{ } & \multicolumn{1}{c}{ } & \multicolumn{1}{c}{ } & \multicolumn{2}{c}{Conditional on $\hat{S} = S$} & \multicolumn{2}{c}{Conditional on $\hat{S} \supseteq S$} \\
\cmidrule(l{3pt}r{3pt}){5-6} \cmidrule(l{3pt}r{3pt}){7-8}
Outcome & Subgroup & Treatment & Uncond & BY2001 & RW2005 & BY2001 & RW2005\\
\midrule
\endfirsthead
\multicolumn{1}{c}{ } & \multicolumn{1}{c}{ } & \multicolumn{1}{c}{ } & \multicolumn{1}{c}{ } & \multicolumn{2}{c}{Conditional on $\hat{S} = S$} & \multicolumn{2}{c}{Conditional on $\hat{S} \supseteq S$} \\
\cmidrule(l{3pt}r{3pt}){5-6} \cmidrule(l{3pt}r{3pt}){7-8}
Outcome & Subgroup & Treatment & Uncond & BY2001 & RW2005 & BY2001 & RW2005 \\
\midrule
\endhead
\midrule
\multicolumn{3}{l}{\textsl{continued on the next page\ldots}}\\*
\endfoot
\bottomrule
\endlastfoot

Amount change & blue county in blue state & 1:1 vs. control & \makecell{-6.38\\ (-25.42, 12.66)} & - & - & - & -\\
Amount change & blue county in blue state & 2:1 vs. control & \makecell{-0.06\\ (-0.46, 0.34)} & - & - & - & -\\
Amount change & blue county in blue state & 3:1 vs. control & \makecell{0.14\\ (-0.24, 0.52)} & - & - & - & -\\
Amount change & blue county in red state & 1:1 vs. control & \makecell{11.29\\ (-23.62, 46.20)} & - & - & - & -\\
Amount change & blue county in red state & 2:1 vs. control & \makecell{11.46\\ (-23.45, 46.37)} & - & - & - & -\\
\addlinespace
Amount change & blue county in red state & 3:1 vs. control & \makecell{11.53\\ (-23.38, 46.44)} & - & - & - & -\\
Amount change & red county in blue state & 1:1 vs. control & \makecell{0.11\\ (-0.35, 0.58)} & - & - & - & -\\
Amount change & red county in blue state & 2:1 vs. control & \makecell{-0.04\\ (-0.54, 0.47)} & - & - & - & -\\
Amount change & red county in blue state & 3:1 vs. control & \makecell{0.06\\ (-0.40, 0.52)} & - & - & - & -\\
Amount change & red county in red state & 1:1 vs. control & \makecell{0.22\\ (-0.24, 0.68)} & - & - & - & -\\
\addlinespace
Amount change & red county in red state & 2:1 vs. control & \makecell{0.07\\ (-0.38, 0.51)} & - & - & - & -\\
Amount change & red county in red state & 3:1 vs. control & \makecell{-0.01\\ (-0.51, 0.49)} & - & - & - & -\\
Dollars given incl match & blue county in blue state & 1:1 vs. control & \makecell{0.09\\ (-0.01, 0.20)} & \makecell{-0.16\\ (-1.04, 0.11)} & \makecell{0.08\\ (-0.05, 0.15)} & \makecell{0.00\\ (-0.01, 0.11)} & \makecell{0.08\\ (-0.00, 0.15)}\\
Dollars given incl match & blue county in blue state & 2:1 vs. control & \makecell{0.23\\ (0.07, 0.40)} & \makecell{0.22\\ (0.07, 0.32)} & \makecell{0.23\\ (0.13, 0.34)} & \makecell{0.22\\ (0.07, 0.32)} & \makecell{0.23\\ (0.13, 0.32)}\\
Dollars given incl match & blue county in blue state & 3:1 vs. control & \makecell{0.32\\ (0.13, 0.50)} & \makecell{0.30\\ (0.15, 0.42)} & \makecell{0.32\\ (0.21, 0.46)} & \makecell{0.30\\ (0.15, 0.41)} & \makecell{0.32\\ (0.21, 0.42)}\\
\addlinespace
Dollars given incl match & blue county in red state & 1:1 vs. control & \makecell{0.10\\ (-0.06, 0.27)} & - & - & - & -\\
Dollars given incl match & blue county in red state & 2:1 vs. control & \makecell{0.35\\ (0.04, 0.65)} & \makecell{0.29\\ (-0.08, 0.57)} & \makecell{0.40\\ (0.05, 0.85)} & \makecell{0.28\\ (0.01, 0.50)} & \makecell{0.32\\ (0.04, 0.50)}\\
Dollars given incl match & blue county in red state & 3:1 vs. control & \makecell{0.56\\ (0.12, 1.00)} & \makecell{0.60\\ (0.15, 1.13)} & \makecell{-0.11\\ (-2.63, 1.30)} & \makecell{0.53\\ (0.14, 0.79)} & \makecell{0.11\\ (0.04, 0.66)}\\
Dollars given incl match & red county in blue state & 1:1 vs. control & \makecell{0.12\\ (-0.02, 0.26)} & - & \makecell{0.08\\ (-0.14, 0.20)} & - & \makecell{0.08\\ (-0.01, 0.19)}\\
Dollars given incl match & red county in blue state & 2:1 vs. control & \makecell{0.26\\ (0.05, 0.47)} & \makecell{0.28\\ (0.07, 0.54)} & \makecell{0.26\\ (0.12, 0.39)} & \makecell{0.24\\ (0.07, 0.37)} & \makecell{0.26\\ (0.12, 0.37)}\\
\addlinespace
Dollars given incl match & red county in blue state & 3:1 vs. control & \makecell{0.26\\ (0.05, 0.48)} & \makecell{0.28\\ (0.04, 0.54)} & \makecell{0.25\\ (0.08, 0.38)} & \makecell{0.24\\ (0.04, 0.38)} & \makecell{0.26\\ (0.11, 0.38)}\\
Dollars given incl match & red county in red state & 1:1 vs. control & \makecell{0.18\\ (0.02, 0.35)} & \makecell{0.21\\ (-0.06, 0.52)} & \makecell{0.23\\ (-0.10, 0.64)} & \makecell{0.15\\ (-0.07, 0.27)} & \makecell{0.13\\ (-0.04, 0.26)}\\
Dollars given incl match & red county in red state & 2:1 vs. control & \makecell{0.31\\ (0.10, 0.52)} & \makecell{0.23\\ (-0.19, 0.54)} & \makecell{0.55\\ (0.21, 1.40)} & \makecell{0.21\\ (-0.19, 0.40)} & \makecell{0.30\\ (0.15, 0.42)}\\
Dollars given incl match & red county in red state & 3:1 vs. control & \makecell{0.40\\ (0.18, 0.61)} & \makecell{0.35\\ (0.03, 0.60)} & \makecell{0.51\\ (0.26, 0.98)} & \makecell{0.34\\ (0.03, 0.50)} & \makecell{0.39\\ (0.23, 0.51)}\\
Dollars given not incl match & blue county in blue state & 1:1 vs. control & \makecell{-0.00\\ (-0.06, 0.06)} & - & - & - & -\\
\addlinespace
Dollars given not incl match & blue county in blue state & 2:1 vs. control & \makecell{0.01\\ (-0.06, 0.07)} & - & - & - & -\\
Dollars given not incl match & blue county in blue state & 3:1 vs. control & \makecell{0.00\\ (-0.06, 0.06)} & - & - & - & -\\
Dollars given not incl match & blue county in red state & 1:1 vs. control & \makecell{0.00\\ (-0.09, 0.10)} & - & - & - & -\\
Dollars given not incl match & blue county in red state & 2:1 vs. control & \makecell{0.05\\ (-0.07, 0.17)} & - & - & - & -\\
Dollars given not incl match & blue county in red state & 3:1 vs. control & \makecell{0.07\\ (-0.06, 0.20)} & - & - & - & -\\
\addlinespace
Dollars given not incl match & red county in blue state & 1:1 vs. control & \makecell{-0.00\\ (-0.09, 0.09)} & - & - & - & -\\
Dollars given not incl match & red county in blue state & 2:1 vs. control & \makecell{0.00\\ (-0.08, 0.09)} & - & - & - & -\\
Dollars given not incl match & red county in blue state & 3:1 vs. control & \makecell{-0.03\\ (-0.10, 0.05)} & - & - & - & -\\
Dollars given not incl match & red county in red state & 1:1 vs. control & \makecell{0.05\\ (-0.04, 0.14)} & - & - & - & -\\
Dollars given not incl match & red county in red state & 2:1 vs. control & \makecell{0.05\\ (-0.03, 0.13)} & - & - & - & -\\
\addlinespace
Dollars given not incl match & red county in red state & 3:1 vs. control & \makecell{0.04\\ (-0.03, 0.11)} & - & - & - & -\\
Response rate & blue county in blue state & 1:1 vs. control & \makecell{0.00\\ (-0.06, 0.07)} & - & - & - & -\\
Response rate & blue county in blue state & 2:1 vs. control & \makecell{0.01\\ (-0.06, 0.07)} & - & - & - & -\\
Response rate & blue county in blue state & 3:1 vs. control & \makecell{0.03\\ (-0.04, 0.09)} & - & - & - & -\\
Response rate & blue county in red state & 1:1 vs. control & \makecell{0.02\\ (-0.08, 0.12)} & - & - & - & -\\
\addlinespace
Response rate & blue county in red state & 2:1 vs. control & \makecell{0.06\\ (-0.05, 0.17)} & - & - & - & -\\
Response rate & blue county in red state & 3:1 vs. control & \makecell{0.08\\ (-0.03, 0.19)} & - & \makecell{-0.19\\ (-1.12, 0.19)} & - & \makecell{-0.08\\ (-0.09, 0.09)}\\
Response rate & red county in blue state & 1:1 vs. control & \makecell{0.00\\ (-0.08, 0.09)} & - & - & - & -\\
Response rate & red county in blue state & 2:1 vs. control & \makecell{0.02\\ (-0.07, 0.11)} & - & - & - & -\\
Response rate & red county in blue state & 3:1 vs. control & \makecell{-0.02\\ (-0.11, 0.06)} & - & - & - & -\\
\addlinespace
Response rate & red county in red state & 1:1 vs. control & \makecell{0.06\\ (-0.02, 0.13)} & - & \makecell{0.03\\ (-0.16, 0.19)} & - & \makecell{0.01\\ (-0.14, 0.09)}\\
Response rate & red county in red state & 2:1 vs. control & \makecell{0.08\\ (-0.00, 0.15)} & \makecell{0.02\\ (-0.22, 0.14)} & \makecell{0.14\\ (0.00, 0.37)} & \makecell{0.01\\ (-0.22, 0.10)} & \makecell{0.07\\ (-0.01, 0.12)}\\
Response rate & red county in red state & 3:1 vs. control & \makecell{0.08\\ (0.00, 0.16)} & \makecell{0.05\\ (-0.10, 0.16)} & \makecell{0.11\\ (0.01, 0.26)} & \makecell{0.05\\ (-0.10, 0.12)} & \makecell{0.07\\ (0.00, 0.12)}\\

\end{longtable}
\end{landscape}
\endgroup

\section{\label{sec: app multiple testing}Multiple hypothesis testing}
This appendix offers additional discussions on multiple hypothesis testing procedures. Specifically, we address step-down rules not covered in the main text, step-up testing methods, and the implementation details of the procedure by \textcite[][, RW2005]{romano2005stepwise}.

\subsection{\label{sec: app more step-down rules}More step-down rules}
The table below contains a list of some well-known step-down rules.
\begin{table}[h]
\centering
\begin{tabular}{lll} 
\toprule
$\alpha_j$ & Step-down rule & Controls\\ 
\midrule 
$\frac{\alpha}{m}$ & Bonferroni correction & FWER  \\
$1 - (1 - \alpha)^{1/m}$ & {\v{S}}id{\'a}k correction \parencite{vsidak1967rectangular} & FWER  \\
$\frac{\alpha}{m + 1 - j}$ & Bonferroni-Holm correction \parencite{holm1979simple} & FWER  \\
$1 - (1 - \alpha)^{\frac{1}{m + 1 - j}}$ & {\v{S}}id{\'a}k-Holm correction  & FWER  \\
$\frac{(\lfloor \gamma j \rfloor + 1)\alpha}{m + \lfloor \gamma j \rfloor + 1 - j}$ & FDP control \parencite{lehmann2005generalizations} & probability $\text{FDP} > \gamma$ \\
\bottomrule
\end{tabular}
\caption{\label{tab:thresholds_simple_stepdown} Step-down rules with thresholds $\bar{x}_j = \Phi^{-1} \left(1 - \alpha_j\right)$. FDP denotes the False Discovery Proportion, that is, the proportion of false discoveries among the significant effects.}
\end{table}
\subsection{\label{sec: appendix step-up rules}Step-up testing}

Two of the most commonly used step-up procedures are the procedures in \textcite{benjamini1995controlling} and \textcite{benjamini2001control}. These procedures control the false discovery rate (FDR), which is defined as the expected proportion of false positives among the significant effects. Under two-sided testing, the FDR is controlled at level $q$ if    
\begin{align*}
 \text{FDR} = \E \left[ \abs{\hat{S} \cap \{h: \theta_h = 0\}}/\abs{\hat{S}} \right] \leq q.
\end{align*}
The definition of the FDR for one-sided testing is obtained by replacing $\{h: \theta_h = 0\}$ with $\{h: \theta_h \leq 0\}$ in the previous display.

Step-up rules determine the set $\hat{S}^{\mathsf{c}}$ of insignificant treatment effects. The significant effects are then given by $\hat{S} = (\hat{S}^{\mathsf{c}})^{\mathsf{c}} = \{1, \dotsc, m\} \setminus \hat{S}^{\mathsf{c}}$. A one-sided step-up rule compares increasing thresholds with the descending ordering of test statistics.
Let increasing thresholds be denoted by
\begin{align*}
 \bar{x}_1 \leq \bar{x}_2 \leq \dotsm \leq \bar{x}_m.
\end{align*}
For example, for the procedure in \textcite{benjamini1995controlling}, we set 
\begin{align*}
 \bar{x}_j = \Phi^{-1} \left(1 - \frac{m - j + 1}{m}q\right),
\end{align*}
where $q$ is a user-defined parameter that gives the nominal false discovery rate (FDR).
For the procedure in \textcite{benjamini2001control} we set 
\begin{align*}
 \bar{x}_j = \Phi^{-1} \left(1 - \frac{m - j + 1}{m C_m} q \right),
\end{align*}
where $C_m = \sum_{i = 1}^m 1/ i \approx 1/2 + \log m$.
Let $(<, j)$ denote the data-dependent index of the $j$th-smallest test statistic, i.e.,
\begin{align*}
 X_{(<,1)} \leq X_{(<, 2)} \leq \dotsm \leq X_{(<,m)}.
\end{align*}

A one-sided step-up rule proceeds as follows: 
\begin{algorithm}[Step-up testing]
\label{alg:step_up}
\begin{enumerate}[label = (\Alph*), ref = \Alph*]
	\item 
	Initialise the set of insignificant effects $\hat{S}^{\mathsf{c}} \leftarrow \emptyset$ and the step counter $j \leftarrow 1$.
	\item 
	\label{alg:simple:step_down:check}
	If $j = m$ or if $X_{<, j} \geq \bar{x}_j$ then exit the algorithm and return the significant treatment effects $\hat{S} = (\hat{S}^{\mathsf{c}})^{\mathsf{c}}$. Otherwise, add $j$ to $\hat{S}^{\mathsf{c}}$.
	\item Increment the step counter $j \leftarrow j + 1$ and go to Step~\ref{alg:simple:step_down:check}.
\end{enumerate}
\end{algorithm}
A one-sided step-up rule can be turned into a two-sided step-up rule with replacing the test statistics $X_h$ by their absolute values and adjusting the threshold function. 

We now provide a counterpart of Lemma~\ref{lem:geometry generalized step-down} for step-up testing. 
\begin{Lemma}
	\label{lem:geometry support step-up}
	For realised test statistics $X = x$, Algorithm~\ref{alg:step_up} selects $\hat{S} = S$ if and only if, for all $h \in S$, 
	\begin{align*}
		x_h \geq \bar{x}_{m - \lvert S \rvert + 1}
	\end{align*}
	and either $S^{\mathsf{c}} = \emptyset$ or, for all $h \notin S$,
	\begin{align*}
		x_h < \bar{x}_{\sigma(h)}.
	\end{align*}
\end{Lemma}
\begin{Proof}
The proof is analogous to the proof of Lemma~\ref{lem:geometry generalized step-down}. We give it for completeness. The ``only if'' part follows by taking $\sigma = \tilde{\sigma}$ such that $\tilde{\sigma}^{-1} (j) = (<,j)$ for $j = 1, \dotsc, \lvert S^{\mathsf{c}} \rvert$. To prove the if part, we suppose that the sufficient condition holds for some $\sigma_1 \in \mathcal{E}(S^{\mathsf{c}})$ and construct a finite sequence of permutations $\sigma_1, \dotsc, \sigma_{N} = \tilde{\sigma}$ with the property that $\sigma_{n+1}$ satisfies the sufficient condition if $\sigma_{n}$ does. Consider two permutations $\sigma_n$ and $\sigma_{n+1}$ that are identical up to a neighbor swap, i.e.; there is $j \in \{1, \dotsc, \lvert S^{\mathsf{c}} \rvert - 1\}$ and $h, h' \in S^{\mathsf{c}}$ such that $h = \sigma_{n}^{-1} (j) =  \sigma_{n + 1}^{-1} (j + 1)$ and $h' = \sigma_n^{-1} (j + 1) = \sigma_{n+1}^{-1} (j)$, and $\sigma_n^{-1} (i) = \sigma_{n+1}^{-1} (i)$ for all $i = \{1, \dotsc, \lvert S^{\mathsf{c}} \rvert\} \setminus \{j, j+1\}$. Moreover, suppose that under $\sigma_n$, the neighbors are ordered in descending order, i.e., $x_h > x_{h'}$. If $\sigma_n$ satisfies the sufficient condition, so does $\sigma_{n + 1}$. To prove this, we have to verify the two conditions
\begin{align*}
 x_{h} < \bar{x}_{\sigma_{n + 1}(h)} \quad \text{and} \quad
 x_{h'} < \bar{x}_{\sigma_{n + 1}(h')}.
\end{align*}
These conditions are verified by noting that
\begin{align*}
 x_h < \bar{x}_{\sigma_{n}(h)} = \bar{x}_{j} \leq \bar{x}_{j + 1} = \bar{x}_{\sigma_{n + 1}(h)} \quad \text{and} \quad 
 x_{h'} < x_h < \bar{x}_{\sigma_{n}(h)} = \bar{x}_{j} = \bar{x}_{\sigma_{n + 1}(h')}.
\end{align*}
\end{Proof}

\subsection{RW2005 step-down testing for clustered regression using the Wild bootstrap\label{sec:RW bootstrap}}

We discuss the implementation details of the RW2005 procedure with the Wild bootstrap to account for the clustered correlation structure.
It is used in the analysis of multiple outcomes in Section \ref{sec:application list}.

Consider the regression model 
\begin{align*}
 Y_{k, i} = \alpha_k \, \mathbf{x}_{1, k, i} + \theta_j \, \mathbf{x}_{2, k, i} + \epsilon_{k, i}, 
\end{align*}
where $i = 1, \dotsc, n$ is a unit index and $k = 1, \dotsc, K$ is an outcome index, $Y_{k,i}$ is an observed outcome variable and $\mathbf{x}_{k, i} = (\mathbf{x}'_{1, k, i}, \mathbf{x}'_{2, k, i})'$ is a vector of observed covariates, and $\epsilon_{k, i}$ is an error term. The coefficient vector $\theta = (\theta_1', \dotsc, \theta_K')'$ is a vector of effects of interest (treatment effects) and has length $m = \sum_{k = 1}^K \dim(\mathbf{x}_k)$, where $\dim(\mathbf{x}_k)$ is the length of the covariate vector $\mathbf{x}_k$. 

For one-sided significant testing, we want to reject hypotheses 
\begin{align*}
	H_{0, h}: \theta_h \leq 0
\end{align*}
for $h = 1, \dotsc, m$. For two-sided testing, we want to reject hypotheses
\begin{align*}
	H_{0, h}: \theta_h = 0
\end{align*}
for $h = 1, \dotsc, m$.

We use the Wild bootstrap. The fitted model is given by 
\begin{align*}
	{Y}_{k, i} = \hat{\alpha}_k \, \mathbf{x}_{1, k, i} + \hat{\theta}_k \, \mathbf{x}_{2, k, i} + \hat{\epsilon}_{k, i}.
\end{align*}
The bootstrapped outcomes are given by 
\begin{align*}
	Y_{k, i}^* = \hat{\alpha}_k \, \mathbf{x}_{1, k, i} + W_i \, \hat{\epsilon}_{k, i},
\end{align*}
where $W_i$ is a random variable that is independent of $\hat{\epsilon}_{k, i}$ and has mean zero and variance one. For example, $W_i$ can have a Rademacher distribution, i.e., take the value $1$ with probability $1/2$ and the value $-1$ with probability $1/2$. To obtain a bootstrapped estimate $\theta^*$ of $\theta$, we re-estimate the regression on a bootstrapped sample that replaces the observed outcomes $Y_{i, k}$ by the bootstrapped outcomes $Y_{i, k}^*$.
This process also yields a vector of corresponding bootstrapped standard errors $s^*$.
Repeating this procedure for $B$ bootstrap samples yields $B$ bootstrapped estimates $\theta^*_1, \dotsc, \theta^*_B$ and bootstrapped standard errors ${s}^*_1, \dotsc, {s}^*_B$.

For subsets $A \subseteq \{1, \dotsc, m\}$ and a nominal FWER of $\beta$, define the threshold function for one-sided testing as
\begin{align*}
	\bar{x}(A) = \inf \left\{
		t \in \mathbb{R} :\frac{1}{B} \sum_{b = 1}^B \mathbf{1} \left( \max_{a \in A} \theta_{b, a}^* / {s}^*_{b,a} \leq t \right) \geq 1 - \beta
	\right\}.
\end{align*} 
A corresponding threshold function for two-sided testing is defined by replacing $\theta_{b, a}^*$ in the definition above with its absolute value.

\section{\label{sec:more algorithms}Algorithms for computing the conditional support}

This appendix provides various discussions about algorithms for computing the conditional suppositions. We first give a characterisation of the connectedness of the conditional support for a special case. Then, we present an approach to improve the algorithm. Next, we examine how the algorithm is modified to accommodate one-sided rules, a step-up rule, and alternative selection events.
\subsection{\label{sec:appendix bound on n_I}Characterisation of conditional support}

For the case of one-sided step-down rules and the selection event $\hat{S} = S$, we can characterise the conditional support $\mathcal{X}_s(z, S)$ more precisely. The following result gives an upper bound of $n_I$.
\begin{theorem}
    \label{thm:conditional support stepdown}
 Let $S_+ = \{h \in S: \Omega_{h, s} > 0\}$ and $S_- = \{h \in S: \Omega_{h, s} < 0\}$. For a one-sided step-down rule and the selection event $\hat{S} = S$, $n_I \leq (\lvert S_+ \rvert \lvert S_- \rvert + 1)$. 
\end{theorem}
In particular, the conditional support is guaranteed to be a connected interval if there are no treatment effect estimators that are negatively correlated with $X_s$ (that is, $S_- = \emptyset$).

\begin{Proof}[Proof of Theorem~\ref{thm:conditional support stepdown}]
We write $\mathbf{x} (x_s) = a_h x_s + b_h$ with $a_h = \Omega_{h, s}$ and $b_h = z_h$. Let $\sigma^*(x_s)$ denote the permutation that sorts the components of $\mathbf{x} (x_s)$ in descending order so that $\mathbf{x}_{\sigma^{*, -1}(1)} (x_s) \geq \mathbf{x}_{\sigma^{*, -1}(2)} (x_s) \geq \dotsb \geq \mathbf{x}_{\sigma^{*, -1}(m)} (x_s)$. 
We already know that 
\begin{align*}
	\mathcal{X}_s (z, S) = \bigcup_{k = 1}^{n_I} \, [\ell_k, u_k],
\end{align*}
for non-overlapping intervals $[\ell_k, u_k]$, $k = 1, \dotsc, n_I$. 
By Lemma~\ref{lem:geometry generalized step-down}, $x_s \in I$ is contained in the closure of $\mathcal{X}_s(z, S)$ if and only if 
\begin{align}
    \label{eq:condition_selected}   
	\mathbf{x}_h (x_s) \geq \bar{x}_{\sigma^*(x_s), h} \qquad & \text{for all $h \in S$}
	\\
	\label{eq:condition_not_selected}
	\mathbf{x}_h (x_s) \leq \bar{x}_{\sigma^*(x_s), h} \qquad & \text{for all $h \notin S$}.
\end{align}
We have $\ell_k \geq \ell_{\min}$ and $u_k \leq u_{\max}$ for all $k = 1, \dotsc, n_I$ with $\ell_{\min}$ and $u_{\max}$ defined in Section~\ref{sec:appendix initial bounds} above. We have
\begin{align*}
	a_h x_s + b_h \leq \bar{x} (S^{\mathsf{c}}) 
\end{align*}
for all $x_s$ in $[\ell_1, u_{n_I}]$ and $h \notin S$, implying that condition \eqref{eq:condition_not_selected} is satisfied over the whole range. Holes in the support can only occur because of violations of condition \eqref{eq:condition_selected}, i.e., violations of 
\begin{align}
	\label{eq:proof number intervals condition1}
	a_h x_s + b_h \geq \bar{x}_{\sigma(h)}.
\end{align}

Suppose without loss of generality that the intervals are ordered left to right, i.e., $u_k < \ell_{k+1}$ for $k = 1, \dotsc, n_I - 1$ and that there is a ``hole'' $(u_k, \ell_{k+1})$ between the $k$-th and $(k+1)$-th interval. We now bound the number of such holes.

Without loss of generality, we assume that the first $\lvert S \rvert$ hypotheses are the significant ones, i.e., $S = \{1, \dotsc, \lvert S \rvert\}$. For $h \leq \lvert S \rvert$, write $\bar{x}_{\sigma, h} = \bar{x} \left( \sigma^{-1} (\{\sigma (h), \dotsc, \lvert S \rvert\}) \cup S^{\mathsf{c}} \right)$ and $\bar{\mathbf{x}}_{\sigma} = (\bar{x}_{\sigma, 1}, \dotsc, \bar{x}_{\sigma, \lvert S \rvert})$. Let $\sigma^*(x_s)$ denote the permutation that sorts the components of $x = \mathbf{x} (x_s)$ in descending order. 

Consider any hole $(u_k, \ell_{k+1})$. For simplicity, suppose without loss of generality that $k =1$. For there to be a hole, condition \eqref{eq:proof number intervals condition1} fails when we move from $u_1$ to the right and 
\begin{align*}
	\mathbf{x}_{r^*} (u_1 + \epsilon) < \bar{\mathbf{x}}_{\sigma^*(u_1 + \epsilon), r^{*}},
\end{align*}
for some $r^* \leq \lvert S \rvert$ and $\epsilon > 0$ small enough that $x_{\text{left}} + \epsilon < \ell_2$. 
We now argue that, provided $\epsilon$ is small enough, hypothesis $r^*$ must be downward sloping, i.e., $C_{r^*}^{(s)} < 0$ and hence $\mathbf{x}_{r^*} (x_s)$ is decreasing in $x_s$. 

Let $\sigma_{-}^*(x)$ denote the limit of $\sigma^* (x)$ as we approach $x$ from the left, i.e., the limit of a sequence $\sigma^*(x_n)$ with $x_n \to $ with $x_n \in (-\infty, x)$. Let $\sigma_{+}(x)$ denote the correspondingly defined limit right limit. Note that both limits are well-defined since $\sigma$ is a step function. For $\epsilon$ small enough, $\sigma^*(u_1 - \epsilon) = \sigma^*_{-}(u_1)$ and $\sigma^*(u_1 + \epsilon) = \sigma^*_{+}(u_1)$. From now on, assume that $\epsilon$ is small enough that both equalities hold.

Write $\sigma_{-}^* = \sigma_{-}^*(u_1)$ and $\sigma_{+}^* = \sigma_{+}^*(u_1)$ and suppose that $\bar{x}_{\sigma_{-}^*, r^*} = \bar{x}_{\sigma_{+}^*, r^*}$. 
Since we now have 
\begin{align*}
	\mathbf{x}_{r^*} (u_1 - \epsilon) \geq \bar{x}_{\sigma_-^*, r^*} 
	\quad \text{and} \quad
	\mathbf{x}_{r^*} (u_1 + \epsilon) < \bar{x}_{\sigma_+^*, r^*}
	= \bar{x}_{\sigma_-^*, r^*},
\end{align*}
$\mathbf{x}_{r^*}(x)$ must be decreasing in $x$, i.e., $C^{(s)}_{r^*} < 0$.
On the other hand, suppose that $\bar{x}_{\sigma_{-}^*, r^*} \neq \bar{x}_{\sigma_{+}^*, r^*}$. 
By the properties of the threshold function $\bar{x}$, 
\begin{align*}
	\left\{ h = 1, \dotsc, m:
		\sigma_{-}^{*}(h) \leq \sigma_{-}^{*}(r^*) \right\} \neq
	\left\{ h = 1, \dotsc, m:
		\sigma_{+}^{*}(h) \leq \sigma_{+}^{*}(r^*) \right\}.
\end{align*}
This implies the existence of at least one hypothesis $r'$ that intersects $r^*$ in $[u_1 - \epsilon, u_1 + \epsilon]$. 
Letting $\epsilon \to 0$, we can take the hypothesis $r'$ to intersect $r^*$ at exactly $u_1$, i.e., $\mathbf{x}_{r'}(u_1) = \mathbf{x}_{r^*}(u_1)$. Note that $r' \geq \lvert S \rvert$ since $x_{\min} < u_1 < x_{\max}$.
Let $r''$ denote the intersecting hypothesis $r'$ that is sorted the lowest in $\sigma^*_-$, i.e.,
\begin{align*}
	\sigma_{-}^* (r'') =  \min \left\{\sigma^*_- (t) : t \leq \lvert S \rvert, \mathbf{x}_t (u_1) = \mathbf{x}_{r^*} (u_1) \right\}.
\end{align*}
If $\epsilon$ is small enough such that no hypotheses intersect in $(u_1, u_1 + \epsilon)$, then $\bar{x}_{\sigma_{-}^*, r''}$ is an upper bound of $\bar{x}_{\sigma_{+}^*, r^*}$. For $\lambda$ a very small positive number, $r''$ is selected at $u_1 - \lambda$ under $\sigma^*_-$ and therefore $\mathbf{x}_{r''} (u_1 - \lambda) \geq \bar{x}_{\sigma^*_-, r''}$. Letting $\lambda \to 0$, 
$\mathbf{x}_{r^*} (u_1) = \mathbf{x}_{r''} (u_1) \geq \bar{x}_{\sigma_{-}^*, r''} \geq \bar{x}_{\sigma_{+}^*, r^*}$. Therefore, $\mathbf{x}_{r^*} (u_1 + \epsilon) < \bar{x}_{\sigma_{+}^*, r^*}$ only if $\mathbf{x}_{r^*}$ is downward sloping.

Let $\tilde{x} = \mathbf{x}_{r^*} (u_1 + \epsilon)$ and let 
\begin{align*}
	\mathcal{B} (x_s) = \{ t \leq \lvert S \rvert: \mathbf{x}_t (x_s) \leq \tilde{x} \}
\end{align*}
denote the set of all hypotheses that are below $\tilde{x}$ at $X_s = x$. $\mathcal{B}(u_1 + \epsilon)$ contains at least one hypothesis $r$, namely $r^*$,
such that 
\begin{align*}
	\mathbf{x}_r (u_1 + \epsilon) < \bar{x}_{\sigma^*(u_1 + \epsilon), r}.
\end{align*}
For all $t \in \mathcal{B}(\ell_2)$, 
\begin{align} 
	\label{eq:B l2 not rejected}
	\mathbf{x}_t (\ell_2) \geq \bar{x}_{\sigma^*(\ell_2), t}.
\end{align}
This is only possible if there exists a hypothesis $t^*$ such that $t^* \in \mathcal{B}(u_1 + \epsilon)$ and $t^* \notin \mathcal{B}(\ell_2)$. To see why this is true, suppose that no such $t^*$ exists and hence 
\begin{align*}
	\mathcal{B}(\ell_2) \supseteq \mathcal{B}(u_1 + \epsilon) 
\end{align*}
and 
\begin{align*}
	\bar{x} \left( \mathcal{B}(\ell_2 ) \cup S^{\mathsf{c}} \right) \geq \bar{x} \left( \mathcal{B}(u_1 + \epsilon) \cup S^{\mathsf{c}}\right). 
\end{align*}
Then, for any $t \in \mathcal{B}(\ell_2)$, we have 
\begin{align*} 
	\mathbf{x}_t (\ell_2) \leq \tilde{x} = \mathbf{x}_{r^*} (u_1 + \epsilon) < \bar{x}_{\sigma^*(u_1 + \epsilon), r^*} = \bar{x} \left( \mathcal{B}(u_1 + \epsilon) \right),
\end{align*}
where the last equality follows since, at $x_s = u_1 + \epsilon$, $r^*$ is the lowest-ranked hypothesis in $\mathcal{B}(u_1 + \epsilon)$. 
In particular, for all $t \in \mathcal{B}(\ell_2)$, we have
\begin{align*}
	\mathbf{x}_t (\ell_2) < \bar{x} \left( \mathcal{B}(\ell_2) \right).
\end{align*}
This contradicts \eqref{eq:B l2 not rejected}, proving the existence of a hypothesis $t^*$ such that $t^* \in \mathcal{B}(u_1 + \epsilon)$ and $t^* \notin \mathcal{B}(\ell_2)$.
In particular, $\mathbf{x}_{t^*} (u_1 + \epsilon) \leq \tilde{x}$ and $\mathbf{x}_{t^*} (\ell_2) > \tilde{x}$, implying that $t^*$ is upward sloping. 
By definition of $\tilde{x}$, we have $\mathbf{x}_{t^*} (u_1 + \epsilon) \leq \mathbf{x}_{r^*} (u_1 + \epsilon)$ for downward-sloping $r^*$. 
Therefore, there exists a point $x_{\text{cross}} \in [u_1 + \epsilon, \ell_2)$ such that $\mathbf{x}_{t^*} (x_{\text{cross}}) = \mathbf{x}_{r^*} (x_{\text{cross}})$. 

We have  shown, that, for every $(u_k, \ell_{k +1})$, $k = 1, \dotsc, n_I$, there exist a downward-sloping hypothesis $r^*$ and an upward-sloping hypothesis $t^*$ that intersect in $(u_k, \ell_{k+1})$. Since there are at most $\lvert S_+ \rvert \lvert S_- \rvert$ such pairs of hypotheses, there are at most $\lvert S_+ \rvert \lvert S_- \rvert$ holes in the conditional support. This proves that the conditional support is made up of at most $\lvert S_+ \rvert \lvert S_- \rvert + 1$ disconnected intervals. 
\end{Proof}

\subsection{\label{sec:appendix one-sided support}One-sided rules}

For a one-sided rule, we do not have to consider the negative and positive parts of $\mathbf{x}_{z, h} (x_s) = \Omega{h, s} x_s + z_h$ separately. The following algorithm computes the conditional support $\mathcal{X}_s(z, S)$ for a one-sided step-down rule and the selection event $\hat{S} = S$.

{\singlespacing
\begin{algorithm}[Compute $\mathcal{X}_s(z, S)$ for one-sided step-down rule and $\hat{S} = S$]
\label{alg: simple conditional support one-sided}
\begin{enumerate}[label = (\Alph*), ref = \Alph*]
	\item 
	Find the intervals $I$ by computing all intersection points of the linear functions $\mathbf{x}_{z, h} (x_s)$, $h = 1, \dotsc, m$ with each other. 
	\item For each interval $I$:
	\begin{enumerate}[label = \roman*, ref = \roman*]
		\item Let $\tilde{x}$ denote a value in the interior of $I$ and find the unique permutation $\sigma^*_I$ that orders the components of $\{\mathbf{x}_{z, h}(\tilde{x})\}_{h=1, \dotsc, m}$ in descending order. 
			\item Let $H_{\ell}= \{h \in S: \Omega_{h, s} > 0\} \cup \{h \notin S: \Omega_{h, s} < 0\}$, 
			$H_{u} = \{h \in S: \Omega_{h, s} < 0\} \cup \{h \notin S: \Omega_{h, s} > 0\}$, and
			\begin{align*}
				\ell (I) =& \max_{h \in H_{\ell}} \frac{\bar{x}_{\sigma^*_I, h} - z_h}{\Omega_{h, s}}
				\\
				u(I) =& \min_{h \in H_{u, +}} \frac{\bar{x}_{\sigma^*_I, h} - z_h}{\Omega_{h,s}}.
			\end{align*}
	\end{enumerate}
	\item Return $\bigcup_{I} I \cap \left[\ell(I), u(I)\right]$.
\end{enumerate}
\end{algorithm}}

\subsection{\label{sec:appendix initial bounds}Algorithmic improvements using initial bounds}
Here we present refinements of the algorithms for computing the conditional support $\mathcal{X}_s(z, S)$. For simplicity, we focus on the algorithm for one-sided step-down rules and the selection event $\hat{S} = S$. The extension to other cases is straightforward.
Such refinements can be achieved by computing initial bounds and using the Bentley-Ottmann algorithm \parencite{bentley1979algorithms}.

We let $a_h = \Omega_{h, s}$ and $b_h = z_h$. For $x_s \in \mathcal{X}_s(z, S)$, we require
\begin{align*}
	a_h x_s + b_h \leq \bar{x} (S^{\mathsf{c}}) 
\end{align*}
for all $h \notin S$.
This condition is equivalent to $x_s \in [\ell_{\min}, u_{\max}]$, where 
\begin{align*}
	\ell_{\min} = \max_{h \notin S, a_h < 0} \frac{\bar{x} (S^{\mathsf{c}}) - b_h}{a_h} \quad \text{and} \quad u_{\max} = \min_{h \notin S, a_h > 0} \frac{\bar{x} (S^{\mathsf{c}}) - b_h}{a_h},
\end{align*}
where we use the convention that $\max \emptyset = -\infty$ and $\min \emptyset = \infty$. We can then restrict the search for the conditional support to the initial interval $[\ell_{\min}, u_{\max}]$. Moreover, the initial bounds already impose all required restrictions on $h \notin S$, and we only have to check the restrictions imposed by the selection event for $h \in S$.
To find intersections of $\{ \mathbf{x}_{z, h} (x_s)\}_{h \in S}$, we can use the Bentley-Ottmann algorithm \parencite{bentley1979algorithms} that finds all intersections in $\mathcal{O} (\tilde{n}_I \log \lvert S \rvert )$ time, where $\tilde{n}_I$ now is the number of intersections within $[\ell_{\min}, u_{\max}]$. Since $h \notin S$ can be ignored, processing any of the $\tilde{n}_I + 1$ intervals separated by the points of intersection takes $\mathcal{O} (\lvert S \rvert \log \lvert S \rvert)$ time. The total time complexity is therefore $\mathcal{O} (\tilde{n}_I \lvert S \rvert \log \lvert S \rvert)$ with $\tilde{n}_I \leq \lvert S \rvert (\lvert S \rvert - 1) / 2$. 

The initial bounds can be further refined by considering a necessary (but not sufficient) condition for $h \in S$. Let 
\begin{align*}
	\bar{x}_{\text{crit}} = \min_{h \in S} \bar{x} (\{h\} \cup S^{\mathsf{c}}).
\end{align*}
For example, in the case of a one-sided Holm rule, we have $\bar{x}_{\text{crit}} = \Phi^{-1} (1 - \alpha / (m - \lvert S \rvert + 1))$. 
For all $h \in S$, we require 
\begin{align*}
	a_h x_s + b_h \geq \bar{x}_{\text{crit}}.
\end{align*}
This implies the improved initial bounds
\begin{align*}
	\ell_{\min} =& \max \left\{ \frac{\bar{x} (S^{\mathsf{c}}) - b_h}{a_h} : h \notin S, a_h < 0, \frac{\bar{x}_{\text{crit}} - b_h}{a_h} : h \in S, a_h > 0 \right\}
	\\
	u_{\max} =& \min \left\{ \frac{\bar{x} (S^{\mathsf{c}}) - b_h}{a_h} : h \notin S, a_h > 0, \frac{\bar{x}_{\text{crit}} - b_h}{a_h} : h \in S, a_h < 0 \right\}.
\end{align*}

\subsection{\label{sec:appendix step-up support}Conditional support for a step-up rule}
Next, we consider a step-up rule. For simplicity, we focus on the case of one-sided testing and the selection event $\hat{S} = S$. 
For a one-sided step-up rule, and $\sigma \in \mathcal{E}(S^{\mathsf{c}})$ let   
\begin{align*}
	\bar{x}_{\sigma, h} = \begin{cases} 
		\bar{x}_{\sigma(h)} & \text{if $h \notin S$}	
		\\
		\bar{x}_{m - \lvert S \rvert + 1} & \text{if $h \in S$}.
	\end{cases}
\end{align*}
We define search intervals $I$ in the same way as we do for one-sided step-down rules. On interval $I$, it suffices to consider the permutation $\sigma^*_I$ that orders the elements of $\mathbf{x}_z(x_s)$ in ascending order and $x_s \in \mathcal{X}_s (z, S)$ if and only if
\begin{align*}
	a_h x_s + b_h \leq \bar{x}_{\sigma^*_I, h}
\end{align*}
for $h \notin S$ and
\begin{align*}
	a_h x_s + b_h \geq \bar{x}_{\sigma^*_I, h}
\end{align*}
for $h \in S$. From these inequalities, we determine interval-specific bounds $[\ell(I), u(I)]$ that are then aggregated to obtain the conditional support

\subsection{\label{sec:appendix support alternative}Conditional support for alternative selection events}

Lastly, we provide algorithms to compute the conditional support for alternative selection events.
In the main text, we have described an algorithm to compute the conditional support under the selection event $\hat{S} = S$. We now modify this algorithm to accommodate the event $R \subseteq \hat{S}$, where $R \subseteq S$ and $S$ is the set of significant effects taken as fixed. By setting $R = \{s\}$ for $s \in S$ or setting $R = S$, we can cover two of the selection events discussed in Section~\ref{sec:conditioning event}.

We focus on the case of a one-sided step-down rule. The extensions to two-sided and step-up rules are straightforward and similar to the modifications given in Sections \ref{sec:algorithm conditional support} and \ref{sec:appendix step-up support}, respectively.

We split the real line into intervals $I$ as described in Section~\ref{sec:algorithm conditional support} of the main text and define $\sigma^*_I \in \mathcal{E}(\{1, \dotsc, m\})$ as the permutation that orders the elements of $\mathbf{x}_z(x_s)$ in descending order for $x_s \in I$.

We then iterate over all $I$ and find subsets $[\ell(I), u(I)]$ of $I$ that satisfy the condition of Lemma~\ref{lem:geometry step-down alternative event}. To determine $\ell(I)$ and $u(I)$, we use the following algorithm:
\begin{algorithm}[General algorithm to compute $\mathcal{X}_s(z, S)$]
\begin{enumerate}[label = (\Alph*)]
	\item Initialise $\ell(I)$ and $u(I)$ as the lower and upper bounds of $I$.
	\item Determine $\sigma^*_I$ and set $j = 1$.
	\item Loop from $j = 1$ to $j = \max_{h \in R} \sigma^*_I (h)$.
	\begin{enumerate}[label = (\arabic*)]
		\item Let $h = \sigma^{*, -1}_I (j)$.
		\item If $a_h > 0$, update 
		\begin{align*}
			\ell(I) = \max \left\{ \ell(I), \frac{\bar{x}_{\sigma^*_I, h} - b_h}{a_h} \right\}.
		\end{align*}
		\item If $a_h < 0$, update
		\begin{align*}
			u(I) = \min \left\{ u(I), \frac{\bar{x}_{\sigma^*_I, h} - b_h}{a_h} \right\}.
		\end{align*}
		\item If $a_h = 0$ and $\bar{x}_{\sigma^*_I, h} < b_h$, exit the loop and return $[\ell(I), u(I)] = \emptyset$.
		\item If $\ell(I) > u(I)$, exit the loop and return $[\ell(I), u(I)] = \emptyset$.
		\item Increment $j$ and continue the loop.
	\end{enumerate} 
	\item Return $\bigcup_I [\ell(I), u(I)]$.
\end{enumerate}
\end{algorithm}

\section{\label{sec:additional theory}Additional theoretical results}

This appendix collects theoretical results on two distinct issues: we discuss the properties of the Rosenblatt transformation and provide characterisations of alternative selection events.

\subsection{Properties of Rosenblatt transformation}

Here, we state two key properties of the conditional distribution function $F_{s} (x_s \mid z, \theta_s, S)$. First, it is invertible as a function of $\theta_s$.
\begin{Lemma}
\label{lem:Rosenblatt_invertibility}
Let $S \subset \{1, \dotsc, m\}$, $s \in S$, $x \in \mathbb{R}^m$ and $z = x - \Omega_{\bullet, s} x_s$. Suppose that $\mathcal{X}_s (z, S) \neq \emptyset$. Then 
$F_{s} (x_s \mid z, \theta_s, S)$ is strictly decreasing in $\theta_s$. Moreover, 
\begin{align*}
	\lim_{\theta_s \to \infty} F_{s} (x_s \mid z, \theta_s, S) &= 0 \quad \text{and}
\\
	\lim_{\theta_s \to -\infty} F_{s} (x_s \mid z, \theta_s, S) &= 1.
\end{align*}
\end{Lemma}
This property guarantees that \eqref{eq:Rosenblatt equal to p} has a unique solution, ensuring that our median-unbiased estimator and conditional confidence interval are well-defined.

To prove Lemma~\ref{lem:Rosenblatt_invertibility}, we introduce a lemma for general truncated Gaussian distributions. Lemma ~\ref{lem:Rosenblatt_invertibility} immediately follows this lemma.
\begin{Lemma}\parencite[][Lemma A.2]{kivaranovic2021}
	\label{lem: truncated normal}
	Let $F(x ; \mu, \sigma)$ be the cumulative distribution function of a Gaussian random variable $X\sim N(\mu, \sigma^2)$ conditional on $X \in \bigcup_{k=1}^{\bar {k}} [l_k, u_k]$, where $- \infty \leq l_1 < u_1 < l_2 < u_2 < \dots < u_{\bar k} \leq \infty$. It holds that, for any given $x \in \bigcup_{k=1}^{\bar {k}} (l_k, u_k)$ and $\sigma > 0$, 
	\begin{enumerate}
		\item $F(x ; \mu, \sigma)$ is strictly decreasing in $\mu$; 
		\item $\lim_{\mu \to \infty }  F(x ; \mu, \sigma)  = 0$ and $\lim_{\mu \to - \infty }  F(x ; \mu, \sigma)  = 1$.
	\end{enumerate}
\end{Lemma}
The conditioning set in the lemma above is a finite union of closed intervals. The closedness is not essential because the boundary points have measure zero. We can replace the closed intervals with open intervals. The finiteness of the number of intervals is crucial for the proof given in \textcite{kivaranovic2021}. Dealing with an infinite union of intervals would necessitate a different proof, but this is not required for our current purpose and is not explored in this paper.

\begin{Proof}[Proof of Lemma~\ref{lem:Rosenblatt_invertibility}]
	A simple application of Lemma \ref{lem: truncated normal} can show this lemma. We note that $F_s (x_s \mid z, \theta_s, S)$ (as a function of $x_s$) is the cumulative distribution function of a truncated Gaussian distribution with $N (V_{s,s}^{-1/2} \theta_s, 1)$ truncated on $\mathcal{X}_s(z,S)$. Additionally, $\mathcal{X}_s(z,S)$ is a finite union of closed intervals. Since $V_{s,s}^{-1/2}$ is fixed, $V_{s,s}^{-1/2} \theta_s \to \infty$ (or $\to - \infty$) is equivalent to $\theta_s \to \infty$ (or $\to - \infty$). Thus, we obtain the desired results by applying Lemma \ref{lem: truncated normal}. 
\end{Proof}

Secondly, evaluating the conditional distribution function at the observed $x = X_s$ and $z = Z^{(s)}$ (``Rosenblatt transformation'') yields a random variable that is uniformly distributed on the unit interval conditional on the selection event. 
\begin{Lemma}
\label{lem:rosenblatt_uniform_distribution}
Let $S \subset \{1, \dotsc, m\}$ such that $P (\hat{S} = S) > 0$. Then, for $s \in S$,
\begin{align*}
	F_{s} (X_s \mid Z^{(s)}, \theta_s, S) \mid \left\{\hat{S} = S \right\} \sim \mathcal{U}[0, 1],
\end{align*}
where $\mathcal{U} [0, 1]$ denotes the uniform distribution on the unit interval.
\end{Lemma}
The true value $\theta_s$ of the parameter is the only unknown part of $F_{s} (X_s \mid Z^{(s)}, \theta_s, S)$. Thus, the distribution of this random variable is pivotal when we specify $\theta_s$ and its distribution is given by Lemma~\ref{lem:rosenblatt_uniform_distribution}.

\begin{Proof}[Proof of Lemma~\ref{lem:rosenblatt_uniform_distribution}]
	Let 
	\begin{align*}
		f \left(x_s, z\right) \equiv  P_{X_s} \left(
			X_s \leq x_s \mid X_s \in \mathcal{X}_s (z, S)
		\right) 
	\end{align*}
	and write the Rosenblatt transformation as
	\begin{align*}
		F_{s} \left(X_s \mid Z^{(s)}, \theta_s, S\right) = f \left(X_s, Z^{(s)}\right)
	\end{align*}
	For any $z$ such that $\mathcal{X}_s (z, S)$ is measurable and has positive Lebesgue measure, 
	\begin{align*}
		f \left(X_s, z \right) \mid X_s \in \mathcal{X}_s(z, S) \sim \mathcal{U} [0, 1].
	\end{align*}
	This can be shown by standard arguments for probability transforms of continuous random variables.
	Let $G_{z\mid S}$ denote the marginal probability measure of $Z^{(s)}$ conditional on the event $\hat{S} = S$. 
	
	For $u \in [0,1]$, by the law of iterated expectations, 
	\begin{align*}
		& P \left( F_s (X_s \mid Z^{(s)}, \theta_s, S) \leq u \mid \hat{S} = S \right)
		\\
		=& P \left(f(X_s, Z^{(s)}) \leq u \mid \hat{S} = S \right) 
		\\
		=& \int P \left( f(X_s, z) \leq u \mid Z^{(s)} = z, \hat{S} = S \right) \, d G_{z \mid S} (z)
		\\
		=& \int P \left( f(X_s, z) \leq u \mid X_s \in \mathcal{X}_s(z, S) \right) \, d G_{z \mid S} (z)
		= \int u \, d G_{z \mid S} (z)
		= u, 
	\end{align*}
	where the second inequality follows from the law of iterated expectations and the third inequality follows since $X_s$ and $Z^{(s)}$ are uncorrelated and jointly Gaussian and hence independent.
\end{Proof}

\subsection{Alternative selection events}
Here, we provide characterisations of alternative selection events. We have discussed the algorithm to compute these alternative selection events (Algorithms~\ref{alg:generalized_step_down} and ~\ref{alg:step_up}). We give characterisations of the sets of $X$ corresponding to these events.
Let $S$ denote the observed set of significant effects taken as fixed, and let $R \subseteq S$. 

\paragraph{Step-down rules} The following lemma gives a geometric characterisation of the event $\hat{S} \supseteq R$ for step-down rules by Algorithm~\ref{alg:generalized_step_down}. 

\begin{Lemma}
\label{lem:geometry step-down alternative event}
For realised test statistics $X = x$, Algorithm~\ref{alg:generalized_step_down} selects $\hat{S} \supseteq R$ 
if and only if there exists $\sigma \in \mathcal{E}(\{1, \dotsc, m\})$ such that, for all $h = 1, \dotsc, m$ with $\sigma(h) \leq \max_{s \in R} \sigma(s)$,   
\begin{align*}
	x_{h} \geq \bar{x} \left(\sigma^{-1}(\{\sigma (h), \dotsc, m \}) \right).
\end{align*}
\end{Lemma}
\begin{Proof}
The result follows by taking the union over the conditions in Lemma~\ref{lem:geometry generalized step-down} for all $S' \supseteq R$.
\end{Proof}

\paragraph{Step-up rules} The following lemma gives a geometric characterisation of the event $\hat{S} \supseteq R$ for step-up rules by Algorithm~\ref{alg:step_up}.
\begin{Lemma}
\label{lem:geometry step-up alternative event}
For realised test statistics $X = x$, Algorithm~\ref{alg:step_up} selects $\hat{S} \supseteq R$ if and only if there exist $k \leq m - \lvert S \rvert$ and $\sigma \in \mathcal{E} (\{1, \dotsc, m\})$ such that
\begin{align*}
	S \subseteq \sigma^{-1} (\{k+1, \dotsc, m\})
\end{align*}
and 
\begin{align*}
	x_{\sigma^{-1} (j)} < \bar{x}_{j} \quad \text{for all $j = 1, \dotsc, k$}
\end{align*}
and
\begin{align*}
	x_{\sigma^{-1} (j)} \geq \bar{x}_{k + 1} \quad \text{for all $j = k + 1, \dotsc, m$}.
\end{align*}
\end{Lemma}
\begin{Proof}
The result follows by taking the union over the conditions in Lemma~\ref{lem:geometry support step-up} for all $S' \supseteq S$.
\end{Proof}

\section{Mathematical proofs of results in the main text}\label{S: A: proof}
\begin{Proof}[Proof of Lemma~\ref{lem:geometry generalized step-down}]
The ``only if'' part follows by taking $\sigma = \tilde{\sigma}$ such that $\tilde{\sigma}^{-1} (j) = (>, j)$ for $j = 1, \dotsc, \lvert S \rvert$. To prove the ``if'' part, we use mathematical induction, and suppose that the sufficient condition holds for some $\sigma_1 \in \mathcal{E}(S)$ and construct a finite sequence of permutations $\sigma_1, \dotsc, \sigma_{N} = \tilde{\sigma}$ with the property that $\sigma_{n+1}$ satisfies the sufficient condition if $\sigma_{n}$ does. Consider two permutations $\sigma_n$ and $\sigma_{n+1}$ that are identical up to a neighbor swap, i.e., there is $j \in \{1, \dotsc, \lvert S \rvert - 1\}$ and $h, h' \in S$ such that $h = \sigma_{n}^{-1} (j) =  \sigma_{n + 1}^{-1} (j + 1)$ and $h' = \sigma_n^{-1} (j + 1) = \sigma_{n+1}^{-1} (j)$, and $\sigma_n^{-1} (i) = \sigma_{n+1}^{-1} (i)$ for all $i = \{1, \dotsc, \lvert S \rvert \} \setminus \{j, j+1\}$. Moreover, suppose that under $\sigma_n$ the neighbors are ordered in ascending order, that is, $\lvert x_{h'} \rvert > \lvert x_{h} \rvert$. If $\sigma_n$ satisfies the sufficient condition, so does $\sigma_{n + 1}$. To show this, we have to verify the two conditions 
\begin{align*}
	\lvert x_{h} \rvert \geq & \bar{x} \left(
		\sigma_{n + 1}^{-1} \left(\left\{ \sigma_{n + 1} (h), \dotsc, \lvert S \rvert  \right\} \right) \cup S^{\mathsf{c}}
	\right) \quad \text{and}
\\
	\lvert x_{h'} \rvert \geq & \bar{x} \left(
		\sigma_{n + 1}^{-1} \left(\left\{\sigma_{n + 1} (h'), \dotsc,  \lvert S \rvert \right\} \right)  \cup S^{\mathsf{c}}
	\right).
\end{align*}
To verify the first condition, note that 
\begin{align*}
	\sigma_{n + 1}^{-1} \left(\left\{\sigma_{n + 1} (h), \dotsc, \lvert S \rvert \right\}\right)
	= &
	\sigma_{n + 1}^{-1} \left(\left\{j + 1, \dotsc, \lvert S \rvert \right\}\right)
\\
	=& \sigma_{n}^{-1} \left(\left\{j, \dotsc,\lvert S \rvert \right\}\right) \setminus \{h'\}
	= \sigma_{n}^{-1} \left(\left\{\sigma_{n} (h), \dotsc, \lvert S \rvert \right\}  \right)\setminus \{h'\}
\end{align*}
and therefore 
\begin{align*}
	\bar{x} \left(\sigma_{n + 1}^{-1} \left(\sigma_{n + 1} (h), \dotsc, \lvert S \rvert \right) \cup S^{\mathsf{c}} \right) \leq 
	\bar{x} \left(\sigma_{n }^{-1} \left(\sigma_{n } (h), \dotsc, \lvert S \rvert \right) \cup S^{\mathsf{c}} \right) \leq \lvert x_h \rvert.
\end{align*}
To verify the second condition, note that 
\begin{align*}
	\sigma_{n + 1}^{-1} \left(\left\{\sigma_{n + 1} (h'), \dotsc, \lvert S \rvert\right\}\right)
	= \sigma_{n}^{-1} \left(\left\{\sigma_{n} (h), \dotsc, \lvert S \rvert\right\}\right)
\end{align*}
and therefore 
\begin{align*}
	\bar{x} \left( \sigma_{n + 1}^{-1} \left(\left\{\sigma_{n + 1} (h'), \dotsc, \lvert S \rvert\right\}\right)\cup S^{\mathsf{c}} \right)
	=
	\bar{x} \left( \sigma_{n }^{-1} \left(\left\{\sigma_{n } (h), \dotsc, \lvert S \rvert\right\}\right) \cup S^{\mathsf{c}}\right)
	\leq \lvert x_h \rvert < \lvert x_{h'} \rvert.
\end{align*}
It remains to argue that there is a sequence, $\sigma_1, \dotsc, \sigma_{N} = \tilde{\sigma}$, that transforms $\sigma_1$ into $\tilde{\sigma}$. This transformation involves multiple steps, and each step swaps the order of a neighboring pair that is ordered in ascending order. Such a sequence can be constructed using the bubble sort algorithm.
\end{Proof}
\begin{Proof}[Proof of Lemma~\ref{lem:geometry step-down alternative event}]
The result follows by taking the union over the conditions in Lemma~\ref{lem:geometry generalized step-down} for all $S' \supseteq S$.
\end{Proof}
\begin{Proof}[Proof of Theorem~\ref{thm: validity algorithm}]
Let $\sigma^*(x_s)$ denote the permutation that orders the elements of $\{\lvert \mathbf{x}_h (x_s) \rvert \}_{h \in S}$ in descending order. Let $\clo \mathcal{X}_s(z, S)$ denote the closure of the conditional support $\mathcal{X}_s(z, S)$.
By Lemma~\ref{lem:geometry generalized step-down}, to check $x_s \in \mathcal{X}_s(z, S)$ it suffices to check that $S$ is selected by the step-down rule if we observe $t$-statistics $x = \mathbf{x}_z(x_s)$. This amounts to replacing the union over all $\sigma \in \mathcal{E}(S)$ in \eqref{eq:uncond support X} by the union over the singleton set $\{\sigma^*(x_s)\}$. Thus, $x_s \in \clo \mathcal{X}_s(z, S)$ if and only if
\begin{align}
	\label{eq:proof validity algorithm two inequalities}
	\begin{aligned}
	x_h &\geq \bar{x}_{\sigma^*(x_s), h} \quad \text{or} \quad x_h \leq -\bar{x}_{\sigma^*(x_s), h} \quad \text{for all $h \in S$}
	\\
	\text{and} \quad
	x_h &\leq \bar{x}_{\sigma^*(x_s), h} \quad \text{or}  \quad x_h \geq -\bar{x}_{\sigma^*(x_s), h} \quad \text{for all $h \notin S$}.
	\end{aligned}
\end{align}
As $x_s$ changes, the vector $\mathbf{x}_z(x_s)$ changes, potentially affecting how we order the vector components when we sort them by their absolute values, that is, $\sigma^* (x_s)$ may change as $x_s$ changes. The vector $\mathbf{x}_z(x_s)$ contains $m$ linear functions of $x_s$ that intersect at most $m(m-1)/2$ times with each other and at most $m$ times with the horizontal axis. The ordering of the vector components by their absolute values can change only at these intersection points. In particular, $\sigma^*(x_s)$ takes the same value for all $x_s$ in the interior of an interval $I$ that is defined by two adjacent intersection points. 
From now on we work with a generic interval $I$ and let $\tilde{x}$ denote the midpoint of $I$. We let $\sigma^*_I = \sigma^*(\tilde{x})$, noting that this is relevant ordering not only for $\tilde{x}$ but for all interior points of $I$. To check $x_s \in \mathcal{X}_s(z, S)$ for $x \in I$ we can therefore replace $\sigma^*(x_s)$ by $\sigma^*_I$ in \eqref{eq:proof validity algorithm two inequalities}. 
By construction, $\mathbf{x}_z(x_s)$ does not change sign over $I$. Thus, for $h=1, \dotsc, m$, $\sign \mathbf{x}_{z, h}(x_s) = \sign \mathbf{x}_{z, h}(\tilde{x})$, for all $x_s \in I$. 	
For $h \in S$ such that $\mathbf{x}_{z, h}(\tilde{x}) > 0$, we have to check
\begin{align*}
	x_h = \mathbf{x}_h (x_s) = \Omega_{h, s} x_s + z_h \geq \bar{x}_{\sigma^*_I, h}.
\end{align*}
If $\Omega_{h, s} > 0$, this is equivalent to 
\begin{align*}
	x_s \geq \frac{\bar{x}_{\sigma^*_I, h} - z_h}{\Omega_{h, s}}.
\end{align*}
If $\Omega_{h, s} < 0$, this is equivalent to
\begin{align*}
	x_s \leq \frac{\bar{x}_{\sigma^*_I, h} - z_h}{\Omega_{h, s}}.
\end{align*}
For $h \in S$ such that $\mathbf{x}_{z, h}(\tilde{x}) < 0$, we have to check
\begin{align*}
	x_h = \mathbf{x}_h (x_s) = \Omega_{h, s} x_s + z_h \leq -\bar{x}_{\sigma^*_I, h}.
\end{align*}
If $\Omega_{h, s} > 0$, this is equivalent to
\begin{align*}
	x_s \leq \frac{-\bar{x}_{\sigma^*_I, h} - z_h}{\Omega_{h, s}}.
\end{align*}
If $\Omega_{h, s} < 0$, this is equivalent to
\begin{align*}
	x_s \geq \frac{-\bar{x}_{\sigma^*_I, h} - z_h}{\Omega_{h, s}}.
\end{align*}
Hypotheses $h \in S$ with $\Omega_{h, s} = 0$ do not add any restrictions. For these $h$, $\lvert z_h \rvert \geq \bar{x}_{\sigma^*_I, h}$ whenever we observe $S$. 
For $h \notin S$ such that $\mathbf{x}_{z, h}(\tilde{x}) > 0$, we have to check
\begin{align*}
	x_h = \mathbf{x}_h (x_s) = \Omega_{h, s} x_s + z_h \leq \bar{x}_{\sigma^*_I, h}.
\end{align*}
If $\Omega_{h, s} > 0$, this is equivalent to
\begin{align*}
	x_s \leq \frac{\bar{x}_{\sigma^*_I, h} - z_h}{\Omega_{h, s}}.
\end{align*}
If $\Omega_{h, s} < 0$, this is equivalent to
\begin{align*}
	x_s \geq \frac{\bar{x}_{\sigma^*_I, h} - z_h}{\Omega_{h, s}}.
\end{align*}
For $h \notin S$ such that $\mathbf{x}_{z, h}(\tilde{x}) < 0$, we have to check
\begin{align}
	x_h = \mathbf{x}_h (x_s) = \Omega_{h, s} x_s + z_h \geq -\bar{x}_{\sigma^*_I, h}.
\end{align}
If $\Omega_{h, s} > 0$, this is equivalent to	
\begin{align*}
	x_s \leq \frac{-\bar{x}_{\sigma^*_I, h} - z_h}{\Omega_{h, s}}.
\end{align*}
If $\Omega_{h, s} < 0$, this is equivalent to 
\begin{align*}
	x_s \geq \frac{-\bar{x}_{\sigma^*_I, h} - z_h}{\Omega_{h, s}}.
\end{align*}
Hypotheses $h \notin S$ with $\Omega_{h, s} = 0$ do not add any restrictions. For these $h$, $\lvert z_h \rvert < \bar{x}_{\sigma^*_I, h}$ whenever we observe $S$.
Above, we have derived four lower bounds on $x_s$ and four upper bounds on $x_s$. The theorem defines the minimum of the upper bounds as $u(I)$, and the maximum of the lower bounds as $\ell(I)$. A value $x_s \in I$ satisfies all bounds if and only if $\ell(I) \leq x_s \leq u(I)$.

For each interval $I$, the algorithm finds the $x_s \in I$ that satisfy these restrictions. This is equivalent to finding the $x_s \in I$ for which $x_s \in \clo \mathcal{X}_s(z, S)$. Since the intervals $I$ partition the real line and since the algorithm iterates over all $I$, the algorithm finds all $x_s$ in $\clo \mathcal{X}_s(z, S)$. This proves that the algorithm computes the closure of the conditional support. 

Turning to the time complexity of the algorithm, we note that the number of intersection points is $m(m-1)/2$ for intersections of pairs of linear functions $\mathbf{x}_{z, h}$ and $\mathbf{x}_{z, h'}$ and $m$ for the intersections of a linear function with the horizontal axis. Therefore, there are total of $m(m-1)/2 + m = m(m+1)$ intersection points and $m(m+1)/2 + 1$ intervals defined by two adjacent intersection points. At each such interval $I$, we have to compute $\sigma^*_I$. This amounts to ordering the components of $\lvert \mathbf{x}_z(\bar{x}) \rvert $ and has complexity $\mathcal{O}(m \log m)$. Computing the minima and maxima that give $\ell(I)$ and $u(I)$ has complexity $\mathcal{O}(m)$. The total complexity is therefore 
\begin{align*}
	\left(m(m+1)/2 + 1 \right) \mathcal{O} \left( m \log m + m \right) = \mathcal{O}(m^3 \log m).
\end{align*}
\end{Proof}

\begin{Proof}[Proof of Theorem~\ref{thm:convergence to unconditional CI}]
We only prove the result for 1-sided step-down rules. The proof for 2-sided step-down rules is analogous.
For simplicity, we operate in $t$-statistic space and prove the corresponding results for confidence intervals for $\mu_{n, s}$. Let $c_{\alpha} = \Phi^{-1} (1 - \alpha/2)$. We prove that for all $\epsilon > 0$,
\begin{align*}
	P_{\mu_n, \Omega_n} \left( d_H \left( [X_s - c_{\alpha}, X_s + c_{\alpha}], \left[ \tilde{\theta}_s^{(1 - \alpha/2)} / \sqrt{\mathbf{v}_s}, \tilde{\theta}_s^{(\alpha/2)} / \sqrt{\mathbf{v}_s}\right] \right) > \epsilon \mid \hat{S} = S \right) \to 0.
\end{align*}
To show this, we show that the left and right endpoints of the two confidence intervals converge.
For the left endpoint of the two confidence intervals to not converge, there must exist $\kappa > 0$ and a subsequence on which
\begin{align*}
	P_{\mu_n, \Omega_n} \left( \left\lvert 
		X_s - c_{\alpha} - \tilde{\theta}_s^{(1 - \alpha/2)} / \sqrt{\mathbf{v}_s}
	\right\rvert > \epsilon \mid \hat{S} = S \right) \geq 2 \kappa. 
\end{align*}
This in turn implies that there exists a subsequence on which either 
\begin{align*}
	& P_{\mu_n, \Omega_n} \left( X_s - c_{\alpha} - \epsilon < \tilde{\theta}_s^{(1 - \alpha/2)} / \sqrt{\mathbf{v}_s} \mid \hat{S} = S \right) 
	\\
	=& P_{\mu_n, \Omega_n} \left( F_s \left( X_s \mid Z^{(s)}, X_s - c_{\alpha} - \epsilon, S \right) < 1 - \alpha/2 \mid \hat{S} = S \right) \geq \kappa 
\end{align*}
or 
\begin{align*}
	& P_{\mu_n, \Omega_n} \left( X_s - c_{\alpha} + \epsilon > \tilde{\theta}_s^{(\alpha/2)} / \sqrt{\mathbf{v}_s} \mid \hat{S} = S \right) 
	\\
	=& P_{\mu_n, \Omega_n} \left( F_s \left( X_s \mid Z^{(s)}, X_s - c_{\alpha} + \epsilon, S \right) > 1 - \alpha/2 \mid \hat{S} = S \right) \geq \kappa.
\end{align*}
We show that this is not possible. The argument for the convergence of the right endpoint is analogous, and we omit it for brevity.

Given that $P_{\mu_n, \Omega_n} ( \hat{S} = S )$ is bounded away from zero, it suffices to show that there do not exist subsequences on which
\begin{align}
	\label{eq:proof below left endpoint}
	P_{\mu_n, \Omega_n} \left(\{\hat{S} = S \} \wedge \{F_s \left( X_s, \mid Z^{(s)}, X_s - c_{\alpha} - \epsilon, S \right) < 1 - \alpha/2 \}\right) \geq & \kappa 
\\
	\label{eq:proof above right endpoint}
	\text{and} \quad P_{\mu_n, \Omega_n} \left( \{\hat{S} = S \} \wedge \{ F_s \left( X_s \mid Z^{(s)}, X_s - c_{\alpha} - \epsilon, S \right) > 1 - \alpha/2\}\right) \geq & \kappa.
\end{align}
Assume that there exists a subsequence on which either \eqref{eq:proof below left endpoint} or \eqref{eq:proof above right endpoint} holds. We pass to this subsequence. Define 
\begin{align*}
	\nu_n = (\Phi(\mu_{n, 1}), \dotsc, \Phi(\mu_{n, m})).
\end{align*}
The choice of $\Phi$ to define this mapping is not essential, and we could have used any other bijection from the real line onto a compact space. The sequence $(\nu_n, \Omega_n)$ is a sequence of random vectors on a compact space, implying that there exists a convergent subsequence. We pass to this subsequence and denote the limit as $(\nu, \Omega)$. 

Write $Z^{(s)}$ as $Z^{(s)} = M_s X$, where $M_s = \mathbf{I}_m - \Omega \mathbf{P}_s$, with $\mathbf{P}_s$ the projection matrix onto the $s$-th coordinate, i.e., the $m \times m$ matrix with a one at position $(s, s)$ and zeros elsewhere. Let $\mu_n = \diag(\mathbf{v})^{-1/2} \hat{\theta}_n$ and $\tilde{Z}^{(s)} = M_s (X - \mu_n)$.

Letting $\tilde{z} = z - M_s \mu_n$ for $z \in \mathbb{R}^m$, we have $x_s \in \mathcal{X}_s(z, S)$ if and only if 
\begin{align*}
	\Omega_{n, \bullet, s} x_s + z \in \mathcal{X} (S) 
\quad \Longleftrightarrow \quad \Omega_{n, \bullet, s} (x_s - \mu_{n, s}) + \mu_n + \tilde{z} \in \mathcal{X} (S). 
\end{align*}
In particular, $\xi + \tilde{x}_s + \mu_{n, s} + c \in \mathcal{X}_s (z, S)$ if and only if
\begin{align*}
	\Omega_{n, \bullet, s} \left( \xi + \tilde{x}_s + c\right) + \mu_n + \tilde{z} \in \mathcal{X} (S). 
\end{align*}
For a one-sided step-down rule, $\xi + \tilde{x}_s + \mu_{n, s} + c \in \mathcal{X}_s (z, S)$ if and only if
\begin{align*}
	\xi \in \mathcal{Y}_c (\tilde{x}_s,  \tilde{z}, \nu, \Omega_n), 
\end{align*}
where 
\begin{align*}
	\mathcal{Y}_c (\tilde{x}_s, \tilde{z}, \nu, \Omega) = \bigcup_{\sigma \in \mathcal{E}(S)} \Bigg\{
	\begin{aligned}[t]
		& \bigcup_{h \in S} \left\{ \xi \in \mathbb{R}: \Omega_{h, s} (\xi + \tilde{x}_s + c) + \Phi^{-1} \left(\nu_{h}\right) + \tilde{z}_h \geq \bar{x}_{\sigma, h} \right\} \cap
		\\
		& \left\{
			\xi \in \mathbb{R}: \Omega_{h, s} (\xi + \tilde{x}_s + c) + \Phi^{-1} \left(\nu_{h}\right) + \tilde{z}_h < \bar{x}_{\sigma, h}	
		\right\}
		\Bigg\}
	\end{aligned},
\end{align*}
and we omitted the dependence on $s$ for brevity.
Moreover, $X \in \mathcal{X}(S)$ if and only if 
\begin{align*}
	X_s \in \mathcal{Y}_0 (0, \tilde{Z}^{(s)}, \nu_n, \Omega_n).
\end{align*}
With this notation, we can write for given $\nu$ and $\Omega$ and $\tilde{x}_s = x_s - \Phi^{-1}(\nu_{s})$ 
\begin{align*}
	F_s \left(x_s \mid z, x_s + c, S\right) = 
	\frac{
		\int \mathbf{1} \left\{ \xi \in \mathcal{Y}_c (\tilde{x}_s, \tilde{z}, \nu, \Omega) \right\}  
		\mathbf{1} \left\{\xi \leq -c \right\} \, d\Phi(\xi)
	}{
		\int \mathbf{1} \left\{ \xi \in \mathcal{Y}_c (\tilde{x}_s, \tilde{z}, \nu, \Omega) \right\} \, d\Phi(\xi) 
	}.
\end{align*}
Define the functions  
\begin{align*}
	g_{c, 1}(\tilde{x}_s, \tilde{z}, \nu, \Omega) =&
	\int \mathbf{1} \left\{ \xi \in \mathcal{Y}_c (\tilde{x}_s, \tilde{z}, \nu, \Omega) \right\}  
	\mathbf{1} \left\{\xi \leq -c \right\} \, d\Phi(\xi),
\\
	g_{c, 2}(\tilde{x}_s, \tilde{z}, \nu, \Omega) =&
	\int \mathbf{1} \left\{ \xi \in \mathcal{Y}_c (\tilde{x}_s, \tilde{z}, \nu, \Omega) \right\} \, d\Phi(\xi). 
\end{align*}
Both functions are continuous. This can be seen by considering any converging sequence 
\begin{align*}
	\left(\tilde{x}_s^{(k)}, \tilde{z}^{(k)}, \nu^{(k)}, \Omega^{(k)}\right) \overset{k \to \infty}{\longrightarrow} (\tilde{x}_s, \tilde{z}, \nu, \Omega)
\end{align*}
and appealing to the dominated convergence theorem. Moreover, any sequence 
\begin{align*}
	\left(\tilde{x}_s^{(k)}, \tilde{z}^{(k)}, \nu^{(k)}, \Omega^{(k)}\right)
\end{align*}
such that $g_{c, 2}(\tilde{x}_s^{(k)}, \tilde{z}^{(k)}, \nu^{(k)}, \Omega^{(k)}) \to 0$ must have $\mathcal{Y}_c (\tilde{x}_s, \tilde{z}, \nu, \Omega) = \emptyset$ for the limit and 
\begin{align*}
	\mathbf{1} \left\{ \xi \in \mathcal{Y}_c (\tilde{x}_s^{(k)}, \tilde{z}^{(k)}, \nu^{(k)}, \Omega^{(k)}) \right\} \to 0
\end{align*} 
for all $\xi \in \mathbb{R}$. In particular, this holds for $\xi = -c$, implying that 
\begin{align*}
	\mathbf{1} \left\{ \tilde{x}_s^{(k)} \in \mathcal{Y}_0 (0, \tilde{z}^{(k)}, \nu^{(k)}, \Omega^{(k)}) \right\} \to 
	\mathbf{1} \left\{ \tilde{x}_s \in \mathcal{Y}_0 (0, \tilde{z}, \nu, \Omega) \right\} =
	0.
\end{align*}
Therefore, for any convergent sequence 
\begin{align*}
	\left(\tilde{x}_s^{(k)}, \tilde{z}^{(k)}, \nu^{(k)}, \Omega^{(k)}\right) \overset{k \to \infty}{\longrightarrow} (\tilde{x}_s, \tilde{z}, \nu, \Omega)
\end{align*}
we have for an interval $I$ 
\begin{align*}
	& \lim_{k \to \infty}
	\mathbf{1} \left\{ \tilde{x}_s^{(k)} \in \mathcal{Y}_0 (0, \tilde{z}^{(k)}, \nu^{(k)}, \Omega^{(k)}) \right\}
	\mathbf{1} \left(
	\frac{
		 g_{c, 1} \left( \tilde{x}_s^{(k)}, \tilde{z}^{(k)}, \nu^{(k)}, \Omega^{(k)} \right)
	}
	{
		g_{c, 2}\left(\tilde{x}_s^{(k)}, \tilde{z}^{(k)}, \nu^{(k)}, \Omega^{(k)}\right)
	} \in I \right)
	\\
	= & 
	\mathbf{1} \left\{ \tilde{x}_s \in \mathcal{Y}_0 (0, \tilde{z}, \nu, \Omega) \right\}
	\mathbf{1} 
	\left(
		\frac{g_{c, 1} \left( \tilde{x}_s, \tilde{z}, \nu, \Omega \right)}{g_{c, 2}(\tilde{x}_s, \tilde{z}, \nu, \Omega)} \in I
	\right). 
\end{align*}
This establishes the continuity of the mapping 
\begin{align*}
	f_{c, I}: (\tilde{x}_s, \tilde{z}, \nu, \Omega) \mapsto
	\mathbf{1} \left\{ \tilde{x}_s \in \mathcal{Y}_0 (0, \tilde{z}, \nu, \Omega) \right\}
	\mathbf{1}
	\left(
		\frac{g_{c, 1} \left( \tilde{x}_s, \tilde{z}, \nu, \Omega \right)}{g_{c, 2}(\tilde{x}_s, \tilde{z}, \nu, \Omega)} \in I	
	\right).
\end{align*}

We now return to our convergent subsequence 
\begin{align*}
	(\nu_n, \Omega_n) \overset{n \to \infty}{\longrightarrow} (\nu, \Omega).
\end{align*}
Along this subsequence and for given $\tilde{x}_s$ and $\tilde{z}$, we consider the limits $g_{c, 1}(\tilde{x}_s, \tilde{z}, \nu, \Omega)$ and $g_{c, 2}(\tilde{x}_s, \tilde{z}, \nu, \Omega)$ more closely. Suppose that $\tilde{x}_s \in \mathcal{Y}_0 (0, \tilde{z}, \nu, \Omega)$. For $\xi \in \mathcal{Y}_c \left(\tilde{x}_s, \tilde{z}, \nu, \Omega\right)$ we require that there exists a $\sigma \in \mathcal{E}(S)$ such that for all $h \in S$ we have
\begin{align}
	\label{eq:proof cond in S}
	\Omega_{n, h, s} \left(
		\xi + \tilde{x}_s + c	
	\right) + \Phi^{-1} \left(\nu_{h}\right) + \tilde{z}_h \geq \bar{x}_{\sigma, h}
\end{align}
and for all $h \notin S$ we have
\begin{align}
	\label{eq:proof cond not in S}
	\Omega_{n, h, s} \left(
		\xi + \tilde{x}_s + c	
	\right) + \Phi^{-1} \left(\nu_{h}\right) + \tilde{z}_h < \bar{x}_{\sigma, h}.
\end{align}
For $h \in S$ with $\Omega_{h, s} = 0$, either inequality \eqref{eq:proof cond in S} holds for all $\xi \in \mathbb{R}$ or $\mathcal{Y}_0 (0, \tilde{z}, \nu, \Omega) = \emptyset$. Similarly, for $h \notin S$ with $\Omega_{h, s} = 0$, either inequality \eqref{eq:proof cond not in S} is satisfied for all $\xi \in \mathbb{R}$ or $\mathcal{Y}_0 (0, \tilde{z}, \nu, \Omega) = \emptyset$. For $h \in S$ with $\mu_{h} = \Phi^{-1}(\nu_h) = \infty$, inequality \eqref{eq:proof cond in S} is satisfied for all $\xi \in \mathbb{R}$. For $h \notin S$ with $\mu_{h} = \Phi^{-1}(\nu_h) = -\infty$, inequality \eqref{eq:proof cond not in S} is satisfied for all $\xi \in \mathbb{R}$. Since $\tilde{x}_s \in \mathcal{Y}_0 (0, \tilde{z}, \nu, \Omega)$ implies that $\mathcal{Y}_0 (0, \tilde{z}, \nu, \Omega)$ is non-empty, we have
\begin{align*}
	\mathcal{Y}_c (\tilde{x}_s, \tilde{z}, \nu, \Omega) = \mathbb{R}.
\end{align*}
Along with the continuity of $f_{c, I}$ this implies that 
\begin{align*}
	f_{c, I}(\tilde{x}_s, \tilde{z}, \nu_n, \Omega_n) \overset{n \to \infty}{\longrightarrow} \mathbf{1} \left\{
		\tilde{x}_s \in \mathcal{Y}_0\left(0, \tilde{z}, \nu, \Omega \right)	
	\right\}
	\mathbf{1} \left\{
		\int_{\xi \leq - c} \, d\Phi(\xi) \in I
	\right\}.
\end{align*}
By the continuous mapping theorem, we now have 
\begin{align*}
	&P_{\nu_n, \Omega_n} 
	\left(
	\hat{S} = S \wedge
	F_s \left(X_s  \mid Z^{(s)}, X_s - c_{\alpha} - \epsilon, S \right) < 1 - \alpha / 2 \right)
	\\
	=& \E_{\nu_n, \Omega_n} f_{-c_{\alpha} - \epsilon, \alpha/2} (\tilde{X}_s, \tilde{Z}^{(s)}, \nu_n, \Omega_n) 
	\\
	\overset{n \to \infty}{\longrightarrow} \quad &
	\E_{\nu, \Omega} f_{-c_{\alpha} - \epsilon, \alpha/2} (\tilde{X}_s, \tilde{Z}^{(s)}, \nu, \Omega)
	\\
	=&
	\E_{\nu, \Omega} \left[
		\mathbf{1} \left\{ \hat{S} = S \right\}
		\mathbf{1} \left\{
			\int_{\xi \leq c_{\alpha} + \epsilon} \, d\Phi(\xi) < 1 - \alpha / 2
		\right\}
	\right] = 0,
\end{align*}
contradicting inequality \eqref{eq:proof below left endpoint}.
Similarly, 
\begin{align*}
	P_{\nu_n, \Omega_n} \left( \hat{S} = S \wedge F_s \left(X_s \mid Z^{(s)}, X_s - c_{\alpha} + \epsilon, S \right) > 1 - \alpha / 2 \right) \to 0
\end{align*}
as $n \to \infty$, contradicting inequality \eqref{eq:proof above right endpoint}. 
\end{Proof}

\begin{Proof}[Proof of Theorem~\ref{thm:validity median-unbiased oracle}]
	We have 
	\begin{align*}
		 P \left( \tilde{\theta}^{\text{ub}}_s \geq \theta_{s} \mid \hat{S} = S \right) 
		 = &
		 P \left( \tilde{\theta}_s^{(0.5)} \leq \theta_{s} \mid \hat{S} = S \right) 
	\\
		 = & 
		 P \left( 
			 F_s \left(X_s \mid Z^{(s)}, \theta_s, S \right) 
			 \leq F_s \left(X_s \mid Z^{(s)}, \tilde{\theta}_s^{(0.5)}, S \right)
			 \mid \hat{S} = S
		 \right)
	\\
		 = & 
		 P \left( 
		 	F_s \left(X_s \mid Z^{(s)}, \theta_s, S \right)	
			 \leq 0.5
			 \mid \hat{S} = S
		 \right) = 0.5, 
	\end{align*}
	where the second equality follows from Lemma~\ref{lem:Rosenblatt_invertibility} and the last equality follows from Lemma~\ref{lem:rosenblatt_uniform_distribution}.
\end{Proof}
\begin{Proof}[Proof of Theorem~\ref{thm:validity CS oracle}]
	We have 
	\begin{align*}
		& P \left(\theta_s \in \text{CCI}_\alpha (\theta_s \mid S)\mid \hat{S} = S \right)	
	\\
		=& P \left( \tilde{\theta}_s^{(1 - \alpha/2)} \leq \theta_s \leq \tilde{\theta}_s^{(\alpha/2)} \mid \hat{S} = S\right)
	\\
		=&P \left( \theta_s \leq \tilde{\theta}_s^{(\alpha/2)} \mid \hat{S} = S \right) 
		- P \left( \mu_s \leq \tilde{\theta}_s^{(1 - \alpha/2)} \mid \hat{S} = S \right)
	\\
		=& P \left( F_{s} (X_s \mid Z^{(s)}, \theta_s, S) \geq \alpha/2 \right) 
		- P \left( F_{s} (X_s \mid Z^{(s)}, \theta_s, S) \geq 1 - \alpha/2 \right) 
	\\
		=& 1 - \alpha/2 - \alpha /2 = 1 - \alpha,
	\end{align*}
	where the third equality follows from Lemma~\ref{lem:rosenblatt_uniform_distribution}. 
\end{Proof}

\begin{Proof}[Proof of Theorem~\ref{thm:asymp median-unbiased}]
	Follows immediately from Lemma~\ref{lem:asymp}.
\end{Proof}

\begin{Proof}[Proof of Theorem~\ref{thm:asymp CS validity}]
	Follows immediately from Lemma~\ref{lem:asymp}.
\end{Proof}

\section{Technical lemmas\label{sec:lemmas}}

\begin{Lemma}
	\label{lem:asymp}
    Suppose that Assumption \ref{assume:asymp} holds. Let $S \subseteq \{1, \dotsc, m\}$ and $s \in S$. For all $p \in (0, 1)$,
    \begin{align*}
    \lim_{n \to \infty} \sup_{P \in \mathcal{P}_n} \left\lvert P \left( \hat{\theta}^{(p)}_s > \theta_s (P) \mid \hat{S} = S \right) - p \right\rvert P \left( \hat{S} = S \right) = 0.
    \end{align*}
\end{Lemma}
\begin{Proof}
Throughout the proof, for a sequence of random vectors $W_n$ say that $W_n$ vanishes in probability along $P_n$, or  
\begin{align*}
    W_n = o_{p, P_n} (1),
\end{align*}
if, for all $\epsilon > 0$,
\begin{align*}
    P_n \left( \lVert W_n \rVert > \epsilon \right) \to 0.
\end{align*}
For a threshold function $\bar{x}$ and $\sigma \in \mathcal{E}(S)$, let 
\begin{align*}
    \bar{x}_{\sigma, h} = \begin{cases}
        \bar{x} \left(\sigma^{-1} \left(\{h, \dotsc, \lvert S \rvert\} \right) \cup S^{\mathsf{c}} \right) & \text{if $h \in S$},
        \\
        \bar{x} \left(S^c \right) & \text{if $h \notin S$}.
    \end{cases}
\end{align*}
Write $\hat{Z}$ as $\hat{Z} = \widehat{M}_s \hat{X}$, where $\widehat{M}_s = \mathbf{I}_m - \hat{\Omega} \mathbf{P}_s$, with $\mathbf{P}_s$ the projection matrix onto the $s$-th coordinate, i.e., the $m \times m$ matrix with a one at position $(s, s)$ and zeros elsewhere.
Let $\hat{X} = \diag^{-1/2} (\hat{\mathbf{v}}) \hat{\theta}$, $\theta_n = \theta (P_n)$ and
\begin{align*}
    \mathbf{v}_n = \mathbf{v} (P_n) = (V_{11}(P_n), V_{22}(P_n), \dotsc, V_{m,m}(P_n))'.
\end{align*}
The selection event $\hat{S} = S$ realises if and only if
\begin{align*}
    \hat{X}_s - \hat{v}_s^{-1/2} \theta_{n, s} \in \mathcal{Y}_s \left(\widehat{\Omega}, \diag^{-1/2}(\hat{\mathbf{v}}) \theta_{n}, \widehat{M}_s \diag^{-1/2} (\hat{\mathbf{v}}) (\hat{\theta} - \theta_{n}), \hat{\bar{x}} \right),
\end{align*}
where
\begin{align*}
    \mathcal{Y}_s \left(\Omega, \mu, \tilde{Z}, \bar{x} \right) = 
    \begin{aligned}[t]
    \bigcup_{\sigma \in \mathcal{E}(S)} \bigg(
    & \bigcap_{h \in S} \left\{ \xi \in \mathbb{R} : \Omega_{h, s} \xi + \mu_h + \tilde{Z}_h \geq \bar{x}_{\sigma, h} \right\} \cap
\\
    & \bigcap_{h \notin S} \left\{ \xi \in \mathbb{R} : \Omega_{h, s} \xi + \mu_h + \tilde{Z}_h < \bar{x}_{\sigma, h} \right\}
    \bigg).
    \end{aligned}
\end{align*}
We prove 
\begin{align*}
    \lim_{n \to \infty} \sup_{P \in \mathcal{P}_n} \left\lvert P \left( 
        \hat{\theta}_s^{(p)} > \theta_s(P) \mid \hat{S} = S
    \right) - p \right\rvert P \left( \hat{S} = S \right) = 0.
\end{align*}
The proof for the other claim of the lemma is almost identical.
Suppose that \eqref{eq:asymp lemma claim} does not hold. Then, there exists $\kappa > 0$ and a sequence $P_n \in \mathcal{P}_n$ such that along a subsequence,
\begin{align}
    \label{eq:asymp lemma claim}
    \left\lvert P_n \left(\hat{\theta}_s^{(p)} > \theta_s(P_n) \mid \hat{S} = S \right) - p \right\rvert P_n \left( \hat{S} = S \right) \geq \kappa.
\end{align}
Pass to this subsequence. Our proof aims at finding a subsequence with certain properties. Starting from our original subsequence, we will pass to subsubsequences and subsubsubsequences and so forth to find this subsequence. As we pass down to a subsequence, we replace the original sequence by the subsequence without changing the notation. Therefore, the meaning of, e.g., $P_n$ changes throughout the proof. Keeping track of subsequence indices would make the notation cumbersome. 
We write $\theta_n = \theta(P_n)$, $\mathbf{v}_n = \mathbf{v}(P_n)$, $\bar{x}_n = \bar{x}(P_n)$, $V_n = V(P_n)$, $\Omega_n = \Omega(P_n)$ and $\mu_n = \diag^{-1/2} (\mathbf{v}_n) \theta_n$. Along the subsequence, $P_n ( \hat{S} = S ) \geq \kappa$. 

We now establish the convergence of different quantities along appropriately chosen subsequences. 
We first consider the sequence of threshold function $\bar{x}_n$. The domain $\mathcal{D}$ of $\bar{x}_n$ is the space of all subsets of $\{1, \dotsc, m\}$, which has cardinality $2^m$. Therefore, we can think of $\bar{x}_n$ as a finite-dimensional vector that is contained in the $2^m$-dimensional hypercube $[-C_{\bar{x}}, C_{\bar{x}}]^{2^m}$. This is a compact set. Therefore, $\bar{x}_n$ has a convergent subsequence. Pass to this subsequence and denote the limit by $\bar{x}^*$. We now show that $\bar{x}^*$ is a threshold function, i.e., for subsets $A \subseteq B$ of $\{1, \dotsc, m\}$, $\bar{x}^*(A) \leq \bar{x}^*(B)$. Suppose that this is not the case. Then, there exist $A \subseteq B$ and $b> 0$ such that $\bar{x}^*(A) - \bar{x}^*(B) \geq b$. Therefore, 
\begin{align*}
    b <& \bar{x}^*(A) - \bar{x}^*(B) 
    \\
    =& \bar{x}_n(A) - \bar{x}_n(B) + o(1), 
\end{align*}
implying that $\bar{x}_n(A) \geq \bar{x}_n(B) + b/2$ eventually, contradicting the fact that $\bar{x}_n$ is a threshold function. 
This establishes that there is a threshold function $\bar{x}^*$ such that 
\begin{align*}
    \hat{\bar{x}} = \bar{x}^* + o_{p, P_n} (1).
\end{align*}
All $\Omega_n$ are correlation matrices. The space of correlation matrices is a compact space. Therefore, there exists a subsequence along which $\Omega_n$ converges to a limiting correlation matrix $\Omega^*$. Pass to this subsequence.  
For a random vector $W$, write $W = O_{p, P_n} (1)$ if there is a constant $C$ such that 
\begin{align*}
    P_n \left( \lVert W \rVert > C \right) \to 0.
\end{align*}
Let $\widehat{D}_n = \diag^{-1/2} (\hat{\mathbf{v}}_n) \diag^{-1/2} (\mathbf{v}_n)$. Note that $\Omega_n = \diag^{-1/2} (\mathbf{v}_n) V_n \diag^{-1/2} (\mathbf{v}_n)$. Decompose, 
\begin{align*}
    \widehat{\Omega} - \Omega^* =& \widehat{D}_n \diag^{-1/2} (\mathbf{v}_n) \left(\hat{V} - V_n \right) \diag^{-1/2} (\mathbf{v}_n) \widehat{D}_n + \widehat{D}_n \left( \Omega_n - \Omega^* \right) \widehat{D}_n 
\\
    & + \left(\widehat{D}_n - \mathbf{I}_m \right) \Omega^* \widehat{D}_n + \Omega^* \left( \widehat{D}_n - \mathbf{I}_m \right).
\end{align*}
By Assumption~\ref{assume:asymp}.\ref{assume:asymp_local:vhat}, $\widehat{D}_n - \mathbf{I}_m = o_{p, P_n} (1)$, $\widehat{D}_n = O_{p, P_n} (1)$ and $\hat{V} - V_n = o_{p, P_n} (1)$. From above, $\Omega_n - \Omega^* = o(1)$. Plugging these asymptotic results into the decomposition above, we have $\widehat{\Omega} - \Omega^* = o_{p, P_n} (1)$.
Let $M^* = \mathbf{I}_m - \Omega^* P_s$. We now prove, 
\begin{align}
    \label{eq:asymp lemma normality}
    \left(X_s - \hat{v}_s^{-1/2} \theta_{n, s}, \widehat{M}_s \diag^{-1/2} (\hat{\mathbf{v}}) (\hat{\theta} - \theta_n)\right) \Rightarrow N \left(0,
    \begin{bmatrix}
        1 & \mathbf{0}_{1 \times 0} 
        \\
        \mathbf{0}_{m \times 1} & M_s^* \Omega^* M_s^*
    \end{bmatrix}
    \right)
\end{align}
along $P_n$ in the usual sense of weak convergence.
Let $f \in BL_{1}$ and $\xi_{\Omega} \sim N (0, \Omega)$. We have $\Omega_{n} \to \Omega^*$ and therefore $\xi_{\Omega_{n}} \Rightarrow \xi_{\Omega^*}$. This is obvious from the pointwise convergence of the characteristic functions. 
Therefore, 
\begin{align*}
	\big\vert 
		\E \left[ f \left(\xi_{\Omega_n}\right) \right] 
		- \E \left[ f (\xi_{\Omega^*}) \right] 
	\big\rvert \to 0. 
\end{align*}
Let $\mu_n = \diag^{-1/2} (\mathbf{v}_n) \theta_n$. By Assumption~\ref{assume:asymp}.\ref{assume:asymp_local:normality}, 
\begin{align*}
	\left\vert 
		\mathbb{E}_{P_n} \left( f \left(\diag^{-1/2} (\mathbf{v}_n) \theta_n - \mu_n \right) \right)
		- \E \left[ f \left(\xi_{\Omega_n}\right) \right]
	\right\rvert \to 0. 
\end{align*}
Therefore, by the triangle inequality,
\begin{align*}
	\left\vert 
		\mathbb{E}_{P_n} \left( f \left(\diag^{-1/2} (\mathbf{v}_n) \theta_n - \mu_n \right) \right)
		- \E \left[ f \left(\xi_{\Omega^*}\right) \right]
	\right\rvert \to 0. 
\end{align*}
Since $f$ was chosen arbitrarily this holds for all $f \in BL_1$ and 
\begin{align*}
	\diag^{-1/2} (\mathbf{v}_n) \theta_n - \mu_n \Rightarrow N(0, \Omega^*).
\end{align*}
Noting that $\widehat{D}_n - \mathbf{I}_m = o_{p, P_n} (1)$ and $\widehat{M}_s - M_s^* = o_{p, P_n} (1)$, Slutzky's lemma implies \eqref{eq:asymp lemma normality}.

Let $\lVert \cdot \rVert_2$ denote the matrix norm induced by the Euclidean norm. 
Define the events
\begin{align*}
    \mathcal{E}_{1, \eta} =& \Big\{
    \max_{h = 1, \dotsc, m} \left\lvert \hat{v}_h / v_{n, h} - 1 \right\rvert \leq \eta, 
    \left\lvert \hat{\bar{x}} - \bar{x}^* \right\rvert_\infty \leq \eta,
    \lVert \widehat{\Omega} - \Omega^* \rVert_2 \leq \eta
    \Big\},
    \\
    \mathcal{E}_{2, b}=& \Big\{
    \lVert \widehat{M}_s \diag^{-1/2} (\hat{\mathbf{v}}) (\hat{\theta} - \theta_n) \rVert_{\max} \leq b 
    \Big\},
\end{align*}
where $\eta > 0$ is a small number. For all our arguments, we can take $\eta = 1$, but we write $\eta$ to emphasise that the bounds that we derive hold because $\eta$ represents an upper bound on the distance between different convergent sequences and their limits.

Suppose that $\mu_{n, h'}$ is unbounded from below for some $h' \in S$.
Then, there exists a sequence $b_n \to \infty$ such that, passing to a subsequence if necessary, $\mu_{n, h'} \leq -3 b_n$. On the event $\mathcal{E}_{1, \eta}$ for $\eta$ small enough, we have $\hat{v}_{n, h'}^{-1/2} \theta_{n, h'} \leq -2 b_n$. Therefore, on the event $\mathcal{E}_{1, \eta} \cap \mathcal{E}_{2, b_n}$, 
\begin{align*}
    & \widehat{\Omega}_{h', s} \xi +  \hat{v}_{h'}^{-1/2} \theta_{n, h'} + \left(\widehat{M}_s \diag^{-1/2} (\hat{\mathbf{v}}) (\hat{\theta} - \theta_n) \right)_{h'} - \hat{\bar{x}}^*_{\sigma, h'} 
\\
    \leq & (1 + \eta) \lvert \xi \rvert  - b_n - \min_{\sigma, h} \bar{x}^*_{\sigma, h} + \eta.
\end{align*}
Since the right-hand side does not depend on $\sigma$, we can bound the probability of the selection event by 
\begin{align*}
    & P_n \left( \hat{S} = S \right)
    \\
    \leq & P_n \left( \widehat{X}_s - \hat{v}_s^{-1/2} \theta_{n, s} \in \left\{ \xi \in \mathbb{R}: (1 + \eta) \lvert \xi \rvert - b_n - \min_{\sigma, h} \bar{x}^*_{\sigma, h} + \eta \geq 0 \right\} \wedge \mathcal{E}_{1, \eta} \wedge \mathcal{E}_{2, b_n} \right)
    \\
    & + P_n (\mathcal{E}_{1, \eta}^{\mathsf{c}}) + P_n (\mathcal{E}_{2, b_n}^{\mathsf{c}})
\end{align*}
Let $\xi \sim N(0, 1)$ and let $\epsilon > 0$. By monotone convergence, for $b$ large enough, 
\begin{align*}
    P \left( \xi \in \left\{ \xi \in \mathbb{R}: (1 + \eta) \lvert \xi \rvert - b - \min_{\sigma, h} \bar{x}^*_{\sigma, h} + \eta \geq 0 \right\} \right) \leq \epsilon / 4.
\end{align*}
The probability on the right-hand side decreases in $b$. By \eqref{eq:asymp lemma normality}, for $n$ large enough, 
\begin{align*}
    & P_n \left( \widehat{X}_s - \hat{v}_s^{-1/2} \theta_{n, s} \in \left\{ \xi \in \mathbb{R}: (1 + \eta) \lvert \xi \rvert - b_n - \min_{\sigma, h} \bar{x}^*_{\sigma, h} + \eta \geq 0 \right\} \right)
    \\
    \leq & P \left( \xi \in \left\{ \xi \in \mathbb{R}: (1 + \eta) \lvert \xi \rvert - b_n - \min_n{\sigma, h} \bar{x}^*_{\sigma, h} + \eta \geq 0 \right\} \right) + \epsilon / 4 \leq \epsilon / 2.
\end{align*}
For $n$ large enough such that $P_n (\mathcal{E}_{1, \eta}) \leq \epsilon / 4$ and $P_n (\mathcal{E}_{2, b_n}) \leq \epsilon / 4$, we have $P_n ( \hat{S} = S ) \leq \epsilon$. Since $\epsilon$ is an arbitrarily small positive number, this contradicts $P_n(\hat{S} = S) \geq \kappa$. Similarly, we can arrive at a contradiction if we assume that $\mu_{n, h'}$ is unbounded from above for some $h' \notin S$.

Now suppose that $\mu_{n, h'}$ is unbounded from above for some $h' \in S$. Arguing as above, there is a sequence $b_n \to \infty$ such that, passing to a subsequence if necessary, $\hat{v}_{n, h'}^{-1/2} \theta_{n, h} \geq 2 b_n$ on $\mathcal{E}_{1, \eta}$. 
On the event $\mathcal{E}_{1, \eta} \cap \mathcal{E}_{2, b_n}$,
\begin{align*}
    & \widehat{\Omega}_{h',s} \xi + \hat{v}_{h'}^{-1/2} \theta_{n, h'} + \left(\widehat{M}_s \diag^{-1/2} (\hat{\mathbf{v}}) (\hat{\theta} - \theta_n) \right)_{h'} - \hat{\bar{x}}_{\sigma, h'}
    \\
    \geq & - (1 + \eta) \lvert \xi \rvert + b_n - \max_{\sigma, h} \bar{x}^*_{\sigma, h} - \eta.
\end{align*}
Let 
\begin{align*}
    \mathcal{Y}_{s, H} \left( \Omega, \mu, \tilde{z}, \bar{x} \right) =
    \begin{aligned}[t]
    \bigcup_{\sigma \in \mathcal{E}(S)} \Big( & \bigcap_{h \in S \setminus H} \left\{ \xi \in \mathbb{R} : \Omega_{h, s} \xi + \mu_h + \tilde{z}_h \geq \bar{x}_{\sigma, h} \right\}
    \\
    & \cap \bigcap_{h \in S^{\mathsf{c}} \setminus H} \left\{ \xi \in \mathbb{R} : \Omega_{h, s} \xi + \mu_h + \tilde{z}_h < \bar{x}_{\sigma, h} \right\} \Big).
    \end{aligned}
\end{align*}
On $\mathcal{E}_{1, \eta} \cap \mathcal{E}_{2, b_n}$, we have the sandwich
\begin{align*}
    & \mathcal{Y}_{s, \{h'\}} \left( \widehat{\Omega}, \diag^{-1/2} (\hat{\mathbf{v}}) \theta_{n}, \widehat{M}_s \diag^{-1/2} (\hat{\mathbf{v}}) (\hat{\theta} - \theta_n), \hat{\bar{x}} \right)
    \\
    \subseteq & \mathcal{Y}_s \left( 
        \widehat{\Omega}, \diag^{-1/2} (\hat{\mathbf{v}}) \theta_n, \widehat{M}_s \diag^{-1/2} (\hat{\mathbf{v}}) (\hat{\theta} - \theta_n), \hat{\bar{x}}
    \right) 
    \\
    \subseteq &
    \mathcal{Y}_{s, \{h'\}} \left( 
        \widehat{\Omega}, \diag^{-1/2} (\hat{\mathbf{v}}) \theta_n, \widehat{M}_s \diag^{-1/2} (\hat{\mathbf{v}}) (\hat{\theta} - \theta_n), \hat{\bar{x}}
    \right) \cup A_{1, n},
\end{align*}
where
\begin{align*}
    A_{1, n} = \left\{ \xi \in \mathbb{R} : - (1 + \eta) \lvert \xi \rvert + b_n - \max_{\sigma, h} \bar{x}^*_{\sigma, h} - \eta \geq 0 \right\}.
\end{align*}
We can argue a similar sandwich with $A_{1, n}$ replaced by 
\begin{align*}
    A_{2, n} = \left\{ \xi \in \mathbb{R} : (1 + \eta) \lvert \xi \rvert - b_n - \min_{\sigma, h} \bar{x}^*_{\sigma, h} + \eta < 0 \right\}, 
\end{align*}
for $h' \notin S$ that are unbounded from below. Let $A_n = A_{1, n} \cup A_{2, n}$. 

We have so far established that $h \in S$ with $\mu_{n, h}$ unbounded from below and $h \notin S$ with $\mu_{n, h}$ unbounded from above cannot occur along the subsequence $P_n$ that we are considering. Gather all $h \in S$ with $\mu_{n, h}$ unbounded from above and all $h \notin S$ with $\mu_{n, h}$ unbounded from below in the set $H$. For all $h \notin H$, we have $\mu_{n, h}$ bounded from above and below and hence contained in a compact set. Therefore, $\mu_{n, h}$ converges to a finite limit $\mu_h^*$ for all $h \notin H$ along a subsequence. Pass to this subsequence. Write $\mu^*$ for the limit of the vector $\mu_n$, allowing individual components $s$ and $h \notin H$ to be infinite.

Define the family of functions, 
\begin{align*}
    \xi \mapsto f_{\mathcal{Y}, H} (\xi \mid \Omega, \mu, \tilde{z}, \bar{x}) = 1 \left\{ \xi \in \mathcal{Y}_{s, H} (\Omega, \mu, \tilde{z}, \bar{x}) \right\}.
\end{align*}
Consider a generic convergent sequence 
\begin{align*}
    (\Omega^{(j)}, \mu^{(j)}, \tilde{z}^{(j)}, \bar{x}^{(j)}) \to (\Omega', \mu', \tilde{z}', \bar{x}')
\end{align*}
and note that $f_{\mathcal{Y}, H} (\cdot \mid \Omega^{(j)}, \mu^{(j)}, \tilde{z}^{(j)}, \bar{x}^{(j)})$ converges pointwise to $f_{\mathcal{Y}, H} (\cdot \mid \Omega', \mu', \tilde{z}', \bar{x}')$ except possibly on a set of Lebesgue measure zero. Since the standard Gaussian measure $d\Phi$ is absolutely continuous with respect to the Lebesgue measure, $f_{\mathcal{Y}, H} (\cdot \mid \Omega^{(j)}, \mu^{(j)}, \tilde{z}^{(j)}, \bar{x}^{(j)})$ converges $d\Phi$-almost everywhere. By the dominated convergence theorem, 
\begin{align*}
    \int_\xi f_{\mathcal{Y}, H} (\xi \mid \Omega^{(j)}, \mu^{(j)}, \tilde{z}^{(j)}, \bar{x}^{(j)}) \, d\Phi (\xi) \to \int_\xi f_{\mathcal{Y}, H} (\xi \mid \Omega', \mu', \tilde{z}', \bar{x}') \, d\Phi (\xi) > 0
\end{align*}
as $j \to \infty$. The inequality on the right-hand side follows from the fact that $\mathcal{Y}_{s, H} (\Omega', \mu', \tilde{z}', \bar{x}')$ contains nonempty intervals and that the Gaussian distribution is supported on the entire real line.
Let 
\begin{align*}
    f_{x}(\xi \mid \tilde{x}) = 1 \left\{ \xi \leq \tilde{x} \right\}.
\end{align*}
For any convergent sequence 
\begin{align*}
    (\Omega^{(j)}, \mu^{(j)}, \tilde{z}^{(j)}, \bar{x}^{(j)}, \tilde{x}^{(j)}) \to (\Omega', \mu', \tilde{z}', \bar{x}', \tilde{x}')
\end{align*}
it can be argued similarly as above that 
\begin{align*}
    & \int_\xi f_{\mathcal{Y}, H} (\xi \mid \Omega^{(j)}, \mu^{(j)}, \tilde{z}^{(j)}, \bar{x}^{(j)}) f_{x} (\xi \mid \tilde{x}^{(j)}) \, d\Phi (\xi)
    \\
    \to & \int_\xi f_{\mathcal{Y}, H} (\xi \mid \Omega', \mu', \tilde{z}', \bar{x}') f_{x} (\xi \mid \tilde{x}') \, d\Phi (\xi).
\end{align*}
This establishes that the function 
\begin{align*}
    g_{H, L}: (\mathbb{R}^{m \times m}_+, \mathbb{R}^m, \mathbb{R}^m, \mathbb{R}^m, \mathbb{R}_{m \lvert \mathcal{E}(S) \rvert}, \mathbb{R}_+) \to \mathbb{R}_+
\end{align*}
defined by 
\begin{align*}
    g_{H, \ell} (\Omega, \mu, \tilde{z}, \bar{x}, \tilde{x}, a)
    = \frac{\int_\xi f_x(\xi \mid \tilde{x}) f_{\mathcal{Y}, H} (\xi \mid \Omega, \mu, \tilde{z}, \bar{x}) \, d\Phi (\xi)}{a + \int_\xi f_{\mathcal{Y}, H} (\xi \mid \Omega, \mu, \tilde{z}, \bar{x}) \, d\Phi (\xi)}
\end{align*}
is continuous.
Let $(\Xi, \mathcal{Z})$ denote an $\mathbb{R}^{1 + m}$-random vector with distribution equal to Gaussian distribution on the right-hand side of \eqref{eq:asymp lemma normality}.
Let $a_n = \int_\xi 1 \left\{ \xi \in A_n \right\} \, d\Phi (\xi)$ and note that $a_n$ is a positive sequence that vanishes as $n \to \infty$.
We have 
\begin{align*}
    \left( \hat{X}_s - \hat{v}_s^{-1/2} \theta_{n, s}, \widehat{M}_s \diag^{-1/2} (\hat{\mathbf{v}}) (\hat{\theta} - \theta_n), \widehat{\Omega}, \mu_n, \hat{\bar{x}}, a_n \right) \Rightarrow \left(\Xi, \mathcal{Z}, \Omega^*, \mu^*, \bar{x}^*, 0 \right)
\end{align*}
and hence by the continuous mapping theorem,
\begin{align*}
    & g_{H, \ell} \left( \widehat{\Omega}, \mu_n, \widehat{M}_s \diag^{-1/2} (\hat{\mathbf{v}}) (\hat{\theta} - \theta_n), \hat{\bar{x}}, \hat{X}_s - \hat{v}_s^{-1/2} \theta_{n, s}, a_n \right)
    \\
    \Rightarrow & g_{H, \ell} \left( \Omega^*, \mu^*, \mathcal{Z}, \bar{x}^*, \Xi, 0 \right).
\end{align*}
An almost identical argument proves that for
\begin{align*}
    g_{H, u} \left(\Omega, \mu, \tilde{z}, \bar{x}, \tilde{x}, a \right) = \frac{a + \int_\xi f_x(\xi \mid \tilde{x}) f_{\mathcal{Y}, H} (\xi \mid \Omega, \mu, \tilde{z}, \bar{x}) \, d\Phi (\xi)}{\int_\xi f_{\mathcal{Y}, H} (\xi \mid \Omega, \mu, \tilde{z}, \bar{x}) \, d\Phi (\xi)},
\end{align*}
we have 
\begin{align*}
    & g_{H, u} \left(\widehat{\Omega}, \mu_n, \widehat{M}_s \diag^{-1/2} (\hat{\mathbf{v}}) (\hat{\theta} - \theta_{n, s}), \hat{\bar{x}}, \hat{X}_s - \hat{v}_{n, s}^{-1/2} \theta_s, a_n \right)
    \\
    \Rightarrow & g_{H, u} \left(\Omega^*, \mu^*, \mathcal{Z}, \bar{x}^*, \Xi, 0 \right).
\end{align*}
The Rosenblatt transform is given by 
\begin{align*}
    \widehat{F}_s \left(\hat{X}_s \mid \hat{Z}, \theta_s, S\right)
    =& F_s \left( \hat{X}_s \mid  \hat{Z}, \theta_s, \hat{\bar{x}}, \hat{v}_s, \widehat{\Omega}, S\right) 
    \\
    =& g_{\emptyset, \ell} \left(\widehat{\Omega}, \mu_n, \widehat{M}_s \diag^{-1/2} (\hat{\mathbf{v}}) (\hat{\theta} - \theta_n), \hat{\bar{x}}, \hat{X}_s - \hat{v}_{n, s}^{-1/2} \theta_{n, s}, 0 \right).
\end{align*}
Its asymptotic counterpart is 
\begin{align*}
    F_s \left( \Xi  \mid \mathcal{Z}, \theta_s, S \right) =
    F_s \left( \Xi \mid \mathcal{Z}, \theta_s, \bar{x}^*, v_s^*, \Omega^*,  S \right) = 
    g_{\emptyset, \ell} \left(\Omega^*, \mu^*, \mathcal{Z}, \bar{x}^*, \Xi, 0 \right).
\end{align*}
The sandwich
\begin{align*}
    & \mathcal{Y}_{s, H} \left( \widehat{\Omega}, \mu_n, \widehat{M}_s \diag^{-1/2} (\hat{\mathbf{v}}) (\hat{\theta} - \theta_n), \hat{\bar{x}} \right)
    \\  
    \subseteq & \mathcal{Y}_s \left( \widehat{\Omega}, \mu_n, \widehat{M}_s \diag^{-1/2} (\hat{\mathbf{v}}) (\hat{\theta} - \theta_n), \hat{\bar{x}} \right) 
    \\
    \subseteq & \mathcal{Y}_{s, H} \left( \widehat{\Omega}, \mu_n, \widehat{M}_s \diag^{-1/2} (\hat{\mathbf{v}}) (\hat{\theta} - \theta_n), \hat{\bar{x}} \right) \cup A_n,
\end{align*}
bounds
\begin{align*}
    & g_{H, \ell} \left( \widehat{\Omega}, \mu_n, \widehat{M}_s \diag^{-1/2} (\hat{\mathbf{v}}) (\hat{\theta} - \theta_n), \hat{\bar{x}}, \hat{X}_s - \hat{v}_s^{-1/2} \theta_{n, s}, a_n \right)
    \\
    \leq & g_{\emptyset, \ell} \left( \widehat{\Omega}, \mu_n, \widehat{M}_s \diag^{-1/2} (\hat{\mathbf{v}}) (\hat{\theta} - \theta_n), \hat{\bar{x}}, \hat{X}_s - \hat{v}_s^{-1/2} \theta_{n, s}, 0 \right)
    \\
    \leq & g_{H, u} \left( \widehat{\Omega}, \mu_n, \widehat{M}_s \diag^{-1/2} (\hat{\mathbf{v}}) (\hat{\theta} - \theta_n), \hat{\bar{x}}, \hat{X}_s - \hat{v}_s^{-1/2} \theta_{n, s}, a_n \right) 
\end{align*}
Therefore, 
\begin{align*}
    & \left(
        \widehat{F}_s (\hat{X}_s \mid \hat{Z}, \theta_s, S), 
        \hat{X}_s - \hat{v}_s^{-1/2} \theta_{n, s},
        \widehat{M}_s \diag^{-1/2} (\hat{\mathbf{v}}) (\hat{\theta} - \theta_n),
        \widehat{\Omega}, \mu_n, \hat{\bar{x}}
    \right) 
\\
    \Rightarrow &
    \left(
        F_s (\Xi \mid \mathcal{Z}, \theta_s, S), \Xi, \mathcal{Z}, \Omega^*, \mu^*, \bar{x}^*
    \right).
\end{align*}
Let $Q_n$ denote the probability distribution of the random vector on the left-hand side and let $Q$ denote the probability distribution of the random vector on the right-hand side. For $\mu_n$ we implicitly consider only the components that are bounded. We operate on a subsequence where $Q_n \Rightarrow Q$ in the usual sense of weak convergence. We have the following nesting of events:
\begin{align*}
    & \left\{ \hat{X}_s - \hat{v}_s^{-1/2} \theta_{n, s} \in \mathcal{Y}_{s, H} \left(\widehat{\Omega}, \mu_n, \widehat{M}_s \diag^{-1/2} (\hat{\mathbf{v}}) (\hat{\theta} - \theta_n), \hat{\bar{x}} \right) 
    \right\} 
    \\
    \subseteq & \left\{ \hat{X}_s - \hat{v}_s^{-1/2} \theta_{n, s} \in \mathcal{Y}_s \left(\widehat{\Omega}, \mu_n, \widehat{M}_s \diag^{-1/2} (\hat{\mathbf{v}}) (\hat{\theta} - \theta_n), \hat{\bar{x}} \right) 
    \right\} 
    \\
    \subseteq &  \left\{ \hat{X}_s - \hat{v}_s^{-1/2} \theta_{n, s} \in \mathcal{Y}_{s, H} \left(\widehat{\Omega}, \mu_n, \widehat{M}_s \diag^{-1/2} (\hat{\mathbf{v}}) (\hat{\theta} - \theta_n), \hat{\bar{x}} \right) 
    \right\}
    \\ & \cup \left\{ \hat{X}_s - \hat{v}_s^{-1/2} \theta_{n, s} \in A_n \right\} \cup \mathcal{E}_{1, \eta} \cup \mathcal{E}_{2, b_n}, 
\end{align*}
where $b_n$ is an increasing sequence that diverges to infinity such that 
\begin{align*}
    \left\lVert \diag^{-1/2} (\mathbf{v}_n) \theta_n \right\rVert_{\max} \leq b_n.
\end{align*}
The left-hand side of the inclusion corresponds to the $S$ event in Lemma~\ref{lemma:conditional_weak_convergence} under the probability measure $Q_n$. The middle part of the inclusion corresponds to the $S_n$ event in Lemma~\ref{lemma:conditional_weak_convergence} also under the probability measure $Q_n$. Lemma~\ref{lemma:conditional_weak_convergence} requires 
\begin{align*}
    Q_n \left(S \, \triangle \, S_n \right) \to 0.
\end{align*}
By the inclusion above, this is implied by 
\begin{align*}
    Q_n \left( \hat{X}_s - \hat{v}_s^{-1/2} \theta_{n, s} \in A_n \right) \to 0, \quad Q_n \mathcal{E}_{1, \eta} \to 0 \quad \text{and} \quad Q_n \mathcal{E}_{2, b_n} \to 0.
\end{align*}
We have already verified these three conditions above. 
Lemma~\ref{lemma:conditional_weak_convergence} requires also that $Q (\partial S) = 0$, where $\partial S$ denotes the boundary of $S$. Bound 
\begin{align*}
    \partial S \subseteq  \bigcup_{\sigma \in \mathcal{E}(S)} \bigcup_{h \in \left\{1, \dotsc, m\right\} \subset H} \left\{ (\xi, z) \in \mathbb{R}^{1 + m}: \Omega^*_{h, s} \xi + \mu^*_h + z_h = \bar{x}^*_{\sigma, h} \right\}
\end{align*}
and note that the set on the right-hand side has Lebesgue measure zero and hence measure zero under the limiting Gaussian distribution. 
Now, applying Lemma~\ref{lemma:conditional_weak_convergence} yields
\begin{align*}
    \widehat{F}_s \left( \hat{X}_s \mid \hat{Z}, \theta_s, S \right) \mid S_n \Rightarrow F_s \left( \Xi  \mid \mathcal{Z}, \theta_s, S \right) \mid S.
\end{align*}
Therefore, by Lemma~\ref{lem:rosenblatt_uniform_distribution}, we have 
\begin{align*}
    \widehat{F}_s \left( \hat{X}_s \mid \hat{Z}, \theta_s, S \right) \mid S_n \Rightarrow \mathcal{U}[0, 1].
\end{align*}
By Lemma~\ref{lem:Rosenblatt_invertibility} for $U \sim \mathcal{U}[0, 1]$, 
\begin{align*}
    \left\lvert P_n \left(\hat{\theta}_s^{(p)} > \theta_{n, s} \mid \hat{S} = S \right) - p \right\rvert
=& \left\lvert P_n \left( \widehat{F}_s \left( \hat{X}_s \mid \hat{Z}, \theta_{n, s}, S \right) < p \mid \hat{S} = S \right) - p \right\rvert
\\
=& P \left(U < p \right) - p + o(1) = o(1)
\end{align*}
on an appropriately chosen subsequence.
This contradicts inequality \eqref{eq:asymp lemma claim} and concludes the proof.
\end{Proof}
\begin{Lemma}[Convergence of conditional probability measures]
\label{lemma:conditional_weak_convergence}
Let $P$ and $P_n$, $n \in \mathbb{N}$, denote probability measures on a metric space $M$ that is equipped with its Borel-algebra. Suppose that
\begin{align*}
 	P_n \Rightarrow P, 
\end{align*} 
where the symbol ``$\Rightarrow$'' represents weak convergence. Let $S$ and $S_n$, $n \in \mathbb{N}$, denote Borel subsets of $M$. Assume that $S$ is $P$-continuous, i.e., $P (\partial S) = 0$, where $\partial S$ is the boundary of $S$, $P S = p^{*} > 0$, and
\begin{align*}
	P_n (S \bigtriangleup S_n) \to 0,
\end{align*}
as $n \to \infty$, where $S \bigtriangleup S_n$ denotes the symmetric set difference between $S$ and $S_n$.
Then, for $X \sim P$ and $X_n \sim P_n$
\begin{align*}
	X_n \mid S_n \Rightarrow X \mid S, 
\end{align*}
\end{Lemma}
\begin{Proof}
For $X_n \sim P_n$, write $Q_n$ for the probability measure of $X_n \mid S_n$. For $X \sim P$, write $Q$ for the probability measure of $X \mid S$. 
First note that, since $S$ is $P$-continuous, $P_n S \to P (S) = p^* > 0$. 
The inclusions, $S_n \subset S \cup (S_n \cap S^{\mathsf{c}})$ and 
$S \subset S_n \cup (S \cap S_n^{\mathsf{c}})$ and the union bound for probabilities imply 
\begin{align*}
	P_n S - P_n \left(S \cap S^{\mathsf{c}}_n \right) 
	\leq 
	P_n S_n 
	\leq 
	P_n S + P_n \left(S_n \cap S^{\mathsf{c}} \right)
\end{align*}
and thus 
\begin{align*}
	\abs{P_n S_n - p^*} \leq \abs{P_n S - P S} + P_n \left(S \bigtriangleup S_n \right).
\end{align*}
The first term on the right-hand side vanishes by weak convergence of $P_n$ to $P$ since $S$ is $P$-continuous. The second term on the right-hand side vanishes by assumption. Therefore, $P_n S_n \to p^* > 0$ and it is without loss of generality to assume that $P_n S_n$ is bounded away from zero. Thus, for $A \subset M$, $Q_n A = P_n (A \cap S_n) / P_n (S_n)$ is well-defined.
By the Portmanteau lemma, it suffices to show  
\begin{align*}
	Q_n A \to Q A
\end{align*}
as $n \to \infty$ for all $Q$-continuous sets $A \subseteq M$. For $Q$-continuous $A \subseteq M$, 
\begin{align*}
	Q_n A = \frac{P_n (A \cap S_n)}{P_n S_n} = 
	\frac{P_n (A \cap S)}{P_n S_n}
	+ \frac{P_n (A \cap S_n) - P_n (A \cap S)}{P_n S_n} =: p_n + q_n
\end{align*}
The inclusions, 
\begin{align*}
	A \cap S_n \subseteq & \left(A \cap S\right) \cup \left(S_n \cap S^{\mathsf{c}}\right)
\\
	A \cap S \subseteq & \left(A \cap S_n\right) \cup \left(S \cap S_n^{\mathsf{c}}\right)
\end{align*}
and the union bound for probabilities imply
\begin{align*}
	\abs{q_n} = \frac{\abs{ P_n \left(A \cap S_n\right) - P_n \left(A \cap S\right) }}{P_n S_n} \leq \frac{P_n \left(S_n \bigtriangleup S \right)}{P_n S_n} \to 0. 
\end{align*}
Next, we show $p_n \to Q A$, thus proving $Q_n A \to Q A$ and concluding the proof.
For a set $A$ let $\overline{A}$ denote the closure of $A$. For sets $A, B$ it can be shown that 
\begin{align*}
% \label{eq:boundary:intersection}
	\partial (A \cap B) \subseteq \left(\overline{B} \cap \partial A \right) \cup \left(\overline{A} \cap \partial B \right). 
\end{align*}
Applying this inequality with $A = A$ and $B = S$ yields 
\begin{align*}
	\partial (A \cap S) 
	\subseteq \left(\overline{S} \cap \partial A\right) \cup (\partial S)
	\subseteq \left(S \cap \partial A\right) \cup (\partial S)
\end{align*}
and therefore 
\begin{align*}
	P \partial (A \cap S) \leq P \left(S \cap \partial A\right) + P (\partial S)
	= P \left(S \cap \partial A\right) + 0 = 0, 
\end{align*}
where the first equality follows since $S$ is $P$-continuous and the second equality follows since $A$ is $Q$-continuous and therefore 
\begin{align*}
	0 = Q (\partial A) = \frac{P (S \cap \partial A)}{p^*}.
\end{align*}
This implies that $A \cap S$ is $P$-continuous and therefore $P_n (A \cap S) \to P (A \cap S)$ by the Portmanteau lemma. Thus,  
\begin{align*}
	p_n = \frac{P_n (A \cap S)}{P_n S_n} \to \frac{P (A \cap S)}{P S} = Q A. 
\end{align*}
\end{Proof}

\section{\label{sec:further_simu}Further simulations results}

We report the simulation results for the one-sided test below.  The data-generating process follows that in 
Section \ref{sec:simulation} of the main text.  
Several remarks are in order.  
First, in the case of a one-sided test, selection probabilities—the chances that the first parameter is selected ($\{1\} \subseteq \hat{S}$)—range from 3.0\% to 30.6\% across designs and sample sizes.
Second,  the conditional intervals consistently achieve coverage close to the nominal level of 90\%. 
In contrast,  the naive confidence intervals always undercover,  while the Bonferroni intervals tend to be conservative.  
Third,  in terms of interval width,  the conditional intervals tend to be wider than the Bonferroni intervals,  which is in contrast to the results of interval width in the case with two-sided test.
The result of relatively wide conditional intervals for the one-sided test is also in line with \textcite{kivaranovic2021length} and \textcite{dzemski2025location}. 
Fourth,  our conditional median-unbiased estimator mitigates the bias issue of the naive estimator.

\begin{table}[H]
\begin{adjustbox}{width=\columnwidth,center}
    \begin{tabular}{lrrrrrrrrrr}
    \toprule
          & \multicolumn{5}{c}{\textbf{Normal}}                &       \multicolumn{5}{c}{\textbf{Chi-squared}} \\ \hline 
   \hfill $n$   & 100   & 300  & 500  & 700  & 900    & 100   & 300  & 500   & 700  & 900  \\ 
	Sel Prob &  0.064  & 0.132   &  0.177  & 0.242   &  0.306   &  0.030  &  0.091  & 0.138  & 0.146  & 0.286  \\
\\
          & \multicolumn{10}{c}{\textbf{Conditional Coverage}} \\ 
Cond CI  &  0.928  & 0.904  & 0.915  & 0.909 & 0.911  & 0.920  & 0.923  & 0.920  & 0.920  & 0.919  \\
Naive CI     & 0.234  & 0.603  & 0.715  & 0.791  & 0.838  & 0.140  & 0.600   & 0.778   & 0.772   & 0.849   \\
Bonf CI   & 0.821 & 0.915 & 0.943  & 0.952 & 0.966  & 0.926  & 0.949  & 0.965  & 0.965  & 0.972  \\
 \\
           & \multicolumn{10}{c}{\textbf{Conditional Median CI Length}} \\
Cond CI  & 1.211 & 0.631 & 0.503  & 0.383  & 0.317 & 1.855 & 0.806 & 0.541  & 0.496 & 0.327  \\
Naive CI  &  0.327 & 0.189  & 0.147  & 0.124  & 0.109 & 0.319  & 0.187  & 0.146  & 0.123  & 0.109  \\
Bonf CI    & 0.463  & 0.268  & 0.207  & 0.175  & 0.155 & 0.451  & 0.265  & 0.206  & 0.175 & 0.154 \\
 \\ 
          & \multicolumn{10}{c}{\textbf{Conditional Median Bias}} \\
Cond Est  & -0.034  & -0.027  & -0.031   & -0.022  & -0.018  &  -0.159  & -0.051  & -0.039 & -0.020  & -0.019   \\
Naive Est  &  0.141  & 0.076  & 0.051  & 0.039  & 0.031 & 0.155  & 0.085  & 0.054  & 0.056  & 0.030   \\
\bottomrule
    \end{tabular}
\end{adjustbox}
      \caption{Performance under Holm one-sided selection with FWER = $0.1$ and $\alpha=0.1$}

\footnotesize{Note: ``Normal" and ``Chi-squared" denote Designs I and II, respectively. ``Sel Prob" reports the selection probabilities of the first hypothesis under different sample sizes.  ``Cond CI" and ``Cond Est" denote the conditional CI and median unbiased estimator proposed in this paper, respectively.  ``Naive CI" and ``Naive Est" denote the conventional (unconditional) CI and estimator,  respectively,  while ``Bonf CI" denotes the conventional (unconditional) CI with Bonferroni correction. }
        \label{tab:further_simu_1}%  
\end{table}

\printbibliography

@article{leiner2025data,
  title={Data fission: splitting a single data point},
  author={Leiner, James and Duan, Boyan and Wasserman, Larry and Ramdas, Aaditya},
  journal={Journal of the American Statistical Association},
  volume={120},
  number={549},
  pages={135--146},
  year={2025},
  publisher={Taylor \& Francis}
}

@article{ziliak2004size,
	author = {Stepen T. Ziliak and Deirdre N. McCloskey},
	journal = {Journal of Socio-Economics},
	number = {5},
	pages = {527-546},
	title = {Size matters: the standard error of regressions in the American Economic Review},
	volume = {33},
	year = {2004}}

@article{harvey2016and,
	author = {Harvey, Campbell R and Liu, Yan and Zhu, Heqing},
	journal = {The Review of Financial Studies},
	number = {1},
	pages = {5--68},
	publisher = {Oxford University Press},
	title = {{\ldots} and the cross-section of expected returns},
	volume = {29},
	year = {2016}}

@article{heath2023reusing,
	author = {Heath, Davidson and Ringgenberg, Matthew C and Samadi, Mehrdad and Werner, Ingrid M},
	journal = {The Journal of Finance},
	number = {4},
	pages = {2329--2364},
	publisher = {Wiley Online Library},
	title = {Reusing natural experiments},
	volume = {78},
	year = {2023}}

@article{harvey2020evaluation,
	author = {Harvey, Campbell R and Liu, Yan and Saretto, Alessio},
	journal = {The Review of Asset Pricing Studies},
	number = {2},
	pages = {199--248},
	publisher = {Oxford University Press},
	title = {An evaluation of alternative multiple testing methods for finance applications},
	volume = {10},
	year = {2020}}

@article{kivaranovic2024,
	author = {Danijel Kivaranovic and Hannes Leeb},
	doi = {10.1214/24-EJS2232},
	journal = {Electronic Journal of Statistics},
	pages = {1677-1701},
	title = {A (tight) Upper Bound for the Length of Confidence Intervals with Conditional Coverage},
	volume = {18},
	year = {2024}}

@article{kivaranovic2021,
	author = {Kivaranovic ,Danijel and Leeb ,Hannes},
	doi = {10.1080/01621459.2020.1732989},
	isbn = {0162-1459},
	journal = {Journal of the American Statistical Association},
	journal1 = {Journal of the American Statistical Association},
	number = {534},
	pages = {845--857},
	title = {On the Length of Post-Model-Selection Confidence Intervals Conditional on Polyhedral Constraints},
	type = {doi: 10.1080/01621459.2020.1732989},
	url = {https://doi.org/10.1080/01621459.2020.1732989},
	volume = {116},
	year = {2021},
	year1 = {2021}}

@article{duy2023,
	author = {Duy, Vo Nguyen Le and Takeuchi, Ichiro},
	doi = {10.1007/s10463-022-00837-3},
	isbn = {1572-9052},
	journal = {Annals of the Institute of Statistical Mathematics},
	number = {1},
	pages = {127--157},
	title = {Exact statistical inference for the Wasserstein distance by selective inference},
	url = {https://doi.org/10.1007/s10463-022-00837-3},
	volume = {75},
	year = {2023}}

@article{terada2023,
	author = {Terada, Yoshikazu and Shimodaira, Hidetoshi},
	doi = {10.1007/s10463-022-00838-2},
	isbn = {1572-9052},
	journal = {Annals of the Institute of Statistical Mathematics},
	number = {1},
	pages = {99--125},
	title = {Selective Inference after Feature Selection via Multiscale Bootstrap},
	url = {https://doi.org/10.1007/s10463-022-00838-2},
	volume = {75},
	year = {2023}}

@article{watanabe2021,
	author = {Chihiro Watanabe and Taiji Suzuki},
	doi = {10.1214/21-EJS1853},
	journal = {Electronic Journal of Statistics},
	number = {1},
	pages = {3137 -- 3183},
	publisher = {Institute of Mathematical Statistics and Bernoulli Society},
	title = {Selective inference for latent block models},
	url = {https://doi.org/10.1214/21-EJS1853},
	volume = {15},
	year = {2021}}

@article{zhang2022post,
	author = {Zhang, Dongliang and Khalili, Abbas and Asgharian, Masoud},
	doi = {10.1214/22-SS135},
	journal = {Statistics Surveys},
	pages = {86--136},
	title = {Post-Model-Selection Inference in Linear Regression Models: An Integrated Review},
	volume = {16},
	year = {2022}}

@article{lockhart_significance_2014,
	author = {Lockhart, Richard and Taylor, Jonathan and Tibshirani, Ryan J. and Tibshirani, Robert},
	doi = {10.1214/13-AOS1175},
	issn = {0090-5364},
	journal = {Annals of Statistics},
	number = {2},
	pages = {413--468},
	title = {A {Significance} {Test} for the {Lasso}},
	url = {https://www.jstor.org/stable/43556287},
	urldate = {2024-08-05},
	volume = {42},
	year = {2014}}

@article{tibshirani_exact_2016,
	author = {Tibshirani, Ryan J. and Taylor, Jonathan and Lockhart, Richard and Tibshirani, Robert},
	doi = {10.1080/01621459.2015.1108848},
	issn = {0162-1459},
	journal = {Journal of the American Statistical Association},
	number = {514},
	pages = {600--620},
	title = {Exact {Post}-{Selection} {Inference} for {Sequential} {Regression} {Procedures}},
	url = {https://doi.org/10.1080/01621459.2015.1108848},
	urldate = {2024-07-08},
	volume = {111},
	year = {2016}}

@article{andrews_inference_2024,
	author = {Andrews, Isaiah and Kitagawa, Toru and McCloskey, Adam},
	doi = {10.1093/qje/qjad043},
	issn = {0033-5533},
	journal = {The Quarterly Journal of Economics},
	number = {1},
	pages = {305--358},
	title = {Inference on {Winners}*},
	url = {https://doi.org/10.1093/qje/qjad043},
	urldate = {2024-04-17},
	volume = {139},
	year = {2024}}

@misc{fithian_optimal_2017,
	author = {Fithian, William and Sun, Dennis and Taylor, Jonathan},
	doi = {10.48550/arXiv.1410.2597},
	note = {arXiv:1410.2597 [math, stat]},
	title = {Optimal {Inference} {After} {Model} {Selection}},
	url = {http://arxiv.org/abs/1410.2597},
	urldate = {2024-04-17},
	year = {2017}}

@article{tian_selective_2018,
	author = {Tian, Xiaoying and Taylor, Jonathan},
	doi = {10.1214/17-AOS1564},
	issn = {0090-5364},
	journal = {Annals of Statistics},
	number = {2},
	pages = {679--710},
	title = {Selective {Inference} with a {Randomized} {Response}},
	url = {https://www.jstor.org/stable/26542802},
	urldate = {2024-08-05},
	volume = {46},
	year = {2018}}

@article{jewell_testing_2022,
	author = {Jewell, Sean and Fearnhead, Paul and Witten, Daniela},
	doi = {10.1111/rssb.12501},
	issn = {1369-7412},
	journal = {Journal of the Royal Statistical Society Series B: Statistical Methodology},
	number = {4},
	pages = {1082--1104},
	title = {Testing for a {Change} in {Mean} after {Changepoint} {Detection}},
	url = {https://doi.org/10.1111/rssb.12501},
	urldate = {2024-08-05},
	volume = {84},
	year = {2022}}

@article{bachoc_valid_2019,
	author = {Bachoc, Fran{\c c}ois and Leeb, Hannes and P{\"o}tscher, Benedikt M.},
	doi = {10.1214/18-AOS1721},
	issn = {0090-5364},
	journal = {Annals of Statistics},
	number = {3},
	pages = {1475--1504},
	title = {Valid {Confidence} {Intervals} for {Post}-{Model}-{Selection} {Predictors}},
	url = {https://www.jstor.org/stable/26730430},
	urldate = {2024-08-05},
	volume = {47},
	year = {2019}}

@article{berk_valid_2013,
	author = {Berk, Richard and Brown, Lawrence and Buja, Andreas and Zhang, Kai and Zhao, Linda},
	doi = {10.1214/12-AOS1077},
	issn = {0090-5364},
	journal = {Annals of Statistics},
	number = {2},
	pages = {802--837},
	title = {Valid {Post}-{Selection} {Inference}},
	url = {https://www.jstor.org/stable/23566582},
	urldate = {2024-08-05},
	volume = {41},
	year = {2013}}

@article{holm1979simple,
	author = {Holm, Sture},
	journal = {Scandinavian Journal of Statistics},
	pages = {65--70},
	publisher = {JSTOR},
	title = {A simple sequentially rejective multiple test procedure},
	year = {1979}}

@article{romano2005stepwise,
	author = {Romano, Joseph P and Wolf, Michael},
	journal = {Econometrica},
	number = {4},
	pages = {1237--1282},
	publisher = {Wiley Online Library},
	title = {Stepwise multiple testing as formalized data snooping},
	volume = {73},
	year = {2005}}

@article{vsidak1967rectangular,
	author = {{\v{S}}id{\'a}k, Zbyn{\v{e}}k},
	journal = {Journal of the American Statistical Association},
	number = {318},
	pages = {626--633},
	publisher = {Taylor \& Francis},
	title = {Rectangular confidence regions for the means of multivariate normal distributions},
	volume = {62},
	year = {1967}}

@article{lehmann2005generalizations,
	author = {Lehmann, EL and Romano, Joseph P},
	journal = {Annals of Statistics},
	pages = {1138--1154},
	publisher = {JSTOR},
	title = {Generalizations of the Familywise Error Rate},
	year = {2005}}

@article{benjamini1995controlling,
	author = {Benjamini, Yoav and Hochberg, Yosef},
	journal = {Journal of the Royal statistical society: series B (Methodological)},
	number = {1},
	pages = {289--300},
	publisher = {Wiley Online Library},
	title = {Controlling the false discovery rate: a practical and powerful approach to multiple testing},
	volume = {57},
	year = {1995}}

@article{benjamini2001control,
	author = {Benjamini, Yoav and Yekutieli, Daniel},
	journal = {Annals of Statistics},
	pages = {1165--1188},
	publisher = {JSTOR},
	title = {The control of the false discovery rate in multiple testing under dependency},
	year = {2001}}

@article{lee2016exact,
	author = {Lee, Jason D. and Sun, Dennis L. and Sun, Yuekai and Taylor, Jonathan E.},
	doi = {10.1214/15-AOS1371},
	journal = {Annals of Statistics},
	number = {3},
	pages = {907--927},
	publisher = {Institute of Mathematical Statistics},
	title = {Exact post-selection inference, with application to the lasso},
	volume = {44},
	year = {2016}}

@article{bentley1979algorithms,
	author = {Bentley and Ottmann},
	journal = {IEEE Transactions on Computers},
	number = {9},
	pages = {643--647},
	publisher = {IEEE},
	title = {Algorithms for reporting and counting geometric intersections},
	volume = {100},
	year = {1979}}

@article{rosenblatt1952remarks,
	author = {Rosenblatt, Murray},
	doi = {10.1214/aoms/1177729394},
	journal = {The Annals of Mathematical Statistics},
	number = {3},
	pages = {470--472},
	publisher = {JSTOR},
	title = {Remarks on a Multivariate Transformation},
	volume = {23},
	year = {1952}}

@article{benjamini2005false,
	author = {Benjamini, Yoav and Yekutieli, Daniel},
	journal = {Journal of the American Statistical Association},
	number = {469},
	pages = {71--81},
	publisher = {Taylor \& Francis},
	title = {False discovery rate--adjusted multiple confidence intervals for selected parameters},
	volume = {100},
	year = {2005}}

@article{list2019multiple,
	author = {List, John A and Shaikh, Azeem M and Xu, Yang},
	journal = {Experimental Economics},
	pages = {773--793},
	publisher = {Springer},
	title = {Multiple hypothesis testing in experimental economics},
	volume = {22},
	year = {2019}}

@article{karlan2007does,
	author = {Karlan, Dean and List, John A},
	journal = {American Economic Review},
	number = {5},
	pages = {1774--1793},
	publisher = {American Economic Association},
	title = {Does price matter in charitable giving? Evidence from a large-scale natural field experiment},
	volume = {97},
	year = {2007}}

@article{ioannidis2017power,
	author = {Ioannidis, John PA and Stanley, TD and Doucouliagos, Hristos and others},
	journal = {Economic Journal},
	number = {605},
	pages = {236--265},
	publisher = {Royal Economic Society},
	title = {The Power of Bias in Economics Research},
	volume = {127},
	year = {2017}}

@article{couttenier2019violent,
	author = {Couttenier, Mathieu and Petrencu, Veronica and Rohner, Dominic and Thoenig, Mathias},
	journal = {American Economic Review},
	number = {12},
	pages = {4378--4425},
	publisher = {American Economic Association 2014 Broadway, Suite 305, Nashville, TN 37203},
	title = {The violent legacy of conflict: evidence on asylum seekers, crime, and public policy in Switzerland},
	volume = {109},
	year = {2019}}

@article{roth2022pretest,
	author = {Roth, Jonathan},
	journal = {American Economic Review: Insights},
	number = {3},
	pages = {305--322},
	publisher = {American Economic Association 2014 Broadway, Suite 305, Nashville, TN 37203},
	title = {Pretest with caution: Event-study estimates after testing for parallel trends},
	volume = {4},
	year = {2022}}

@article{kasy2019uniformity,
	author = {Kasy, Maximilian},
	journal = {Journal of Econometric Methods},
	number = {1},
	pages = {20180001},
	publisher = {De Gruyter},
	title = {Uniformity and the delta method},
	volume = {8},
	year = {2019}}

@article{giglio2021thousands,
	author = {Giglio, Stefano and Liao, Yuan and Xiu, Dacheng},
	journal = {The Review of Financial Studies},
	number = {7},
	pages = {3456--3496},
	publisher = {Oxford University Press},
	title = {Thousands of alpha tests},
	volume = {34},
	year = {2021}}

@article{goeman2024selection,
	author = {Goeman, Jelle J and Solari, Aldo},
	journal = {Biometrika},
	number = {2},
	pages = {393--416},
	publisher = {Oxford University Press},
	title = {On selection and conditioning in multiple testing and selective inference},
	volume = {111},
	year = {2024}}

@article{mccloskey2024hybrid,
	author = {McCloskey, Adam},
	journal = {Biometrika},
	number = {1},
	pages = {109--127},
	publisher = {Oxford University Press},
	title = {Hybrid confidence intervals for informative uniform asymptotic inference after model selection},
	volume = {111},
	year = {2024}}

@article{zrnic2024locally,
	author = {Zrnic, Tijana and Fithian, William},
	journal = {The Annals of Statistics},
	number = {3},
	pages = {1227--1253},
	publisher = {Institute of Mathematical Statistics},
	title = {Locally simultaneous inference},
	volume = {52},
	year = {2024}}

@article{kvarven2020comparing,
	author = {Kvarven, Amanda and Str{\o}mland, Eirik and Johannesson, Magnus},
	journal = {Nature Human Behaviour},
	number = {4},
	pages = {423--434},
	publisher = {Nature Publishing Group UK London},
	title = {Comparing meta-analyses and preregistered multiple-laboratory replication projects},
	volume = {4},
	year = {2020}}

@article{andrews2019identification,
	author = {Andrews, Isaiah and Kasy, Maximilian},
	journal = {American Economic Review},
	number = {8},
	pages = {2766--2794},
	publisher = {American Economic Association 2014 Broadway, Suite 305, Nashville, TN 37203},
	title = {Identification of and correction for publication bias},
	volume = {109},
	year = {2019}}

@article{fama2015five,
	author = {Fama, Eugene F. and French, Kenneth R.},
	journal = {Journal of Financial Economics},
	number = {1},
	pages = {1--22},
	publisher = {Elsevier},
	title = {A five-factor asset pricing model},
	volume = {116},
	year = {2015}}

@article{benjamini2010simultaneous,
	author = {Benjamini, Yoav},
	journal = {Biometrical Journal},
	number = {6},
	pages = {708--721},
	publisher = {Wiley Online Library},
	title = {Simultaneous and selective inference: Current successes and future challenges},
	volume = {52},
	year = {2010}}

@article{fan2013large,
	author = {Fan, Jianqing and Liao, Yuan and Mincheva, Martina},
	journal = {Journal of the Royal Statistical Society Series B: Statistical Methodology},
	number = {4},
	pages = {603--680},
	publisher = {Oxford University Press},
	title = {Large covariance estimation by thresholding principal orthogonal complements},
	volume = {75},
	year = {2013}}

@article{bai2002determining,
	author = {Bai, Jushan and Ng, Serena},
	journal = {Econometrica},
	number = {1},
	pages = {191--221},
	publisher = {Wiley Online Library},
	title = {Determining the number of factors in approximate factor models},
	volume = {70},
	year = {2002}}

@article{dzemski2025location,
	author = {Dzemski, Andreas and Okui, Ryo and Wang, Wenjie},
	journal = {arXiv preprint arXiv:2502.20917},
	title = {Location Characteristics of Conditional Selective Confidence Intervals via Polyhedral Methods},
	year = {2025}}

@article{kivaranovic2021length,
	author = {Kivaranovic, Danijel and Leeb, Hannes},
	journal = {Journal of the American Statistical Association},
	number = {534},
	pages = {845--857},
	publisher = {Taylor \& Francis},
	title = {On the length of post-model-selection confidence intervals conditional on polyhedral constraints},
	volume = {116},
	year = {2021}}

@manual{Rpoet2016,
	author = {Jianqing Fan and Yuan Liao and Martina Mincheva},
	note = {R package version 2.0},
	title = {POET: Principal Orthogonal ComplEment Thresholding (POET) Method},
	url = {https://CRAN.R-project.org/package=POET},
	year = {2016}}

@article{sarfati2025post,
	author = {Sarfati, Reca and Vilfort, Vod},
	journal = {arXiv preprint arXiv:2510.02507},
	title = {``Post'' Pre-Analysis Plans: Valid Inference for Non-Preregistered Specifications},
	year = {2025}}
\end{refsection}
\end{document}